\documentclass[12pt,preprint]{aastex}

\usepackage{epstopdf}
\usepackage{graphicx}
\shorttitle{}
\shortauthors{Nesvorn\'y et al.}

\begin{document}
\baselineskip 19.pt

\title{Early Bombardment of the Moon: Connecting the Lunar Crater Record to the Terrestrial Planet
  Formation}

\author{David Nesvorn\'y$^1$, Fernando V. Roig$^2$, David Vokrouhlick\'y$^3$, William F. Bottke$^1$, \\ 
Simone Marchi$^1$, Alessandro Morbidelli$^4$, Rogerio Deienno$^1$} 
\affil{(1) Department of Space Studies, 
Southwest Research Institute, 1050 Walnut St., Suite 300,  Boulder, CO 80302, United States}
\affil{(2) Observat\'orio Nacional, Rua Gal. Jose Cristino 77, Rio de Janeiro, RJ 20921-400, Brazil}
\affil{(3) Institute of Astronomy, Charles University, V Hole\v{s}ovi\v{c}k\'ach 2, CZ–18000 Prague
  8, Czech Republic}
\affil{(4) Laboratoire Lagrange, UMR7293, Universit\'e C\^ote d'Azur, CNRS, Observatoire de la 
C\^ote d'Azur, Bouldervard de l'Observatoire, 06304, Nice Cedex 4, France}
\begin{abstract}
  The lunar crater record features $\sim 50$ basins. The radiometric dating of Apollo samples indicates 
  that the Imbrium basin formed relatively late -- from the planet formation perspective
  -- some $\simeq 3.9$ Ga.  Here we develop a dynamical model for impactors in the inner solar system
  to provide context for the interpretation of the lunar crater record. The contribution of cometary
  impactors is found to be insignificant. Asteroids produced most large impacts on the terrestrial worlds in
  the last $\simeq 3$ Gyr. The great majority of early impactors were rocky planetesimals left behind
  at $\sim 0.5$--1.5 au after the terrestrial planet accretion. The population of terrestrial planetesimals
  was reduced by disruptive collisions in the first $t \sim 20$ Myr after the gas disk dispersal. We 
  estimate that there were $\sim 4 \times 10^5$ diameter $d>10$ km bodies when the Moon formed (total 
  planetesimal mass $\sim 0.015$ $M_{\rm Earth}$ at $t \sim 50$ Myr). The early bombardment of the Moon 
  was intense. To accommodate  $\sim 50$ known basins, the lunar basins that formed before
  $\simeq 4.35$--4.41 Ga must have been erased. The late formation of Imbrium occurs with a
  $\sim 15$--35\% probability in our model. About 20 $d>10$-km 
  bodies were expected to hit the Earth between 2.5 and 3.5 Ga, which is comparable to the number of known 
  spherule beds in the late Archean. We discuss implications of our model for the lunar/Martian crater 
  chronologies, Late Veneer, and noble gases in the Earth atmosphere. 
%The Imbrium-era impactors were stored on low-eccentricity orbits in the orbital resonances with Mars 
%(e.g., 6:5, 7:6) before evolving to 1 au.
\end{abstract}

\section{Introduction}

In the standard model of terrestrial planet formation (Wetherill 1990), accretional collisions between
1 to 1000 km planetesimals lead to gradual build up of lunar- to Mars-size protoplanets that gravitationally
interact and further grow during a late stage of giant impacts (Chambers \& Wetherill 1998). The hafnium-tungsten
(Hf-W) isotopic system analysis indicates that the Moon-forming impact on proto-Earth happened relatively 
late, some $\sim 30$--150 Myr after the appearance of the first solar system solids
(Kleine \& Walker 2017; Thiemens et al. 2019, 2021; Barboni et al. 2019; Maurice et al. 2020, Kruijer et al. 2021). 
The new-born Moon was molten, gradually cooled down, and was eventually able to support impact structures 
on its surface (Meyer et al. 2010, Elkins-Tanton et al. 2011, Miljkovi\'c et al. 2021). This is time zero 
for the lunar crater record.

The early lunar bombardment was intense (Hartmann 1966): at least 40 and up to 90 lunar basins (crater
diameter $D > 300$ km) have been recognized or proposed (e.g., Wilhelms et al. 1987, Spudis 1993, Fassett et
al. 2012). Neumann et al. (2015) used gravity anomalies observed by GRAIL to report a complete list of
all $D>200$ km craters on the whole lunar surface (both the nearside and farside). In Table 1 of Neumann et 
al. (2015), there are 43 basins with the main ring diameter $D > 300$ km. From crater scaling laws appropriate 
for the lunar gravity (e.g., Johnson et al. 2016a), a diameter $d \sim 20$ km impactor is needed to excavate a 
$D = 300$ km crater.\footnote{The crater and impactor diameters are denoted $D$ and $d$, respectively.} 
Miljkovi\'c et al. (2016), modeling the GRAIL data from Neumann et al. (2015) with the \texttt{iSALE-2D} hydrocode, 
found that the lunar surface recorded $\simeq 50$ impacts of $d > 20$ km bodies (the impact speed $\sim 20$ 
km s$^{-1}$ is used here to convert the $C$ parameter reported in their Table 2 to the impactor size).
Here we use the results from Neumann et al. (2015) and Miljkovi\'c et al. (2016) as an important constraint 
on our modeling efforts.
 
At least some of the lunar basins formed relatively late. The radiometric dating of impact melts found
in the Apollo samples indicates that the Imbrium basin formed $\simeq 3.92$~Gyr ago (Ga; Zhang et al. 2019), 
that is $\simeq 650$ Myr after the first solar system solids (Bouvier et al. 2007, Burkhardt et al. 2008). 
The Imbrium basin was excavated by a $d \gtrsim 100$ km impactor (Miljkovi\'c et al. 2013, 2016; Schultz \& 
Crawford 2016). Of all the known basins, only Orientale and Schr\"odinger have
a lower accumulated density of superposed craters than Imbrium, and must therefore be (slightly) younger 
than Imbrium. The Orientale basin was produced by a $d = 50$ km (Miljkovi\'c et al. 2016) or
$d = 64$ km (Johnson et al. 2016b) impactor. A smaller, $d \simeq 20$-km body impact is responsible
for the Schr\"odinger basin ($D=312$ km; Miljkovi\'c et al. 2016).

Having (at least) three basin-forming impacts happening at $t \gtrsim 650$ Myr after the first solar system 
solids is unexpected from the planet formation perspective. In the inner solar system, where the accretion
processes have relatively short timescales ($<100$ Myr; Wetherill 1990), the impact flux should
have rapidly declined over time (Bottke et al. 2007). This has motivated the impact spike hypothesis
where it was assumed that the Imbrium-era impacts mark an epoch of enhanced bombardment (Tera et al. 1974;
Ryder 1990, 2002; Cohen et al. 2000; St\"offler \& Ryder 2001; Kring \& Cohen 2002), and prompted search 
for possible causes (e.g., Levison et al. 2001, Gomes et al. 2005). This epoch is often called 
the Late Heavy Bombardment (LHB). We avoid using this term here because LHB has too many strings attached 
to it (Hartmann 2019).

%The scale has recently tilted, however, in favor of a
%(Hartung 1974, Neukum et al. 2001, Hartmann et al. 2007). 
%The support for the no-spike hypothesis comes, most notably, from the lunar impact melts that date back
%to $\simeq 4.21$ and $\simeq 4.33$ Ga, indicating that basin-forming impacts were happening at that
%time (Norman et al. 2007, Fernandes et al. 2013, Norman \& Nemchin 2014). The dynamical models invoking
%the late instability as a cause of the presumed impact spike (Gomes et al. 2005) have serious problems
%with the survival of the terrestrial planets, excessive collisional grinding in the outer planetesimal
%disk, and the delay itself (Agnor \& Lin 2012, Kaib \& Chambers 2016, Nesvorn\'y et al. 2018).  

The goal of this paper is to develop an accurate dynamical model of asteroid, comet and planetesimal 
impactors in the inner solar system. The model for asteroid and comet impactors is taken from
previous publications (Nesvorn\'y et al. 2017a,b). We take advantage of the recent simulations
of terrestrial planet formation (Sect. 2) that faithfully match the masses, orbits and accretion 
histories of planets (Nesvorn\'y et al. 2021). Planetesimals remaining in these simulations after the Moon-forming
impact (``leftovers'') are cloned and followed for additional 1 Gyr. All impacts on the terrestrial
worlds are recorded in this interval. When combined with asteroids and comets, the new impact
chronology model can be compared to the lunar crater record. We find that a steadily declining impact
flux -- in the spirit of Hartung (1974) -- provides an adequate interpretation of available data.

Initial results from this study were reported in Nesvorn\'y et al. (2022). Here we discuss the results 
in detail and, in addition: (1) provide a thorough description of the methodology, simulations and 
impact profiles obtained under different assumptions (e.g., as a function of the number and radial 
extension of terrestrial planetesimals), (2) explain how and why our results differ from those obtained 
in previous studies (e.g., Bottke et al. 2007, Brasser et al. 2020), (3) obtain the lunar and Martian 
impact chronologies, discuss their relationship and implications for ages of different terrains, (4) 
include constraints from highly siderophile elements for the Earth, Mars, and Moon, and (5) highlight comet 
impactors as the source of noble gasses in the Earth atmosphere (Marty et al. 2016).   

\section{Terrestrial planet formation}

We take advantage of our $N$-body simulations of the standard model (Nesvorn\'y et al. 2021).
The simulations included, for consistency, the effects of radial migration and dynamical instability
of the outer planets (Tsiganis et al. 2005). The instability was assumed to have happened early, within
$\sim 10$ Myr after the protoplanetary disk dispersal (Clement et al. 2018). In the specific
  case considered here, the outer planets started in a chain of mean motion resonances (3:2 and 2:1)
  and the instability happened at $t = 5.8$ Myr after the gas disk dispersal.
  We found that the details of outer planet instability do not matter with small Mars\footnote{Mars represents
    only $\simeq 11$\% of Earth's mass. Many terrestrial planet formation models suggest 
    that a more massive planet should form at 1.2-1.8 au. This is called the small Mars problem.}
(potentially) forming
even in the case where the outer planets are placed on their present orbits at the beginning of simulations.
The model with a radially extended disk of terrestrial protoplanets, however, failed to match the tight orbital spacing
of Venus and Earth. To obtain the correct spacing, the terrestrial protoplanets must have started in a narrow
annulus (0.7--1 au; Hansen 2009) and have (about) the Mars mass to begin with (Jacobson \& Morbidelli
2014).\footnote{Meaning that the initial planetary embryos must have been about the mass of Mars.}
The tight radial spacing of Venus and Earth would be particularly easy to understand if terrestrial protoplanets 
convergently migrated toward $\sim 0.7$--1 au during the gas disk stage (Bro\v{z} et al. 2021).

%\footnote{We use the notation where $t$ runs forward in time and $T$ measured
%  looking back from today. Time $t=0$ corresponds to the formation of first solar system solids roughly
%  $T=4.57$ Gyr ago (Williams \& Cieza 2011). Based on the ages of observed protoplanetary disks, the solar
%  system gas nebula was presumably dispersed at $t_1 \sim 3$--10 Myr (Lada \& Lada XX). Later in this
%  article we estimate that the lunar crater record started at $t = t_{\rm zero} \simeq 200$ Myr or
%  $T = T_{\rm zero} \simeq 4.37$ Ga.}

In the W11e/20M model from Nesvorn\'y et al. (2021), which we will utilize here, 20 Mars-mass protoplanets 
were initially ($t=0$; marking the gas nebula dispersal) distributed in a narrow, dynamically cold annulus 
(orbital radius $0.7 < r < 1$ au). In addition, 1000 planetesimals with the total mass of 2 $M_{\rm Earth}$ were placed 
in an extended disk ($0.3 < r < 4$ au). The radial surface density profile of planetesimals 
was set to $\Sigma(r) \propto r^{-1}$. We performed 100 simulations of the W11e/20M
model where different seeds were used to generate slightly different initial conditions. 
This produced 100 growth histories of the terrestrial planets. The results were statistically analyzed using 
different criteria, including the number and mass of planets, their radial mass concentration, the angular 
momentum deficit (AMD) –- a measure of their orbital excitation, etc.\footnote{We found that in about 20\% of 
cases, good Mars analogs formed (defined as planets with mass $0.5 < M < 2$ $M_{\rm Mars}$ and $1.2 < a < 1.8$ au) 
\textit{and} the Earth and Venus ended with approximately correct masses, orbital excitation and separation 
(see Nesvorn\'y et al. (2021) for how this was quantified).}

\begin{figure}
\epsscale{1.0}
%\hspace*{-0.0cm}\plotone{job35.pdf}
\plotone{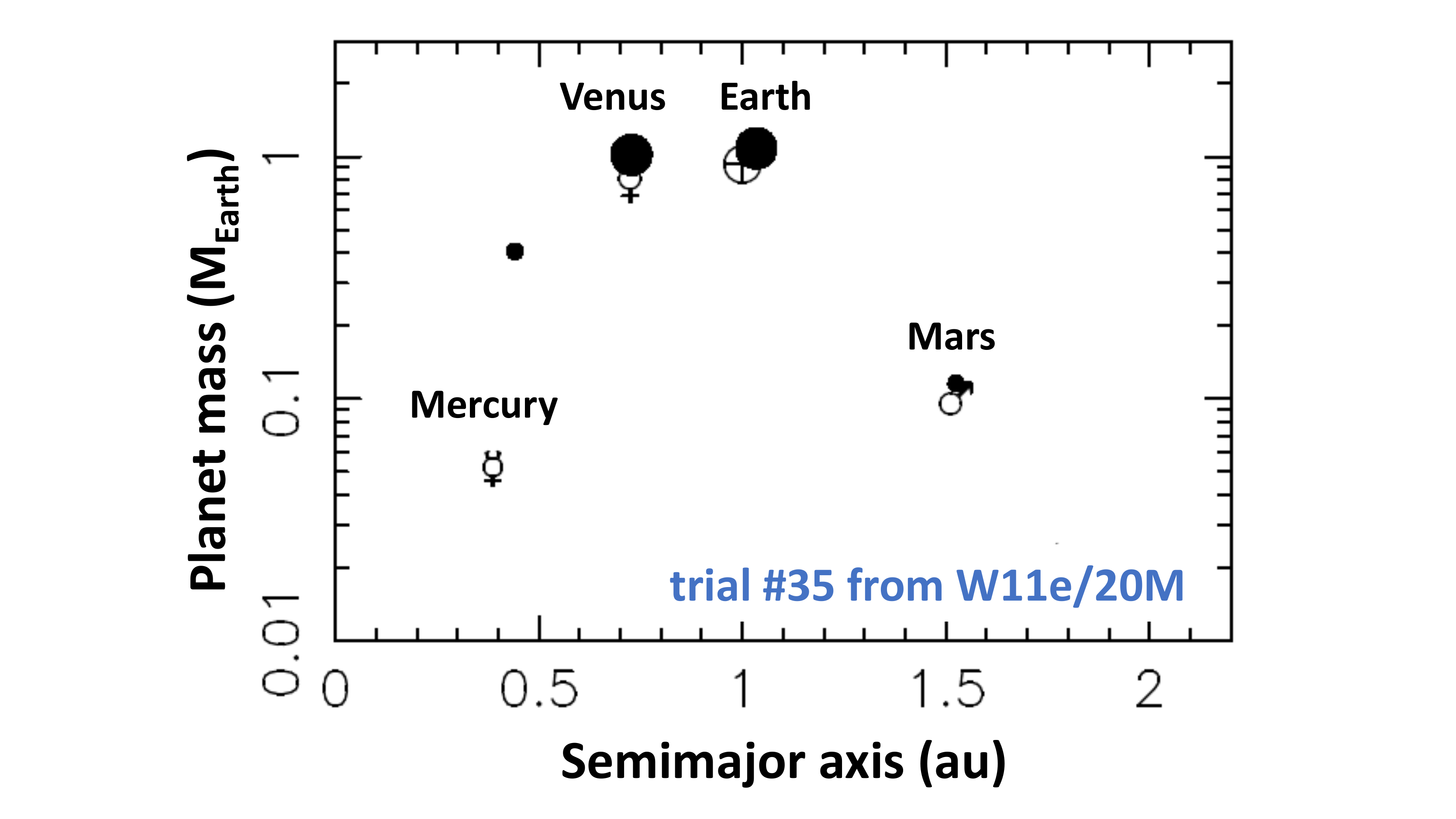}
\caption{The mass and radial distribution of terrestrial planets obtained in the best simulation of
  the W11e/20M model (Nesvorn\'y et al. 2021). The size of a dot is proportional to the planet mass. The real
  terrestrial planets are indicated by planetary symbols. The simulation failed to reproduce the correct mass of
  Mercury, but this is not a major limitation for the present study where we are interested in the impact
  histories of Venus, Earth, Mars, and the Moon.}
\label{job35}
\end{figure}

\begin{figure}
\epsscale{1.6}
%\hspace*{-5.cm}\plotone{job35_orb.pdf}
\hspace*{-5.cm}\plotone{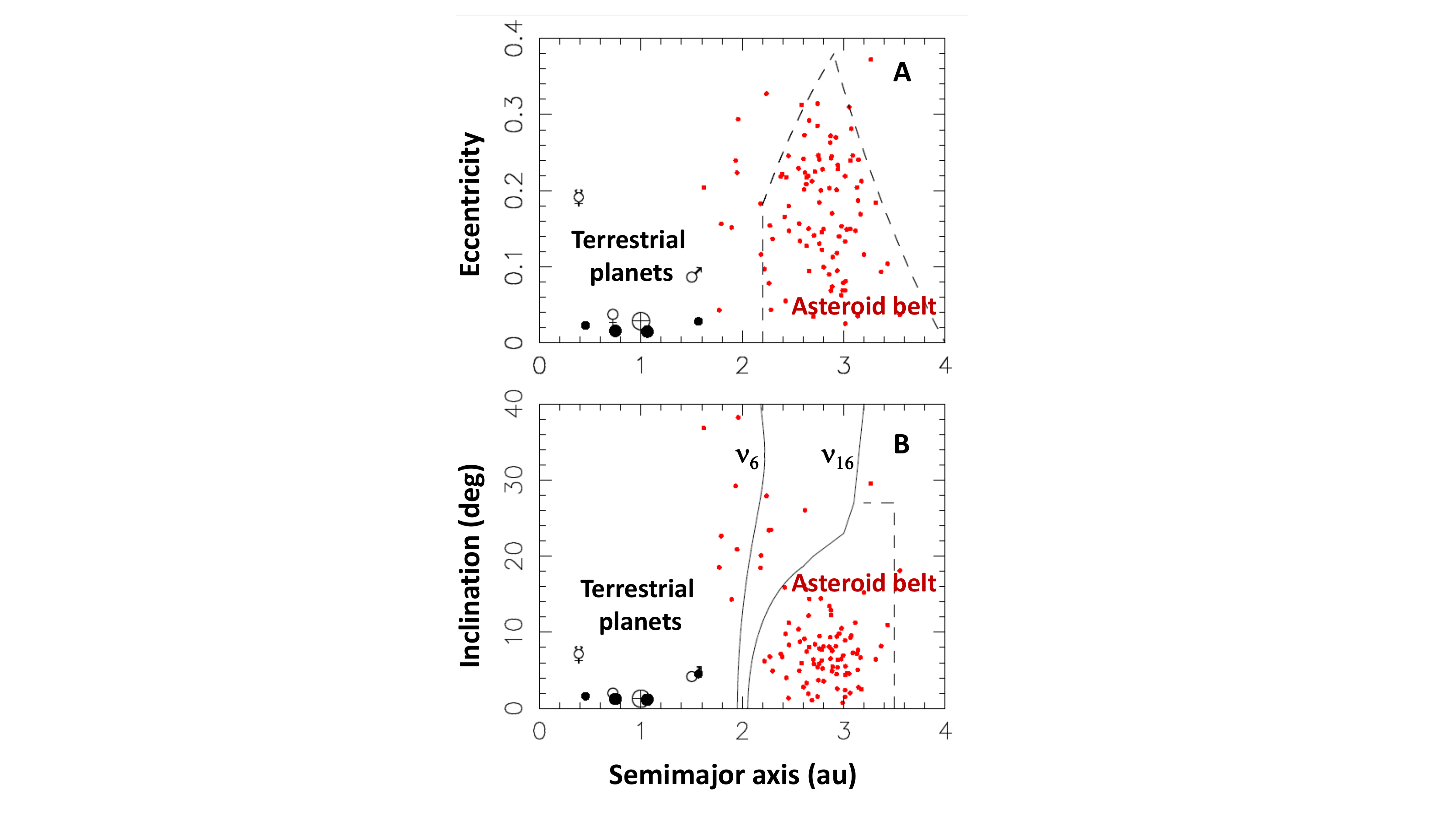}
\caption{The mean orbital elements of planets (black dots) and asteroids (red dots) produced in the best simulation
  of the W11e/20M model (Nesvorn\'y et al. 2021). The lines approximately carve out the asteroid belt
  region. The dashed lines correspond to $a=2.2$~au,
  $q=a(1-e)=1.8$ au, and $Q=a(1+e)=4$~au in panel \textbf{A}, and $a=3.5$ au and $i=25^\circ$ in panel \textbf{B}.
  The solid lines in \textbf{B} are the $\nu_6$ and $\nu_{16}$ secular resonances (plotted here for $e\sim0$).
  The mean orbits of the real terrestrial planets are indicated by planetary symbols. The simulation
  failed to reproduce the excited orbit of Mercury. }
\label{job35_orb}
\end{figure}

Here we adopt a reference case from the W11e/20M model (job \#35) that produced a particularly good match to
the terrestrial planets (Figs. \ref{job35} and \ref{job35_orb}). Venus, Earth and Mars formed in this
simulation with nearly correct masses and semimajor axis: model mass 1.017 $M_{\rm Earth}$ and $a = 0.727$ au
for Venus (0.815 $M_{\rm Earth}$ and $a = 0.723$ au), 1.083 $M_{\rm  Earth}$ and $a = 1.034$ au for Earth
(1 $M_{\rm Earth}$ and $a = 1$ au), 0.116 $M_{\rm Earth}$ and $a = 1.525$ au for Mars (0.107 $M_{\rm Earth}$
and $a = 1.524$ au), where the numbers in parentheses list the real values for reference. The Mercury
analog ended up to be too massive, which is related to the relatively large initial mass of 
protoplanets in W11e/20M and the assumption of inelastic mergers
in Nesvorn\'y et al. (2021). The model orbits of Venus and Earth were only slightly less excited than in reality. 
The model orbit of Mars had a nearly perfect inclination (mean $i = 4.5^\circ$ vs. real
mean $i = 4.4^\circ$) and somewhat lower eccentricity (mean $e = 0.028$ vs. real mean $e = 0.069$).
The orbital structure of the asteroid belt was reproduced.

\begin{figure}
\epsscale{1.1}
%\vspace{-1.cm}\plotone{growth2.pdf}
\vspace{-1.cm}\plotone{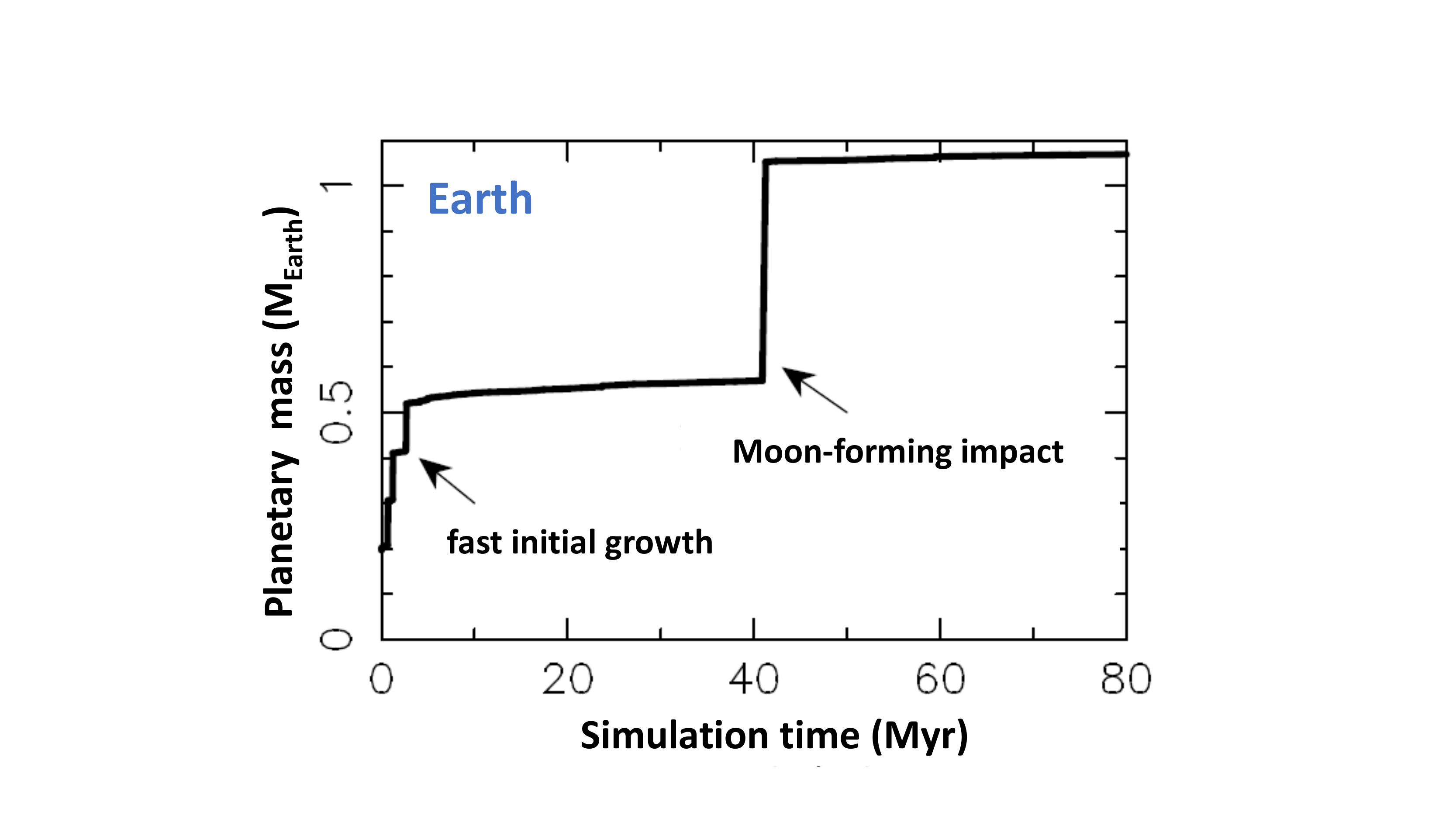}
\caption{The accretional growth of the Earth in the best case from the W11e/20M model (Nesvorn\'y et al. 
2021). The Moon-forming collision between two roughly equal-mass bodies occurred at $t=41.3$ Myr in this 
simulation.}
\label{growth}
\end{figure}

An interesting feature of the case highlighted here is related to the Earth's accretion history and the
Moon-forming impact (Fig. \ref{growth}). The initial growth of proto-Earth was fast. After accreting several
Mars-class protoplanets, the proto-Earth reached mass $\simeq 0.52$ $M_{\rm Earth}$ by $t = 3$~Myr after the 
gas disk removal. A prolonged stage of planetesimal accretion followed during which the proto-Earth modestly 
grew to $\simeq 0.57$ $M_{\rm Earth}$ by $t \simeq 40$ Myr. Then, at $t = 41.3$ Myr, an accretional collision 
between two roughly equal-mass protoplanets occurred, and the Earth mass shot up to $\simeq 1.05$ $M_{\rm Earth}$.
For comparison, Venus grew to 85\% of its final mass in only $\sim 3$~Myr after the start of the
simulation; the remaining 15\% of mass was supplied to Venus by planetesimals over an extended
time interval ($\sim 100$ Myr). This supports the model of Jacobson et al. (2017), who argued that Venus 
has not developed a persistent magnetic field because it did not experience any late energetic impacts 
that would mechanically stir the core and create a long-lasting dynamo.

The accretional growth of the Earth shown in Fig. \ref{growth} would satisfy constraints from the Hf-W and
U-Pb isotope systematics, assuming that the degree of metal-silicate equilibration was intermediate between
the full and 40\% equilibration considered in Kleine \& Walker (2017; their Fig. 8). For example, Thiemens
et al. (2019) found that the $^{182}$W excess in lunar samples can be fully explained by the decay of
now-extinct $^{182}$Hf, and suggested that the Moon formed $\sim 40$--60 Myr after the first solar system
solids.\footnote{Note that $t=0$ in the simulation time corresponds to the time of the protoplanetary gas
  dispersal some $\sim 3$--10 Myr after the first solar system solids (Williams \& Cieza 2011), and the simulation
  time $t = 41.3$~Myr is thus roughly 44--51 Myr after the first solids.} 
The low speed collision (the impact-to-escape speed ratio $v_{\rm imp}/v_{\rm esc} = 1.01$) between two nearly
equal-mass protoplanets (the impactor-to-total mass ratio $\Gamma = 0.46$) falls into the preferred regime
of Moon-forming impacts investigated in Canup (2012). In particular, the collision of two protoplanets with
$\Gamma > 0.4$ would lead to the Moon formation from a well mixed disk and would therefore imply matching
oxygen-isotope compositions of the Earth and Moon (Canup et al. 2021; but see Nakajima \& Stevenson 2014).
%Nakajima & Stevenson (2014)

%The two protoplanets involved in the Moon-forming
%collision were nearly fully formed and on their pre-impact orbits ($a = 0.95$ and 1.11 au; radial separation
%$\simeq 0.16$ au) already by $t \simeq 3$ Myr. Their orbits intermittently crossed each other during orbital
%eccentricity oscillations, but it took nearly 40 Myr before the two protoplanets actually collided.

\section{Chronology of lunar impacts}

The radiometric ages, crater counts, and size distribution extrapolations are the basis of empirical
models for impact cratering in the inner solar system  (Neukum \& Ivanov 1994; Neukum et al. 2001; 
Hartmann \& Neukum 2001; Marchi et al. 2009; Hiesinger et al. 2012, 2020; Robbins et al. 2014; Marchi 
2021). On young lunar surfaces ($T \lesssim 100$ Ma; e.g., Cone, North/South Ray, Tycho; Hiesinger et 
al. 2016), where there is a good statistic only for 
very small, $D \ll 1$~km craters, the crater counts can be extrapolated up with some assumed size distribution. The size 
distribution can be patched together from crater counts on different terrains, which implicitly assumes that the 
size distribution does not change with time, or related -- via the crater scaling laws -- to observations 
of near-Earth asteroids (NEAs; Marchi et al. 2009). The goal is to establish the chronology function, 
$N_1(T)$, defined as the number of $D>1$ km craters per km$^{2}$ (or $10^6$ km$^{2}$; crater density) 
in an area with the crater accumulation age $T$.\footnote{We use the notation where $t$ runs forward 
in time and $T$ is measured looking back from today. Time $t=0$ roughly corresponds to the formation of 
first solar system solids $T\simeq4.57$ Ga (Williams \& Cieza 2011).}  

There are almost no radiometric data for lunar terrains with intermediate ages ($1<T<3$ Ga; but see Li et al. 2021), 
which makes it difficult to establish whether the impact flux on the Moon was essentially constant for $T<3$ Ga, or 
whether it gradually declined from $T=3$ Ga to the present epoch (Hartmann et al. 2007). An independent work 
indicates that the impact flux may have increased by a factor of $\sim 2$--3 in the last $\sim 500$ Myr 
(Culler et al. 2000, Mazrouei et al. 2019; also see Kirchoff et al. 2021). Zellner \& Delano (2015) 
and Huang et al. (2018), however, demonstrated through sample analyses and numerical modeling that the 
reported increase in the lunar impact spherule record (Culler et al. 2000) is a result of sample bias.
The evidence for the impact flux increase is therefore less clear.  

For old lunar surfaces ($T>3$ Ga) with known radiometric ages 
(e.g., lunar maria, Fra Mauro/Imbrium highlands), the counts were traditionally done for large craters ($D > 1$ km), 
and these counts were then extrapolated down to estimate $N_1(T)$. In this sense, $N_1$ establishes
a common reference for young and old lunar terrains. To relieve some of the uncertainties 
arising from the extrapolation method, the crater counts on some selected terrains (e.g., Orientale) 
were performed down to $D \simeq 1$ km, and used as a benchmark (Neukum \& Ivanov 
1994).\footnote{Several studies have concluded that secondary craters are unimportant for the $\sim$ km-scale lunar 
crater population (see Bierhaus et al. 2018 for a review).}

The radiometric ages have been measured for Apollo, Luna and Chang'e samples, and for lunar meteorites (St\"offler \& 
Ryder 2001, Li et al. 2021). The main difficulty in interpreting these ages -- in terms of the cratering and geological 
processes occurring on the Moon -- is how to relate different samples to different events. For example, 
we know that major, basin-forming impacts were happening before 4 Ga (e.g., at $T \simeq 4.21$ 
and 4.34 Ga; Pidgeon et al. 2010, Grange et al. 2013, Merle et al. 2013, Norman et al. 2015), but we do 
not know which specific basins formed at that time. We therefore cannot empirically determine $N_1(T)$ 
for $T>4$~Ga, at least not with the existing data. 

The Imbrium-basin formation has a strong presence in many collected samples. According to the most accurate 
U-Pb geochronological measurements, the Imbrium basin formed at $T \simeq 3.92$ Ga (Zhang et al. 2019). 
The radiometric ages of lunar maria range between 3.1--3.7 Ga (St\"offler \& Ryder 2001). The crater counts 
for Fra Mauro/Imbrium highlands establish that $N_1(T)$ was about a factor of $\sim 10$ higher at 
$T \simeq 3.92$ Ga than during the formation of most lunar maria ($T \simeq 3.1$--3.4 Ga; Robbins et al. 2014). 
There clearly was a substantial increase in the lunar cratering rate at $T \sim 3.5$ Ga.  
 
The basic assumption of many empirical models for impact cratering is that the accumulated number of craters 
larger than $D$ since time $T$, $N(D,T)$, is a separable function of $D$ and $T$; i.e., 
$N(D,T)=N(D) N_1(T)$, where $N(D)$ is the cumulative size distribution of craters (per surface area), the so-called 
\textit{production function}, and $N_1(T)$ is the \textit{chronology function}. Neukum \& Ivanov (1994) 
approximated the production function by a polynomial (Eq. (2) and Table 1 in Neukum et al. 2001). Based on 
extrapolations with this polynomial, the $N_1$ chronology function was given as
\begin{equation}
  N_1(T) = a [\exp(b T) - 1] + c T 
\label{ni94}
\end{equation}
with $a=5.44 \times 10^{-14}$ km$^{-2}$, $b=6.93$ Gyr$^{-1}$, and $c=8.38\times10^{-4}$ Gyr$^{-1}$ km$^{-2}$.
Other published chronology functions used Eq. (\ref{ni94}) with different coefficients. Marchi et al. (2009)
constructed a production function based on the size distribution of NEAs and used it to determine 
$a=1.23 \times 10^{-15}$ km$^{-2}$, $b=7.85$ Gyr$^{-1}$, and $c=1.30\times10^{-3}$ Gyr$^{-1}$ km$^{-2}$.
Hartmann et al. (2007) added a quadratic term to Eq. (\ref{ni94}) to express a slow (linear) decline of 
the impact rate over the last $\sim 3$ Gyr. 

At least some of the differences between different works must arise from different choices of the terrains
where craters were counted and/or the size range of craters for which the counts were performed.
Robbins et al. (2014), for example, carefully carved their terrains of interest around the Apollo and Luna 
landing sites, and performed new crater counts for $D \gtrsim 1$ km craters. This has a notable 
advantage over the previous results because $N_1$ is directly established from the counts themselves 
(no size extrapolation needed). For the oldest sites, however, the crater densities for $D \sim 1$ km 
may be affected by crater saturation (Robbins et al. 2014). 

Fassett et al. (2012) used topography from the Lunar Orbiter Laser Altimeter (LOLA) on the Lunar 
Reconnaissance Orbiter (LRO) to measure the superposed impact crater distributions for 30 lunar 
basins. They reported their counts in terms of $N_{20}(T)$ -- the same as $N_1(T)$ but for $D>20$ km 
craters. For the Imbrium basin, Fassett et al. (2012) measured $N_{20} = 30 \pm 5$ (per $10^6$ km$^2$) 
(later revised to $N_{20} = 26 \pm 5$ in Orgel et al. 2018). This is an important anchor 
of $N_{20}(T)$ for a $T \simeq 3.92$--Gyr old lunar terrain (Zhang et al. 2019).

%$3.79\times10^7$ km$^2$ is the surface of the Moon

\section{Modeling the impact flux in the inner solar system}

There are at least three major source populations of impactors in the inner solar system: (i) leftover 
planetesimals in the terrestrial planet zone (0.3--1.75 au), (ii) main-belt asteroids (1.75--4 au; the 
range given here includes the Extended belt or E-belt for short, Bottke et al. 2012),\footnote{The
  E-belt, representing the inner extension of the main asteroid belt at 1.75--2 au, was hypothesized
  by Bottke et al. (2012) to be an important source of terrestrial and lunar impactors. There are presently
  no asteroids in this region except for Hungarias with $i=15^\circ$--30$^\circ$. The E-belt population on
  low-inclination orbits was presumably cleared during the giant planet migration/instability.}
  and (iii) comets. Here we first discuss 
the results for (ii) and (iii) that were taken from previous work (Sects. 4.1 and 4.2). The model 
for leftover planetesimals is described in Sect. 4.3.

\subsection{Asteroids}

In Nesvorn\'y et al. (2017a), we published a dynamical model for asteroid impactors. The model used the same 
setup for the radial migration and instability of the outer planets as the terrestrial planet formation model 
described in Sect. 2 (the instability happens at $t=5.8$ Myr after the gas disk dispersal). To start with, 
the terrestrial planets were placed on the low-eccentricity and low-inclination 
orbits. The surface density profile of asteroids was assumed to follow $\Sigma(r) \propto r^{-1}$. The radial profile was 
smoothly extended from $r>2$ au, where the model can be calibrated on observations of main belt asteroids 
(see below), to $r=1.75$--2 au. This fixed the initial number of bodies in the now largely extinct E-belt 
(Bottke et al. 2012). The results had large statistics (50,000 model asteroids) and full temporal coverage 
(integration time $t=0$--4.57 Gyr). 

The flux of asteroid impactors was calibrated from today's asteroid belt. We showed that 
the model distribution at the simulated time $t=4.57$ Gyr (i.e., at the present epoch) was a good match to the 
orbital distribution of asteroids. The number of model $d>10$ km asteroids at $t=4.57$ Gyr was 
set to be equal to the number of $d>10$-km main-belt asteroids ($\simeq 8200$), as measured by 
Wide-field Infrared Survey Explorer (WISE; Mainzer et al. 2019). When propagated backward in time -- 
using the simulation results -- this provided the number of asteroids and asteroid impactors 
over the whole solar system history.

The size distribution of main-belt asteroids was used to set up the model size distribution. 
The size distribution is relatively steep for $d\simeq10$--15 km (the cumulative power index $\gamma \simeq -2$), flattens 
for $d=30$--50 km ($\gamma \simeq -1$), and becomes much steeper for $d>100$~km ($\gamma \simeq -3$). 
The adopted size distribution allowed us to scale the absolutely calibrated results for $d>10$ km impactors 
to any impactor size. For example, there are $\simeq 3.5$ times fewer impactors with $d>20$ km, $\simeq8.9$ 
times fewer impactors with $d>50$  km, and $\simeq 26$ times fewer impactors with $d>100$ km. The collisional
evolution of asteroids during the early epochs was not accounted for in the model (see Nesvorn\'y et al. 
2017a for a discussion); here we are mainly interested in the lunar crater record that likely does not 
reach back to these early stages (Sect. 8).
  
%The current size distribution of asteroids is some combination of the initial distribution of planetesimals
%at 1.7-4 au and the effects of collisional grinding (Bottke et al. 2005b, Morbidelli et al. 2009). It is
%possible, for example, that the initial size distribution was defficient in small planetesimals, as expected
%if asteroids formed by the streaming instability (Youdin \& Goodman 2005, Johansen et al. 2007, Morbidelli
%et al. 2009). In this case, the early impact record from asteroids would show a paucity of small impactors as
%well. Whatever the initial size distribution was, however, the effects of collisional grinding are expected
%to be the most important when the asteroid belt was still in its early massive state, within tens of millions
%of years after the gas disk dispersal (Bottke et al. 2005b; early dynamical instability is assumed here).
%They are expected to change the size distribution such that the distribution rapidly reaches a steady state
%similar to the present size distribution of main-belt asteroids. For this reason, to understand the asteroid
%impact record during the LHB, it is probably reasonable to use the current size distribution of main belt
%asteroids as a proxy for the size distribution of asteroid impactors.

Nesvorn\'y et al. (2017a) empirically approximated the impact profiles by a sum of three exponential 
functions, each with a different $e$-folding time. Here we find a simpler functional form 
with fewer parameters that approximates the asteroid impact flux equally well. In terms of the cumulative 
number of Earth impactors with diameters $>d$ at times $>t$, the new chronology function is  
\begin{equation}
  F(d,t) = F_1(d) \exp[ -(t/\tau)^\alpha] + F_2(d) T \ ,
  \label{flux}
\end{equation}
with  $\tau=65$ Myr, $\alpha=0.6$, $T=4570-t$ and $t$ in Myr. This expression should not be used for $t \sim 0$, 
because Eq. (\ref{flux}) does not take into account the effects of early collisional grinding.

The first term in Eq. (\ref{flux}) accounts for the dynamical decline of asteroid impactors during  
early epochs. The second term represents the constant impact flux in the last 3 Gyr. 
There are two size-dependent factors in Eq. (\ref{flux}). $F_1(d)$ follows the size distribution of   
main-belt asteroids (Bottke et al. 2005); the fit to simulation results gives $F_1(10\, {\rm km}) =225$. 
We use the factors mentioned above to scale the impact flux to any $10\leq d \leq 100$ km. Also, from 
the main belt size distribution, $F_1(1\, {\rm km}) = 3.0 \times 10^4$ (Sect. 10). $F_2(d)$ is calibrated  
on modern NEAs. Nesvorn\'y et al. (2021) estimated $\sim 3$ impacts of $d>10$ km NEAs on the Earth per Gyr; 
we thus have $F_2(10\, {\rm km})=3\times10^{-3}$ Myr$^{-1}$. Harris \& D'Abramo (2015) and Morbidelli 
et al. (2020) estimated that $d>1$-km NEAs impact on the Earth on average every 0.75~Myr; we thus have 
$F_2(1\, {\rm km})=1.3$ Myr$^{-1}$. 

%The mean impact speeds at the time of the LHB (3.8--3.9 Ga) are slightly higher: 28, 33, 17, 27 km $s^{-1}$,
%respectively, but such small differences do not have much bearing to issues discussed in this work. 

This fully defines the historical impact flux of asteroids on the Earth.
We use the results of Nesvorn\'y et al. (2017a) to obtain the impact flux for other terrestrial worlds as
well. For the case considered here, Venus received $\simeq1.2$ times more impacts than the Earth, Mars
received $\simeq2.9$ times fewer impacts, and the Moon received $\simeq 20$ times fewer impacts (Mercury is 
not considered in this work). The overall Mars-to-Moon ratio in the number of asteroid impacts is $\simeq 7$ 
(dominated by early impacts; see Sect. 13 for Mars). For reference, Nesvorn\'y et al. (2023) reported
the Mars-to-Moon ratio of $\sim 7.5$ from modeling impacts of large (modern) NEAs.  
We use these factors to re-scale the impact flux from 
the Earth to Venus, Mars and the Moon. This is an excellent approximation of the modeling results obtained 
in Nesvorn\'y et al. (2017a).  The mean impact speeds of asteroids on Venus, Earth, Mars 
and the Moon are 28.6, 23.5, 13.7, and 21.4~km~s$^{-1}$, respectively (gravitational focusing included).

\begin{figure}
\epsscale{1.2}
%\plotone{aster.eps}
%\hspace*{4.cm}\plotone{aster.pdf}
\hspace*{4.cm}\plotone{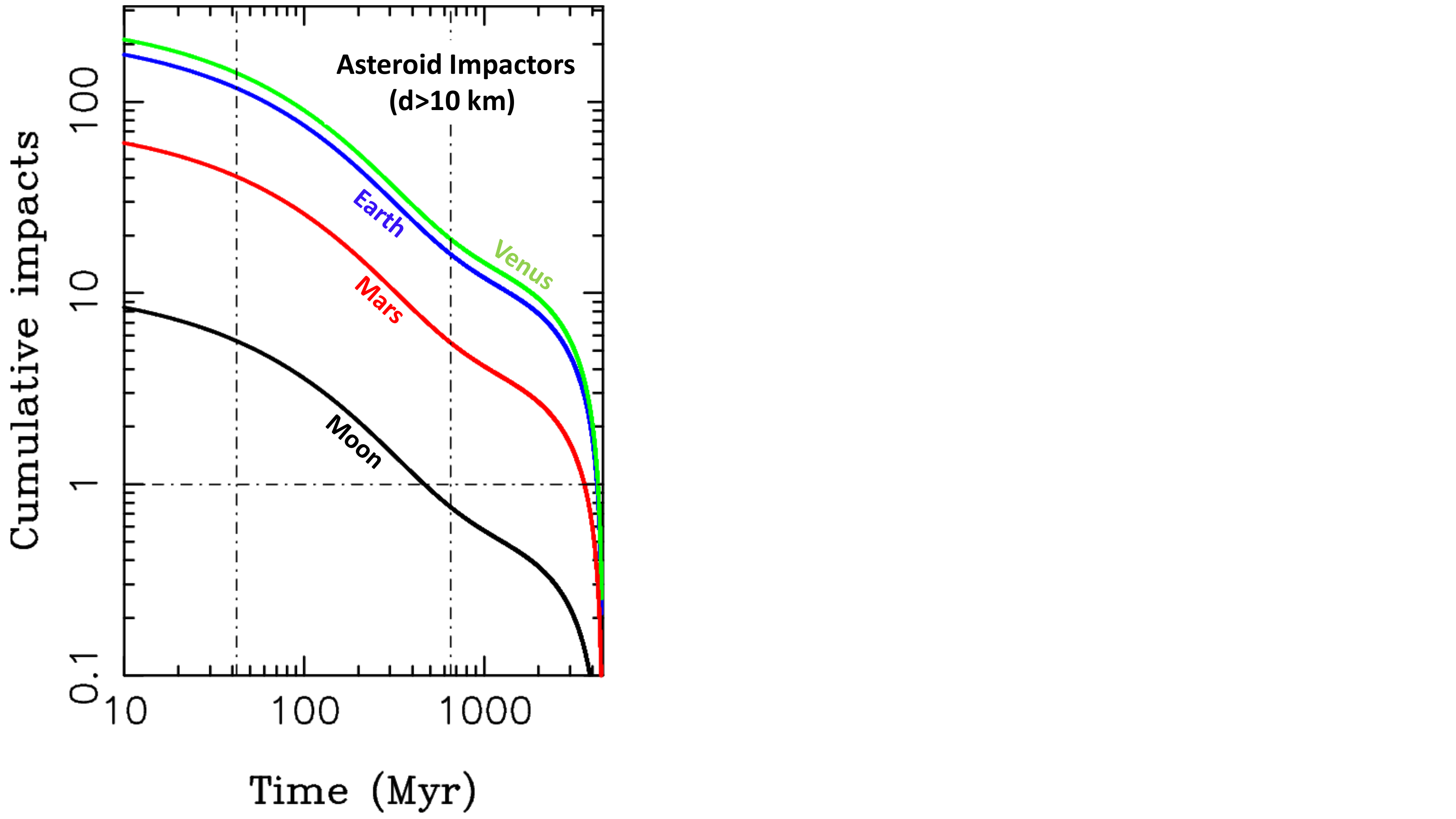}
%gr.impact_aster_figure2.f in LHB
\caption{The impact flux of $d>10$ km asteroids on the Moon (black line), Earth (blue), Venus (green), and Mars
  (red). The plot shows the accumulated number of impacts {\it since} time $t$ (i.e., on the surface with age $T$), 
  where $t=0$ ($T=4.57$ Ga) is the birth of the solar system, and $t \simeq 4.57$ Gyr ($T=0$) is the present time. 
  This is an analytic fit (Eq. \ref{flux}) to the simulation results of Nesvorn\'y et al. (2017a).
  The vertical dash-dotted lines show $t=42$ Myr and $t=650$ Myr for reference.}
\label{aster}
\end{figure}

Figure \ref{aster} shows the historical impact flux of asteroids on the terrestrial worlds. Most impacts happened 
early: for example, $\simeq 90$\% of impacts happened in the first 400 Myr. There were $\simeq 239$ $d>10$ km 
asteroid impacts on the Earth over the whole history of the solar system, $\simeq 118$ $d>10$ km impacts for 
$t>42$ Myr (after the Moon formation in the case considered here), and $\simeq 16$ $d>10$ km impacts for 
$t>650$ Myr (post-Imbrium). These fluxes imply, via the scaling factors mentioned above, only 
$118/3.5/20\sim$1.7 $d>20$ km (i.e., basin-scale) impacts on the Moon for $t>42$ Myr. This implies
that the asteroid contribution to lunar basin record was insignificant. 
\footnote{While this is true 
in the model presented here, where the asteroid depletion is relatively modest ($\lesssim90$\%; 
Nesvorn\'y et al. 2017a; also see Roig \& Nesvorn\'y (2015), Deienno et al. 2018), it has to 
be pointed out that some dynamical models imply much larger depletion ($>90$\%, e.g., Clement et al. 2019). 
The asteroid contribution to the lunar basin record would presumably be more significant in these models.} 
We discuss this issue in detail in Sect. 8.

\subsection{Comets}

A model for cometary impactors was developed in Nesvorn\'y et al. (2017b). To start with, a million 
cometesimals were distributed in a disk beyond Neptune, with Neptune on an initial orbit at 23 au.
The bodies were given low orbital eccentricities, low inclinations, and the surface density $\Sigma(r) 
\propto r^{-1}$. The disk was truncated at 30 au to assure
that Neptune stopped migrating near its current orbital radius at $\simeq$30 au (Gomes et al. 2004).
The simulations were run from $t=0$ (approximately the gas disk dispersal) to the present epoch ($t \simeq 
4.57$ Gyr). The effects of outer planet (early) migration/instability, galactic tides, and perturbations 
from passing stars were accounted for in the model. The results were shown to be consistent with the 
orbital distribution of modern comets, Centaurs and the Kuiper belt (Nesvorn\'y et al. 2017b, 2019, 
2020; Vokrouhlick\'y et al. 2019).

The size distribution of outer disk cometesimals was calibrated from the number of large comets 
observed today, the number of $d>10$ km Centaurs detected by OSSOS (Nesvorn\'y et al. 2019), the size 
distributions of Jupiter Trojans and KBOs, and from the general condition that the initial 
setup leads to plausible migration/instability histories of the outer planets (see Nesvorn\'y 2018 for 
a review). The calibration gives $\sim 6 \times 10^9$ $d>10$~km and $\sim 5 \times 10^7$ $d>100$~km 
cometesimals in the original disk. The size distribution is expected to closely follow a power law 
with the cumulative index $\gamma \simeq -2.1$ for $10<d<100$ km (Morbidelli et al. 2009b), and have a transition 
to a much steeper slope for $d > 100$ km (Bernstein et al. 2004, Fraser et al. 2014). The distribution 
is a product of the initial size distribution that was modified by collisional grinding (Nesvorn\'y 
\& Vokrouhlick\'y 2019).

\begin{figure}
\epsscale{1.2}
%\hspace*{4.cm}\plotone{comet1.pdf}
\hspace*{4.cm}\plotone{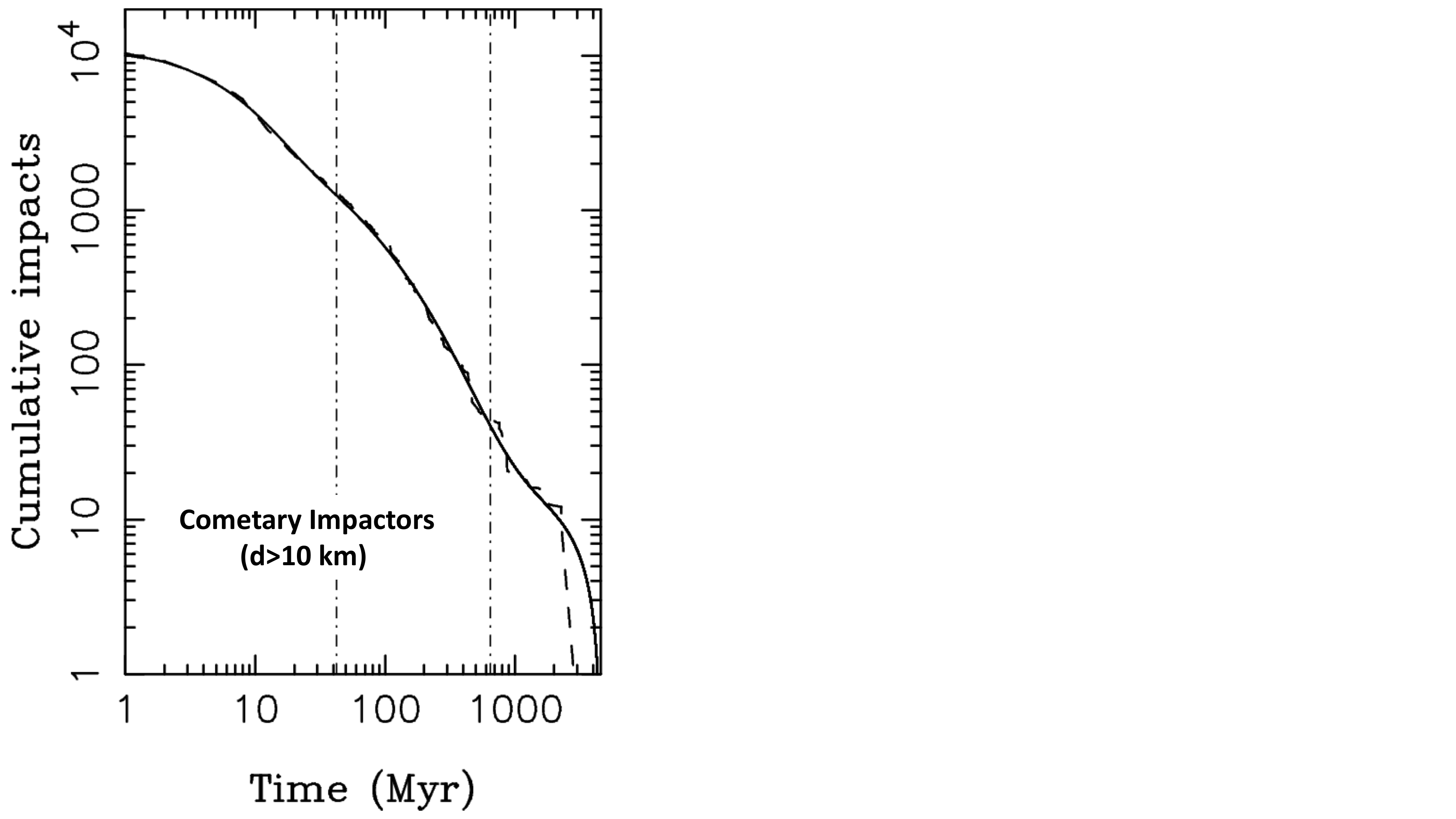}
% gr.impact_comet_figure1.f in LHB 
\caption{The impact flux of $d>10$ km comets on the Earth. The impact profile shown by the dashed line was
  obtained from the model results in Nesvorn\'y et al. (2017b). The solid line shows the analytic fit
  given in Eq. (\ref{flux2}). The fits for Venus and Mars are equally good. Disregarding comet disruptions,
  we estimate that $\sim 4$ impacts of $d>10$-km comets would happen on the Earth in the last Gyr.
  The vertical dash-dotted lines show $t=42$ Myr and $t=650$ Myr for reference.}
\label{comet1}
\end{figure}

The impact flux of comets on the terrestrial worlds was computed with the \"Opik algorithm (Bottke et al. 
1994; there were not enough planetary impacts recorded by the integrator to obtain the
flux directly from the impact statistics). The results of the \"Opik code were normalized to the
initial number of comets in the original disk (see above). The calibrated model gives us the flux of 
cometary impactors over the whole history of the solar system. In total, over 4.57 Gyr, there were 
$\simeq 10^4$ impacts of $d>10$ km comets on the Earth (Fig. \ref{comet1}). The great majority of 
cometary impacts happened early. For example, 90\% of impacts happened in the first $\simeq 55$ Myr, 
and 99\% of impacts happened in the first $\simeq 370$ Myr. The wavy decline of cometary impacts is 
related to the dispersal of the outer cometesimal disk, which was an important source of impactors 
during the early epochs, and the slower decay of the scattered disk population, which is the main 
source of modern comets.

An excellent approximation of the cumulative impact flux of comets on the Earth is
\begin{equation}
  F(d,t) = C_{\rm s}(d) \left \{  F_1 \exp [-(t/\tau_1)^{\alpha_1}] + F_2 \exp [ -(t/\tau_2)^{\alpha_2}] + 
  F_3 (4570-t) \right \}\ ,
  \label{flux2}
\end{equation}
with $F_1=F_2=6.5 \times 10^3$, $\tau_1=7$ Myr, $\alpha_1=1$, $\tau_2=13$ Myr, $\alpha_2=0.44$,
$F_3=4\times10^{-3}$ Myr$^{-1}$, $C_{\rm s}(d)=1$ for $d=10$ km, and $t$ in Myr (Fig. \ref{comet1}).
This expression should not be used for $t \sim 0$, because Eq. (\ref{flux2}) does
not take into account the effects of early collisional grinding. 

The impact profile in Eq. (\ref{flux2}) is anchored to the impact rate of comets at the present
epoch. We find, using the calibration discussed above, that $\sim 4$ $d>10$ km comets should
have impacted on the Earth in the last Gyr;\footnote{This 
would be comparable to the current impact flux of asteroids (Nesvorn\'y et al. 2021b). If comet
disruptions are accounted for, however, the impact flux of $d>10$ km comets on the Earth is 
reduced by a factor of $\sim 6$, a factor of few below the asteroid flux.} 
the average time interval between cometary impacts is $\sim 250$
Myr (this ignores comet disruption; see below). For comparison, the same method yields $\sim 44,000$
yr as the average interval between $d>10$ km impacts on Jupiter. Scaling this down with $\gamma=-2.1$,
we estimate that $d>1$ km comets impact Jupiter on average every $\sim 350$ yr. This is within
the range of historical record of impacts on (and close encounters to) Jupiter (Zahnle et al. 2003, Dones 
et al. 2009). Venus receives $\simeq 0.94$ times the number of Earth impacts,\footnote{There are more
comets with perihelion distances $q \lesssim 1$ au than with $q \lesssim 0.7$ au, and this leads 
to a slight preference for Earth impacts over Venus impacts.} Mars $\simeq 5.0$ times fewer,
and the Moon $\simeq 17$ times fewer.  This gives the Mars-to-Moon ratio of cometary impacts $\simeq 3.4$. 
The mean impact speeds of comets are higher than those of asteroid impactors: 28.2 km s$^{-1}$ for 
the Earth, 25.5 km s$^{-1}$ for the Moon, 36.8 km s$^{-1}$ for Venus, and 19.7 km s$^{-1}$ for Mars 
(focusing included).

%The collisional cross-section ratios of Earth/Moon and Earth/Mars are
%13.5 and 3.5. For comets, Earth impacts are helped by gravitational focusing and the
%impact number ratio is slightly higher than the cross-section ratio (4.7 vs. 3.6). For
%asteroids, the Earth-to-Mars impact number ratio is 2.6, somewhat lower that the cross-section
%ratio. This is probably because asteroids spend more time on Mars-crossing orbit. All this makes
%sense. The impact speeds from comets are 28.2 km/s for the Earth, 25.5 km/s for the Moon, 36.8 km/s
%for Venus, and 19.7 for Mars. These speeds are higher than those of asteroid impactors.

The physical lifetime of comets is shorter than their dynamical lifetime (Levison et al. 1997, 2002),
partly because comets become active and lose mass, but mainly because they spontaneously disrupt
(Chen \& Jewitt 1994, Levison et al. 2006, Di Sisto et al. 2009). The exact mechanism behind the observed
comet disruptions is unknown but rotational spin-up, thermal stresses and volatile-driven outbursts
are all expected to contribute. In Nesvorn\'y et al. (2017b), we tested several disruption laws and 
found that they produce comparable results. In the simplest of these models, the physical
lifetime of comets is limited to a fixed number of perihelion passages $N_{\rm p}(q)$ below $q$.
To fit the orbital inclination distribution of observed $d \sim 1$ km comets, we found $N_{\rm p}(q) \sim 500$
for $q=2.5$ au. The physical lifetime is expected to increase with comet
size. To explain the observed number of $d>10$ km Jupiter-family comets, 
$N_{\rm p}(2.5) \sim$ 5,000 is needed for $d=10$ km. This hints on a roughly linear dependence of 
$N_{\rm p}(2.5)$ on size, $N_{\rm p}(2.5) \sim 500\times(d/1\, {\rm km})$,
and $N_{\rm p}(2.5) \sim$ 50,000 for $d = 100$ km.

\begin{figure}
\epsscale{1.2}
%\hspace*{4.cm}\plotone{comet2.pdf}
\hspace*{4.cm}\plotone{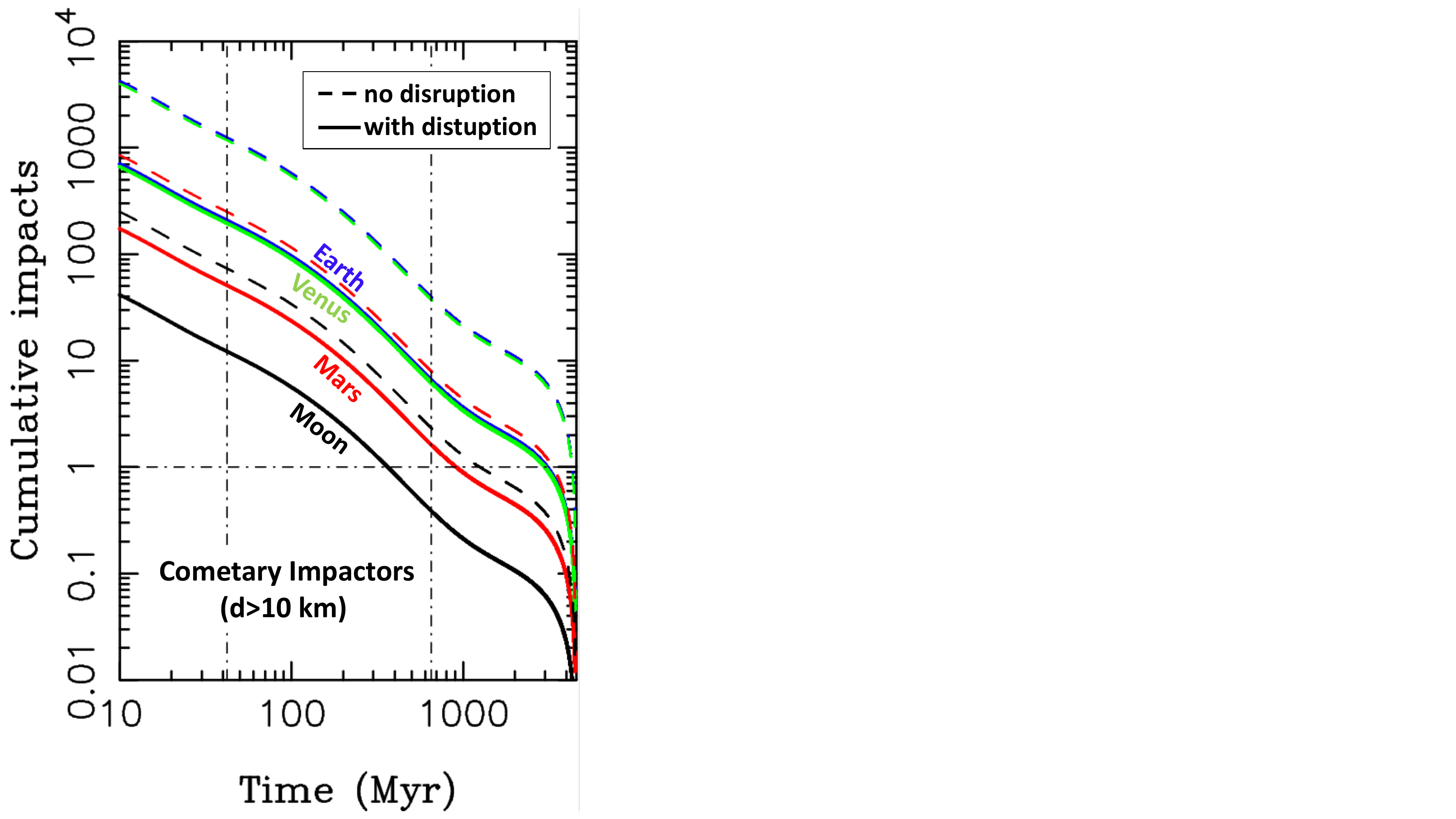}
% gr.impact_comet_figure3.f in LHB 
\caption{The impact flux of $d>10$ km comets on the Moon (black line), Earth (blue), Venus (green), and Mars
  (red). The dashed lines show the analytic fits from Eq. (\ref{flux2}) disregarding comet disruptions. The
  solid lines show the impact flux when comet disruptions are accounted for. The vertical dash-dotted lines 
  show $t=42$ Myr and $t=650$ Myr for reference.}
\label{comet2}
\end{figure}

The impact flux of comets on the terrestrial worlds is reduced, relative to the expectations
discussed above, if spontaneous comet disruptions are taken into account. We include comet disruption,
using the approximate linear dependence described above, in the \"Opik calculation of the
impact flux. The impact flux of comets on the Earth is reduced by a factor of $\sim 59$ for $d>1$ km, 
$\sim 6$ for $d>10$ km, and $\sim 1.8$ for $d>100$ km. The reduction factors for Venus and the Moon are similar.
The reduction factors for Mars are slightly lower, 36, 4.9, and 1.7, respectively. We account
for these factors in Fig. \ref{comet2}. The total number of $d>10$-km comet impacts on the Earth 
is reduced to $\sim$  1,700, of which $\sim 200$ happen for $t>42$~Myr, and $\sim 6$ happen for 
$t>650$ Myr. The Moon is expected to receive $\sim 12$ $d>10$-km comet impacts after its formation 
(assumed here to happen at $t \sim 42$ Myr with the instability at $t \lesssim 10$ Myr). This is a factor 
of $\sim 17$ below what would be needed to explain $\sim 200$ $D>150$ km lunar craters (Bottke \& Norman 
2017; a $d=10$ km impactor is assumed here to produce a $D=150$ km lunar crater). A more powerful 
source of lunar impactors is clearly needed. 

%As for the reduction factor for Jupiter. From comet\_fraction.f I get 0.34
%for 500 and $q=6.0$, but I have previously computed that the reduction factor in this case should
%be 65\% (from gr.impacts.f). This difference is most likely cause by cloning at 5.2 au. The
%comet-fraction code uses clones (the statistic is not good for original particles). But if the
%clones were created earlier, at above 5.5 au, this would add an aditional population of Jupiter
%impactors and the reduction factor would approach 1. I leave it at that.

%What is the impact rate for modern comets? I estimated that the impact probability is 0.011
%for Jupiter and $1.6 \times 10^{-6}$ for the Earth. The impact rate on Jupiter is estimated to
%be 1 $D>1$ km every $10^3$ years (Zahnle argues $10^2$ yr from data but lets ignore that). This
%would imply 1 $D>10$ km impact on the Earth roughly every 1 Gyr, about 3 times lower than what
%I get for asteroids, but once the physical disruption is taken into account, modern asteroids
%win by a factor of $\sim 10$. They win by a $\sim 10$ factor for $D>100$ km as well because the
%asteroid SFD is relatively shallow from 10 to 100 km. 

\subsection{Leftover planetesimals}

\subsubsection{$N$-body simulation setup}

We are interested in the impact flux on the terrestrial worlds after the Moon-forming impact, $t>41.3$ Myr 
in the case described in Sect. 2. With this goal in mind we recorded the orbits of planets and planetesimals 
in the original simulation at $t = 42$ Myr (i.e., shortly after the Moon-forming impact). The planetesimals 
in the asteroid belt region were ignored (see below and Sect. 4.1 for asteroids).
%; Fig. \ref{switch}).
Given that Mercury was too massive in the original simulation (Sect. 2), which could cause problems when examining 
the impact flux, we ignored Mercury in the follow-up simulations. All other planets, Venus to Neptune were
included. We assumed that the terrestrial worlds were fully grown at $t = 42$ Myr and their masses did not 
subsequently~change. The orbits of planets $t = 42$ Myr were taken from the original simulation (i.e., 
were {\it not} reset to the real orbits of the terrestrial planets).    

%This makes the largest difference for Venus that has mass about 20\% higher than it should have. The
%gravitational focusing should not be such a large factor however. [One could correct masses and do
%an additional simulation or one can even plug in planets on their present orbits, potentially
%including Mercury, to check if it makes any difference.]

%We also performed a complimentary simulation where the model planetary system was replaced by the real
%solar system planets (Venus to Neptune, no Mercury).
%They are a few advantages and disadvantage to this. On the plus side of things, the masses and
%orbits in the complimentary simulation are exactly as they should be. On the down side, introducing
%the artificial discontinuity at $t = 42$ Myr is not ideal because -- as planetary orbits change but
%planetesimal orbits do not -- this changes how planets and planetesimals interact. For example,
%the planetesimal orbits in (outer) mean motion resonances with the Earth can be relatively
%long-lived as they are phase-protected from encounters with the Earth, but they could become
%eliminated sooner if they are kicked out from resonances by the discontinuity. This could potentially
%change the impact profiles.

To increase the model statistics, each planetesimal was cloned thousand times by slightly altering the 
velocity vector ($< 10^{-6}$ fractional change). In total, we thus had 128,000 planetesimal clones at 
$t = 42$ Myr. The  $N$-body integrator known as \texttt{swift\_rmvs4} (Levison \& Duncan 1994), which is 
an efficient implementation of the Wisdom-Holman map (Wisdom \& Holman 1991), was used to follow the 
system of planets and (massless) planetesimals over 1 Gyr. The simulation was split over 1280 Ivy-Bridge cores 
of the NASA Pleiades Supercomputer. All impacts of planetesimals on planets were recorded by the 
$N$-body integrator. 

In the model considered here, planetesimals were originally distributed with a flat radial profile, 
$\Sigma(r) \propto r^{-1}$, from 0.3 to 4 au (Sect. 2). The inner part of the planetesimal disk,
$r < 1.5$ au, contained the initial mass $ \simeq 0.65$ $M_{\rm Earth}$. This may or may not
correctly approximate the physical conditions that existed in the inner solar system at $t=0$ (the time of 
gas disk dispersal). We developed a weighting scheme to test implications of different assumptions for 
the impact flux. Each planetesimal was assigned a weight, $0 \leq w \leq 1$, depending on its starting orbit 
and the adopted surface density profile. For example, to test the effects of the planetesimal disk 
truncation at some outer radius, $r_{\rm out}$, the planetesimals with $r<r_{\rm out}$ at $t=0$ were given 
weights $w=1$, and the planetesimals with $r>r_{\rm out}$ were given weights $w=0$. For each assumption, 
the impact record was constructed by monitoring the weights of impactors recorded by the $N$-body 
integrator.

%By assigning weights $w>1$ or $w<1$ we are also able to control whether the initial mass in planetesimals
%was larger or smaller, respectively, relative to the original setup.

%The mass should, in principle, be
%et at $t=0$. We find, however, that if the initial mass is $>0.1$ $M_{\rm Earth}$, the strong collisional
%grinding in the planetesimal disk reduces the mass to $\simeq 0.1$ $M_{\rm Earth}$ before the Moon-forming
%%impact (Bottke et al. 2007, Sect. X). For this reason, at least for the lunar impact record, it really does
%not matter what the initial mass was. What matters, instead, is what the planetesimal mass was \textit{at
%  the time of the Moon-forming impact}, $t\simeq42$ Myr for the case discused here. We moreover find
%that the effects of collisional grinding were minimal for $D>10$ km after the Moon-forming impact, if
%the planetesimal mass was already reduced to $\lesssim 0.1$ $M_{\rm Earth}$ during the preceeding epochs.
%For these reasons, we fix the planetesimal mass to $\simeq0.1$ $M_{\rm Earth}$ at $t\simeq42$ Myr and
%ignore the effects of collisional grinding for $t>42$ Myr.

\subsubsection{Collisional evolution model}

We used the \texttt{Boulder} code (Morbidelli et al. 2009a) to model the collisional evolution of planetesimals 
(Bottke et al. 2007). \texttt{Boulder} employs a statistical particle-in-the-box algorithm  that is capable of 
simulating collisional fragmentation of planetesimal populations. It was developed along the lines of 
other published codes (e.g., Weidenschilling et al. 1997, Kenyon \& Bromley 2001). In brief, for a given impact 
between a projectile and a target body, the algorithm computes the specific impact energy $Q$, defined as the 
kinetic energy of the projectile divided by the total (projectile plus target) mass, and compares it with
critical impact energy, $Q^*_{\rm D}$, defined as the energy per unit mass needed to disrupt and disperse 50\% 
of the target. For each collision, the mass of the largest remnant is computed from the scaling law for 
monolithic basalt (e.g., Benz \& Asphaug 1999). 
%The mass of the largest fragment and the slope of the power-law 
%size distribution of smaller fragments is set as function of $Q/Q^*_{\rm D}$ by empirical fits to the results of 
%published impact models (e.g., Durda et al. 2007).

The main input parameters of the \texttt{Boulder} code are: the (i) initial size distribution of the simulated
populations, (ii) intrinsic collision probability $p_{\rm i}$, and (iii) mean impact speed $v_{\rm i}$. The initial 
size distribution is discussed in  Sect. 5. The probabilities $p_{\rm i}(t)$ and velocities $v_{\rm i}(t)$ of mutual 
collisions between planetesimals were computed from the terrestrial planet simulation described in Sect. 2. We 
used the \"Opik algorithm (Bottke et al. 1994) and considered each pair of planetesimals at a time at each timestep.
The results were averaged over all pairs (at each timestep) to give us a time-dependent description of the 
collisional environment. The leftover planetesimals were modeled as a single population in \texttt{Boulder}
(e.g., we did not consider high-$i$ or high-$i$ orbits separately; simulations with the multi-annulus version 
of \texttt{Boulder} are left for future work). The collisional probabilities were 
initially high ($p_{\rm i} \sim 10^{-16}$ km$^{-2}$ yr$^{-1}$ 
for $r_{\rm out}=1.5$ au and $t<3$ Myr), and quickly decreased  as the planetesimal population dynamically and 
collisionally declined (Sects. 5 and 6). The impact speeds between planetesimals increased from $v_{\rm i} \sim 10$ 
km s$^{-1}$ at $t\lesssim5$~Myr to $v_{\rm i} \sim 20$ km s$^{-1}$ later on. We implemented the impact speed 
dependence of $Q^*_{\rm D}$ from Leinhardt \& Stewart (2012) (the disruption threshold shifts to higher specific 
energies when the impact speeds are higher).

\section{Collisional evolution of planetesimals}

The collisional evolution acted, along with the dynamical decay, to reduce the number of 
planetesimals in the terrestrial zone. We show below that the collisional grinding happened 
very early and produced convergence, where different (assumed) planetesimal populations at 
$t=0$ evolved to similar populations by the time of the Moon-forming impact (both in 
the shape and overall normalization of the size distribution; Bottke et al. 2007).
This result is used to approximately calibrate the number of leftover planetesimals and 
their impact profiles after Moon's formation.

\begin{figure}
  \epsscale{1.2}
  % this is gr.sfd_figure in boulder and test6 (1 Gyr vs 42 Myr)
%\hspace*{4.cm}\plotone{grind.pdf}
\hspace*{4.cm}\plotone{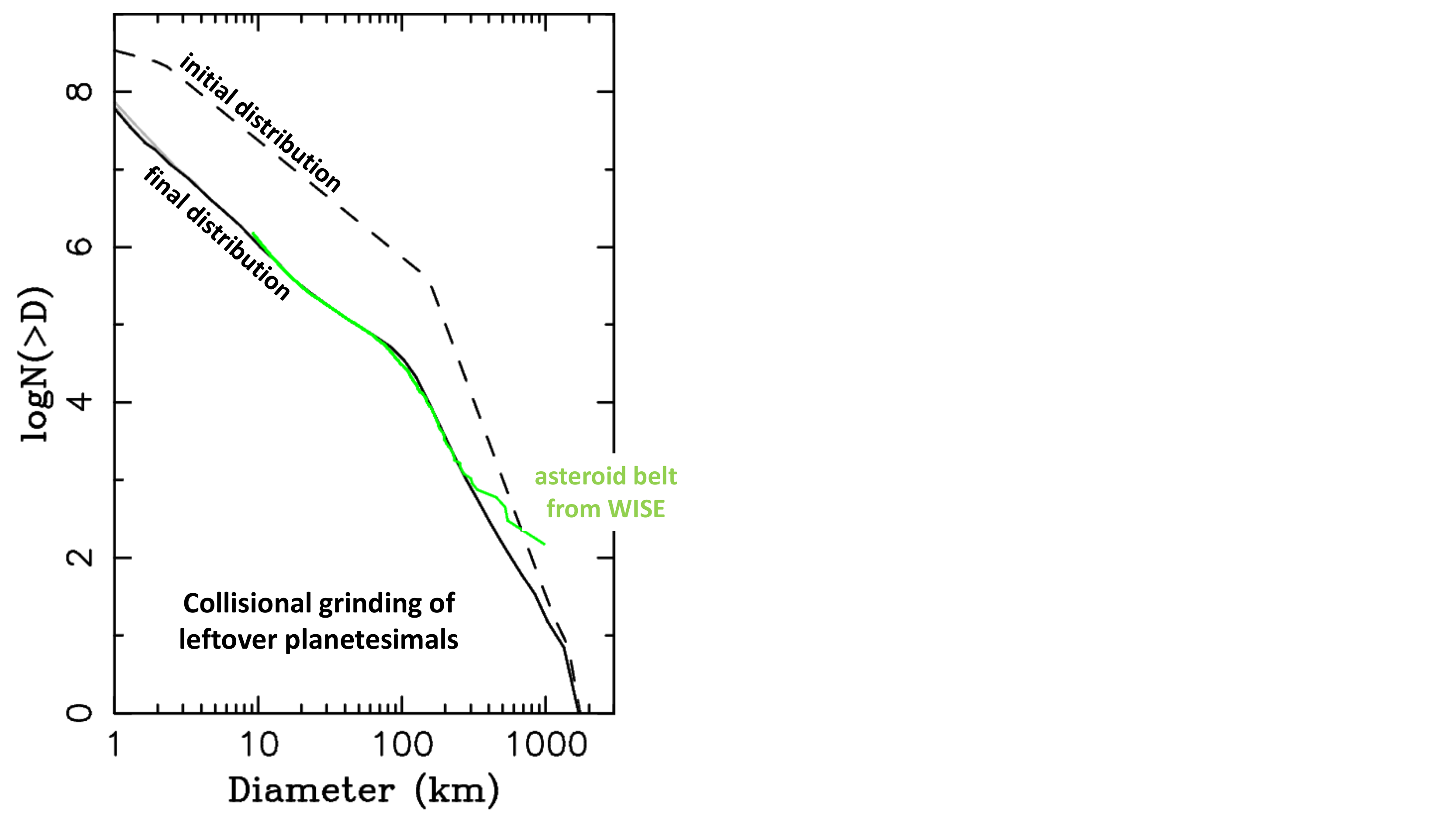}
\caption{The collisional evolution of leftover planetesimals. Initially, at $t=0$, the planetesimal disk at
  0.5--1.5 au was assumed to have the mass $M_0=1$ $M_{\rm Earth}$, and a broken power-law distribution
  with a steeper slope for $d>140$~km and a shallower slope for $d<140$~km (dashed line). 
  The black solid line shows the size distribution after 42~Myr of collisional grinding. 
  The collisional evolution for $t>42$ Myr is negligible. For reference, 
  we also show the size distribution of the main asteroid belt from WISE (green line; Mainzer et al. 2019) 
  that was vertically shifted to plot near the black solid line.}
\label{grind}
\end{figure}

We defined the total mass of leftover planetesimals at $t=0$ ($M_0$) and $t=42$ Myr 
($M_{42}$), and performed collisional simulations with \texttt{Boulder} (Sect. 4.3.2) to understand 
the relationship between $M_0$ and $M_{42}$ for different assumptions.
Figure \ref{grind} shows an example of \texttt{Boulder} run where we placed $M_0 = 1$ $M_{\rm Earth}$
in planetesimals at 0.5--1.5 au. The initial size distribution was assumed to be a broken power-law
$N(>\!\!d) \propto d^{-\gamma}$ with $\gamma=1.5$ for $d < d^*$ and $\gamma=5$ for $d > d^*$, 
and $d^* \sim 100$ km. This setup was inspired by the streaming instability model of planetesimal 
formation (Youdin \& Goodman 2005), where the new-born planetesimals have the characteristic size 
$d \sim 100$~km, and the asteroid size distribution that shows a break at $d \simeq 100$ km 
(Morbidelli et al. 2009a). We do not consider very large, Ceres-class planetesimals that may 
have formed in the terrestrial planet zone, because the lunar impact record does not provide 
direct constraints on their population. According to our tests, however, energetic impacts on 
Ceres-class and larger objects produce fragments and this can influence the size distribution of 
planetesimals in the range relevant for the lunar impact record. Similarly, lunar impactors can 
be produced in collisions between protoplanets (Wishard et al. 2022). We leave these issues 
for future studies. The very large planetesimals may have also been important for the late delivery 
of highly siderophile elements (HSEs; Sect. 14) to the Earth (Bottke et al. 2010, Marchi et al. 
2014) and Mars (Marchi et al. 2020).  

The \texttt{Boulder} code was run from $t=0$ to 1 Gyr. We found that the size distribution of 
planetesimals rapidly changed and reached an equilibrium shape by only $t \sim 20$ Myr. 
The subsequent collisional evolution was insignificant because the planetesimal population
was reduced by a large factor ($\sim 100$ by $t=42$ Myr, collisional and dynamical removals
combined). In this sense, the shape of the size distribution of leftover planetesimals at the time
of the Moon-forming impact, and any time after that, is a fossilized imprint of the intense
collisional grinding that happened in the first $\sim 20$ Myr (Bottke et al. 2007).

The equilibrium size distribution shows a break at $d \simeq 100$ km, a shallow slope for 
$d=20$--100 km, and a slightly steeper slope for $d<20$ km. It is remarkably similar to that of 
the (scaled) asteroid belt (Fig. \ref{grind}). This is an encouraging sign, because an asteroid-like
size distribution is just what is needed to explain the size distribution of ancient lunar craters 
(Str\"om et al. 2005). Minton et al. (2015), however, pointed out that the main-belt--like size 
distribution of impactors would produce too many lunar basins. This could indicate that the size 
distribution of ancient impactors with $d \gtrsim 10$ km was slightly steeper than that of today's 
asteroid belt (Sect. 14). 
%Alternatively, the lunar basins that formed before the lunar magma ocean (LMO) solidified were 
%susceptible to extreme crustal relaxation (Miljkovi\'c et al. 2021), and are not recognized today (Sect. X). 

The number of terrestrial planetesimals was strongly reduced by collisional grinding. With $M_0=1$ 
$M_{\rm Earth}$, and the initial size distribution defined above, there were $\sim 1.9 \times 10^7$ 
$d>10$-km, $6.7 \times 10^6$ $d>20$-km, and $6.9 \times 10^5$ $d>100$-km planetesimals at $t=0$. 
By $t=42$ Myr, the planetesimal population was reduced to $\sim 10^6$ $d>10$-km, $2.6 \times 10^5$ $d>20$-km,
and $3.5 \times 10^4$ $d>100$-km planetesimals ($\sim 19$--26 reduction factors).
The total mass of the planetesimal population dropped by a factor of $\sim 19$ to $M_{42} = 0.053$
$M_{\rm Earth}$. The numbers quoted above only account for the effects of collisional grinding.
In the dynamical simulations, the number of planetesimals was reduced by a factor of $\sim 3.1$
from $t=0$ to 42~Myr. Combining the dynamical and collisional removals we therefore estimate 
the overall reduction of $\sim 59$--81, and $\sim 3.2 \times 10^5$ $d>10$-km, $8.4 \times 10^4$ 
$d>20$-km, and $1.1 \times 10^4$ $d>100$-km planetesimals at $t=42$ Myr.\footnote{An implicit assumption 
here is that the dynamical loss rate is independent of size.}

The effects of collisional grinding depend on the initial mass $M_0$ (Bottke et al. 2007): stronger/weaker 
collisional grinding is expected if $M_0$  was higher/lower. We performed a large number of simulations 
with the \texttt{Boulder} code to characterize this dependence in detail. We found that the stronger 
grinding for larger initial masses leads to a situation where $M_{42}$, and the population leftover 
planetesimals at $t=42$ Myr, do {\it not} sensitively depend on $M_0$; hence the aforementioned convergence. 
For $M_0 > 0.1$ $M_{\rm Earth}$, we estimate $\sim (2.6$--$5.2) \times 10^5$ $d>10$-km planetesimals at $t=42$ Myr. 
The effects of collisional grinding are greatly reduced for $M_0 < 0.1$ $M_{\rm Earth}$, but the number of 
planetesimals at $t=42$ Myr ends up to be smaller in this case (because it was already small initially). 
For $M_0 = 0.03$ $M_{\rm Earth}$, for example, we found $\sim 1.5 \times 10^5$ $d>10$ km planetesimals 
at $t=42$ Myr.

The results depend on other parameters as well, mainly on the shape of the initial size distribution
and on the scaling law adopted in \texttt{Boulder}. Our choice of the initial size distribution described 
above puts most mass in $d \sim 100$-km planetesimals. It is possible that most mass was in smaller or larger 
planetesimals. If so, this would imply a smaller population of $d>100$-km planetesimals at $t=42$ Myr
(Sect. 14). In the collisional simulations described above, we adopted the standard disruption law for 
monolithic basalt (Benz \& Asphaug 1999). Scaling laws that significantly differ from Benz \& Asphaug 
(1999) (e.g., for porous bodies) would presumably produce different results (but note that most
important collisions happen in the gravity regime where the material strength is not that important).   

\section{Impact flux of leftover planetesimals}

In total, in the simulation described in Sect. 4.3.1, 
there were 14,091 impacts of leftover planetesimals on Venus, 9,418 impacts on the Earth, 
and 2,292 impacts on Mars in 1 Gyr. This represents 11\%, 7.4\%, and 1.7\% of the population of leftover
planetesimals in the terrestrial planet zone at the time of the Moon-forming impact ($t=42$ Myr). 
The number of planetesimal impacts on the Moon was obtained from the number of
impacts on the Earth by re-scaling the results to the smaller cross-section and focusing factor
of the Moon. This gives the Earth-to-Moon ratio of $\simeq 20$,
and the overall lunar impact probability of $\simeq 0.4$\%. The impact probabilities are substantially
higher than the ones obtained for asteroids ($\sim 1$\% for Earth impacts from the inner main
belt) and comets ($1.6 \times 10^{-6}$ for Earth impacts if comet disruption is ignored).

\begin{figure}
\epsscale{1.2}
%\hspace*{4.cm}\plotone{left1.pdf}
\hspace*{4.cm}\plotone{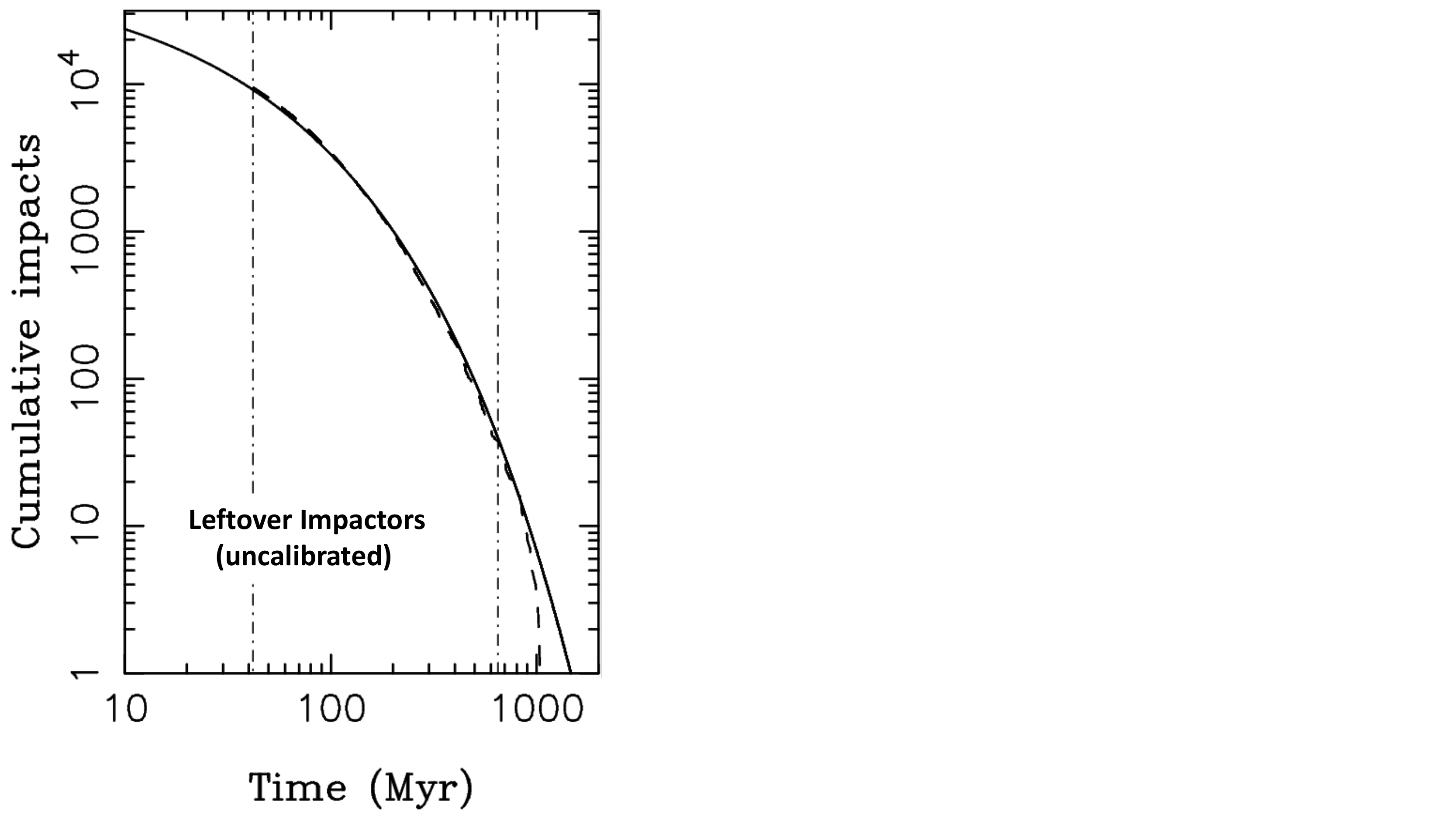}
% gr.impact_leftover3.f in LHB/ on laptop
\caption{The (uncalibrated) impact flux of leftover planetesimals on the Earth. The dashed line is the 
  profile for all impacts on the Earth, 9,418 in total, recorded in the simulation after the Moon-forming 
  impact ($t>42$ Myr). The solid line shows the analytic fit given in Eq. (\ref{flux3}). The model 
  profile drops more steeply for $t=0.8$--1 Gyr than the analytic fit, because the simulation ended 
  at $t=1$~Gyr, and no impacts were recorded for $t>1$ Gyr. The vertical dash-dotted lines 
  show $t=42$ Myr and $t=650$ Myr for reference.} 
\label{left1}
\end{figure}

The raw impact profile -- i.e., the (uncalibrated) impact profile where all planetesimals were given the 
same weight -- is shown in Fig. \ref{left1}. The initial impact flux decline is intermediate between 
those of asteroids and comets. Specifically, 90\% of planetesimal impacts occur within 80~Myr (400 
Myr for asteroids and 55 Myr for comets) and 99\% of impacts occur within 300 Myr. The leftover 
planetesimals are therefore expected to produce a shorter tail of late impacts than asteroids. 
Compared to asteroids, however, the leftover planetesimals represented a much larger 
initial population and are therefore favored, by a large factor, to produce impacts for $t<1$ Gyr 
(see below). 

To a great accuracy, the profile shown in Fig. \ref{left1} can be approximated by a stretched exponential
\begin{equation}
  F(d,t) = F(d) \exp [ -(t/\tau)^\alpha] \ ,
  \label{flux3}
\end{equation}
with $\tau=12.2$ Myr and $\alpha=0.5$. For comparison, Morbidelli et al. (2018) estimated  $\tau=10$ Myr 
and $\alpha=0.5$. Our expression should not be used for $t < 10$~Myr, because Eq. 
(\ref{flux3}) does not take into account the effects of early collisional grinding (Sect. 5). 
We assume that there were $4 \times 10^5$ $d>10$ km planetesimals at $t=42$ Myr (a factor of $\sim 2$
uncertainty; Sect. 5) and use this number to absolutely calibrate of the impact flux for $t>42$ Myr. 
To scale the results to larger impactors, given the results shown in Fig. \ref{grind}, we adopt the 
asteroid belt size distribution as a reference. This gives $F(d)=1.8 \times 10^5$ for $d=10$ km,
$F(d)=5.1 \times 10^4$ for $d=20$ km, and $F(d)=6.9 \times 10^3$ for $d=100$ km. 

Similar results were obtained for Venus and Mars. The functional form in Eq. (\ref{flux3}) provides
an excellent approximation for Venus ($F(d)=2.5 \times 10^5$ for $d=10$ km -- Venus receives 
$\simeq 1.4$ times the number of Earth impacts). The same holds for Mars ($F(d)=4.4 \times 10^4$ 
for $d=10$ km -- Mars receives 4.1 times fewer impacts than the Earth). For Mars, however, $\tau = 
13.5$ Myr fits the model profile better than $\tau=12.2$ Myr. This indicates that the decline of 
Mars impactors was slightly slower, presumably because the planetesimals with $r \sim 1.5$ au 
had longer dynamical lifetimes than the ones below 1 au (e.g., Deienno et al. 2019). For the Moon, 
we adopt the Earth impact profile and scale it down by a factor of $\simeq 20$.
%The model profiles start to drop near 3.8 Ga but this is probably a consequence of the integration
%length (1 Gyr). We should consider extending the simulations. (Check if all jobs finished - looks like
%jobs 1091-1100 are at 750 Myr only.). 20 impacts happened after 700 Myr. 

%The number of planetesimal impacts was overwhelmingly larger than the number of asteroid/comet impacts for 
%$t<1$ Gyr. For example, there were $\sim$18,500 impacts from $d>10$ km leftovers on the Earth after the
%Moon-forming impact at $t=42$ Myr, roughly 57 times more than the combined contribution of asteroids
%and comets for $t>42$ Myr. This era of dominance of the leftover planetesimal impacts lasts $\sim 1$ Gyr.
%For $t>700$ Myr, for example, we expect $\sim 61$ $d>10$ km leftover impacts on the Earth, some $\sim 4$
%and $\sim 10$ times more than the number of asteroid and comet impacts, respectively. The switch from
%leftover to asteroid impacts happens just past $t=1$ Gyr (roughly at $T=3.5$ Ga). For $t>1$ Gyr, there are
%$\sim 14$ leftover, $\sim 12$ asteroid, and $\sim 3.6$ comet impacts ($d>10$ km for the Earth). The flux
%of large asteroid impactors is nearly leveled in the past 3.5 Gyr, with $\sim 3$-4 $d>10$ km impacts on
%the Earth per each Gyr. 

\section{Dependence on the initial radial profile}

The initial surface density profile, $\Sigma(r)$, of planetesimals in the inner solar system is unknown. 
The relatively low mass of the main asteroid belt ($\simeq 4.5 \times 10^{-4}$ $M_{\rm Earth}$; DeMeo \& Carry
2013) probably suggests that the initial planetesimal mass at $r \simeq 2$--4 au was low (see Raymond \& 
Nesvorn\'y 2020 for a review). In the standard model of the terrestrial planet formation (Sect. 2), the initial 
planetesimal mass in the terrestrial planet region ($r<1.5$~au) is usually assumed to be $M_0 > 0.1$ $M_{\rm Earth}$ 
(we adopted this assumption to calibrate the impact model in Sect. 6). How the profile changed from the 
high surface density at $r \sim 1$~au to the low surface density at $r > 2$ au is uncertain. 

To fix the initial number of asteroids in the E-belt ($r=1.75$--2 au), we adopted 
$\Sigma \propto r^{-1}$ for $r>1.75$ au, and used the main belt calibration to extrapolate the 
initial population from $r>2$ au to 1.75--2 au (Sect. 4.1; 
Bottke et al. 2012, Nesvorn\'y et al. 2017a). Here we considered different density profiles for $r<1.75$ au. 
Specifically, employing the weighting scheme described in Sect. 4.3, we explored the effects of the planetesimal 
disk truncation at some outer radius, $r_{\rm out}$. We tested $r_{\rm out} = 1$, 1.25, 1.5 and 1.75 au. 
As for the effects of collisional grinding, the (intrinsic) collisional probabilities were found to be higher 
for disks truncated at smaller outer radii (the intrinsic collisional probability scales as $r^{5/2}$). 
The planetesimal population was dynamically removed faster, however, 
when the disk was truncated at smaller radii. These two effects pull in the opposite directions, and 
partially compensate each other. 

\begin{figure}
\epsscale{1.2}
%\hspace*{4.cm}\plotone{left10.pdf}
\hspace*{4.cm}\plotone{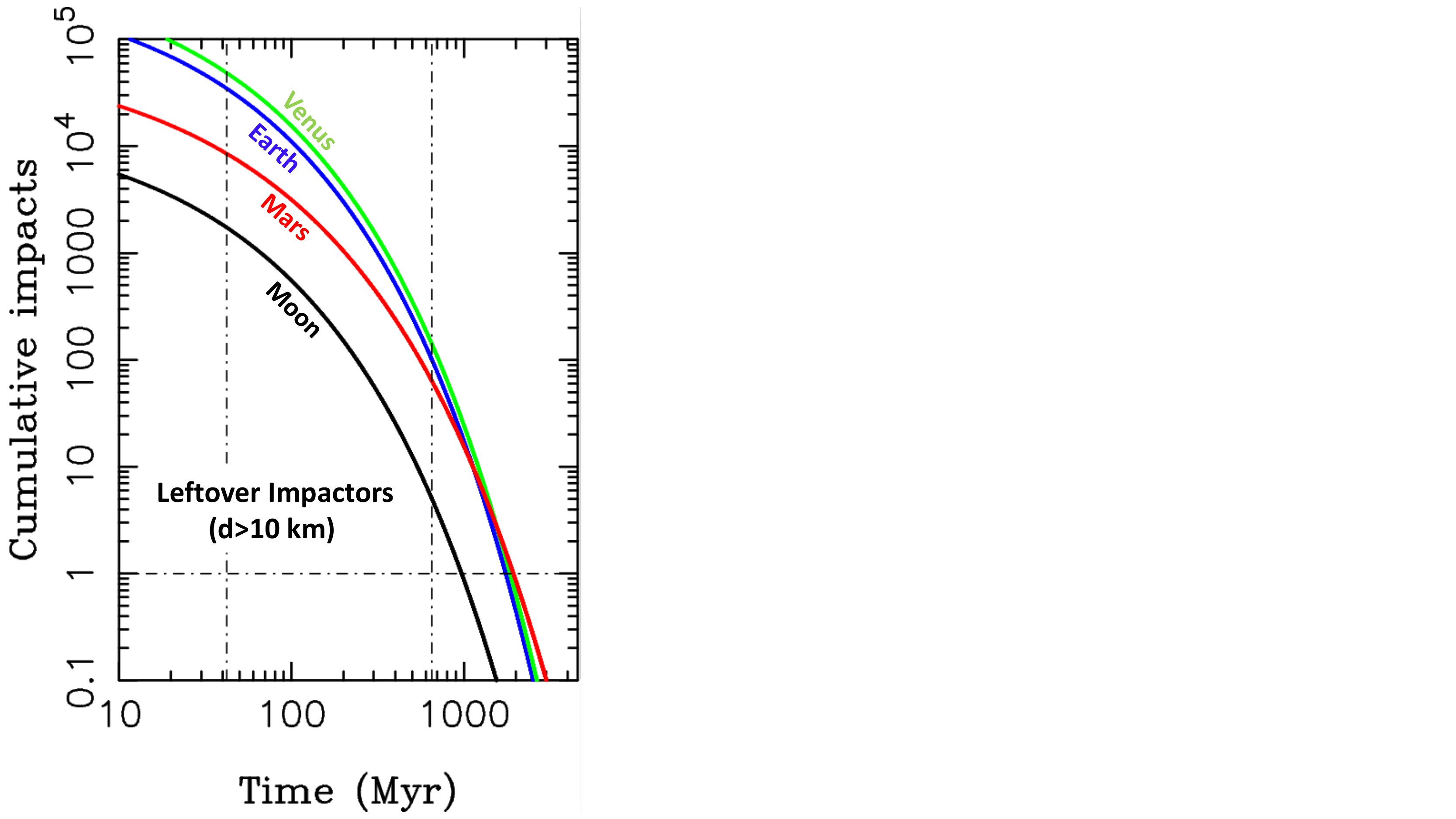}
%\plotone{left100.eps}
% gr.impact_left10.f and gr.impact_left100.f in LHB
\caption{The impact flux of $d>10$ km leftover planetesimals the Moon (black line), Earth (blue), Venus 
  (green), and Mars (red).  The plot shows the accumulated number of impacts {\it since} time $t$ (i.e., 
  on the surface with age $T$), where $t=0$ ($T=4.57$ Ga) is the birth of the solar system, and $t \simeq 4.57$ 
  Gyr ($T=0$) is the present time. The profiles were normalized to having $4 \times 10^5$ $d>10$-km planetesimals 
  at $t=42$ Myr, and $r_{\rm out}=1.5$ au. The vertical dash-dotted lines show $t=42$ Myr and $t=650$ Myr for reference.}
\label{left2}
\end{figure}

As for the impact profiles, the cases with $r_{\rm out} = 1.25$ au and 1.75 au produced results that were very similar 
to those obtained for $r_{\rm out}=1.5$ au. The profiles can be approximated by Eq. (\ref{flux3}) with 
$\tau = 11.5$ Myr, assuming $\alpha=0.5$, which is only a slightly shorter $e$-fold than $\tau =12.2$ Myr 
inferred from the raw profiles. When $\alpha$ was allowed to adjust as well, however, a slightly better fit was 
obtained with $F=3.8\times10^5$, $\tau=6$ Myr and $\alpha=0.45$ ($d>10$~km and Earth). This is the parameterization 
of planetesimal impacts that we adopt in this work (Fig. \ref{left2}). Note that having $\tau=6$ Myr does not
mean that the decline was faster than in the case with $\tau =12.2$ Myr and $\alpha=0.5$, because $\alpha$ 
is now smaller (i.e., the exponential is more stretched). The half-life\footnote{The number of impacts in a short 
interval $\Delta t$ around $t$ can be obtained as ${\rm d}F/{\rm d}t \times \Delta t$. The half-life of impact 
decline is $\delta t_{\rm half} = \tau (t/\tau)^{1-\alpha} / (2 \alpha)$.} of impact flux decline steadily 
increases with time; it is $t_{\rm half}=46$~Myr for $t=200$ Myr and $t=88$ Myr for $t=650$ Myr. The case with 
$r_{\rm out} = 1$ au could be interesting as well, but the planetesimal population declined more steeply in this 
case, and there are not as many late impacts as for $r_{\rm out}=1.25$--1.75 au. 

The impact flux of leftover planetesimals on Venus closely follows the one obtained here for the Earth, 
but the overall impact rate is 1.4 times higher (for $r_{\rm out} = 1.5$ au). The lunar impact flux is 
$\simeq 20$ times lower. The impact profile for Mars is flatter and can adequately be fit with 
$F=8.2\times10^4$ for $d>10$ km, $\tau=6$ Myr and $\alpha=0.42$ (Fig. \ref{left2}). This means that the lunar 
chronology should not strictly be applied to Mars. For example, the Imbrium-age Mars ($t \simeq 650$ Myr or 
$T \simeq 3.92$ Ga) would have accumulated $\simeq 18$ times more planetesimal impacts than 
the same-age Moon, but a $t=42$-Myr old Mars would have accumulated only $\simeq 4.1$ times more 
planetesimal impacts than the Moon (values given here for the whole target surfaces, not per km$^2$). 
The Mars-to-Moon impact ratio is expected to change for $t>1$ Gyr, when the asteroid impacts start to 
dominate (the ratio $\simeq 7$ is expected for asteroids; Sect 4.1). The mean impact speeds of leftover 
planetesimals are 29.0, 23.2, 19.9 and 14.2 km s$^{-1}$ for Venus, Earth, the Moon, and Mars, respectively.
%planetesimals are 21.6, 20.3, 17.2 and 12.7 km s$^{-1}$ for Venus, Earth, the Moon, and Mars, respectively.

\section{Lunar basins}

\begin{figure}
%\epsscale{0.487}
%\plotone{moon1.eps}
%\epsscale{0.49}
%\plotone{moon2.eps}
\epsscale{1.2}
%\hspace*{1.cm}\plotone{moon1.pdf}
\hspace*{1.cm}\plotone{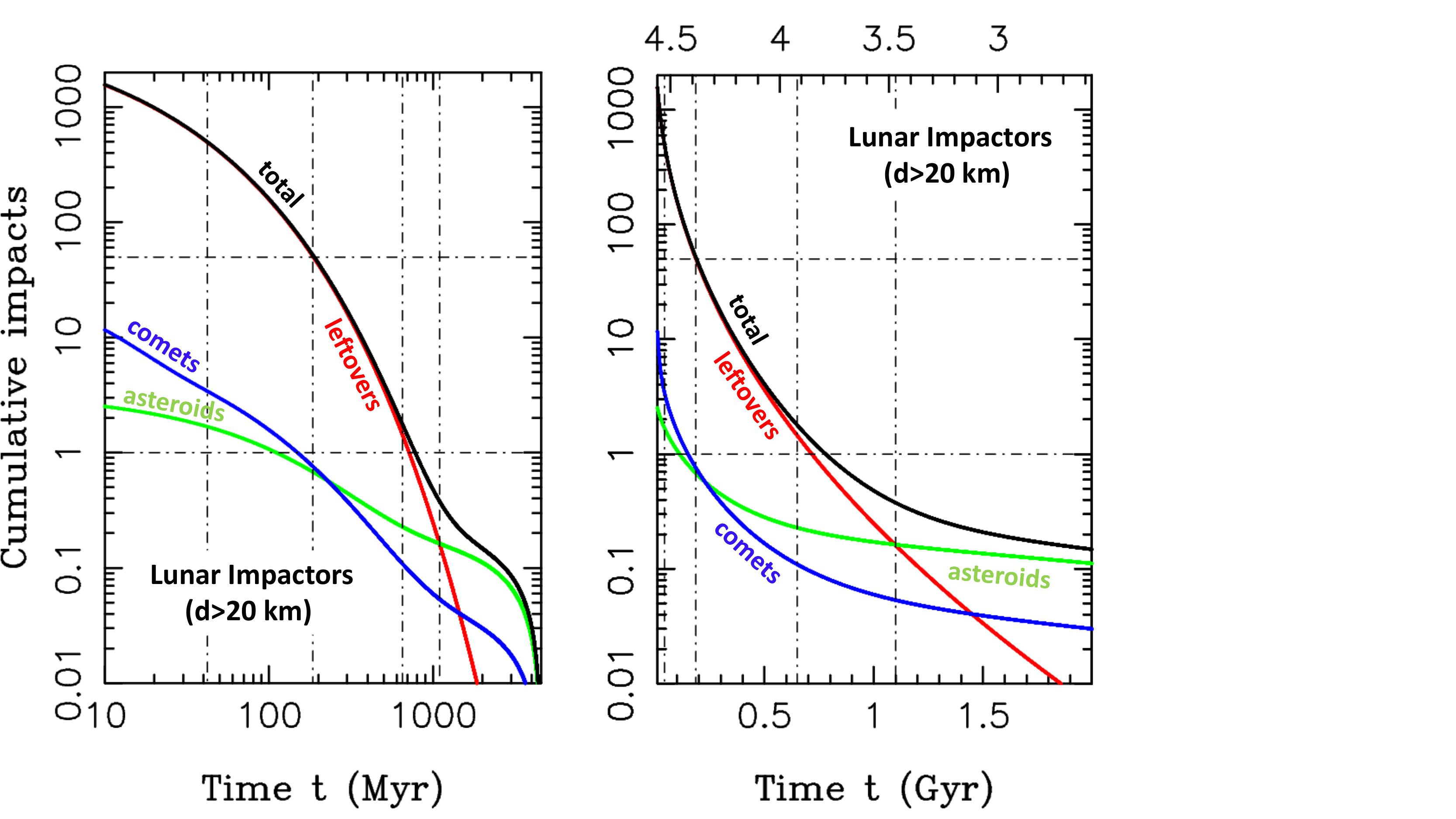}
% gr.impact_moon.f
\caption{Impacts of $d>20$-km bodies on the 
Moon (log time scale in the left panel, linear on the right). The plot shows the accumulated number of impacts 
{\it since} time $t$, where $t=0$ is the birth of the solar system, and $t \simeq 4.57$ Gyr is the present 
time. The planetesimal, asteroid and comet profiles are shows by red, green and blue lines, 
respectively; the black line is the total impact flux. The vertical dash-dotted lines show the Moon 
formation in our model ($t=42$~Myr), estimated start of the known lunar basin record ($t \simeq 190$ or 
$T \simeq 4.38$ Ga), Imbrium formation ($t \simeq 650$ Myr or $T\simeq3.92$ Ga), and transition from the 
planetesimal-dominated to asteroid-dominated impact stages ($t \simeq 1.1$ Gyr or $T \simeq 3.5$ Ga).
The numbers on the upper axis of the left panel indicate time $T$ (in Gyr) measured looking back from today.}
\label{chrono}
\end{figure}

We now turn our attention to lunar impacts. Figure \ref{chrono} shows the integrated history of lunar
impacts including all major sources of impactors in the inner solar system: leftover planetesimals
(Sect. 7), asteroids (Sect. 4.1), and comets (Sect. 4.2). We find that planetesimals dominated
the early impact flux ($t<1.1$ Gyr or $T>3.5$ Ga; Morbidelli et al. 2018). Asteroids took over and produced 
most impacts in the last $\simeq 3.5$ Gyr. The cometary flux was never large enough, in the whole history of 
the inner solar system, to be competitive (the early instability at $t\lesssim10$ Myr is adopted 
here).\footnote{Comets could represent a more significant contribution to the lunar basin record only if 
the instability happened late. For example, to produce more than five lunar basins (i.e., $\gtrsim 10$\% 
of all known basins), the instability would have to happen at $t \gtrsim 170$ Myr.} In the first $\simeq 200$~Myr, 
comets produced more lunar impacts than asteroids, but the number of planetesimal impacts was far greater. Comets 
outpaced leftover planetesimals in the last $\sim 3$ Gyr, but that was when the overall impact flux was 
ruled by asteroids. This would suggest that isotopic and other signatures of comets may be difficult to 
find on the lunar surface.

The overwhelming majority of impacts observed on the lunar surface must date back to $T>3.5$ Ga, when 
the impact flux was orders of magnitude higher than it is today. The model predicts $\simeq500$ 
$d>20$-km lunar impacts for $t>42$ Myr ($T<4.53$ Ga). For comparison, modeling the lunar gravity anomalies 
detected by GRAIL, Miljkovi\'c et al. (2016) found $\simeq50$ impacts of $d > 20$ km bodies (Sect. 2). 
We therefore see that the number of impacts suggested by our model would be excessive, by a factor of 
$\sim 10$, if the Moon surface recorded all large impacts since its formation (Zhu et al. 2019a).\footnote{It 
is possible that some younger basins erased older basins but it is difficult to imagine how this 
could account for the above quoted factor of $\sim10$.}

The Moon was fully molten when it accreted from the debris disk created by the giant impact on proto-Earth 
(see Canup et al. 2021 for a review). The subsequent evolution and solidification of the global 
lunar magma ocean (LMO) was 
controlled by a number of geophysical processes, including tidal heating, formation of an insulating flotation 
crust, etc. (Meyer et al. 2010, Elkins-Tanton et al. 2011). Radiogenic lunar crustal ages span from 4.47 
to 4.31 Ga and suggest a prolonged stage of complete LMO solidification (Shearer et al. 2006, 
Elkins-Tanton et al. 2011, Maurice et al. 2020). The lunar basins that formed while the LMO was still 
present would have been subject to extreme relaxation that would have reduced topographic and crustal 
thickness signatures (Miljkovi\'c et al. 2021). These early basins may be unidentifiable today. 

The long-lived LMO would resolve the problem with the excess of planetesimal impacts on 
the young Moon. Suppose, for example, that the lunar surface started recording basin-scale impacts 
at $t \simeq 190$ Myr ($T \simeq 4.38$ Ga), or roughly $150$ Myr after Moon's formation in our model 
-- the oldest known basins (e.g., South Pole-Aitken) would date back to this time. If so, we would 
expect $\simeq 50$ $d>20$ km impacts to be recorded (Fig. \ref{chrono}), in a close agreement with the 
lunar basin record. Concerning the LMO lifespan, Morbidelli et al. (2018) and Zhu et al. (2019a) 
reached similar conclusions -- they proposed $T \simeq 4.35$ Ga -- based on the HSE constraint 
(Sect. 14). Scaling to smaller impactor sizes, we estimate $\sim 180$ $d>10$ km impacts for $T < 4.38$ Ga, 
in a close agreement with $\sim 200$ $D>150$ km lunar craters inferred in Bottke \& Norman (2017) 
(a $d=10$ km impactor is assumed here to produce a $D = 150$ km lunar crater; Johnson et al. 2016a).

The time of LMO solidification, $t_{\rm LMO}$, inferred here from the lunar basin record is uncertain.
For example, for a slightly higher calibration with $5.2 \times 10^5$ $d>10$ km planetesimals at $t=42$ Myr, 
we obtain $t_{\rm LMO}=215$ Myr ($T_{\rm LMO} \simeq 4.35$ Ga). If, instead, there only were 
$2.6 \times 10^5$ $d>10$~km planetesimals at $t=42$ Myr, then $t_{\rm LMO}=160$ Myr ($T_{\rm LMO} \simeq 4.41$ Ga). 
Turning this argument around, the $T=4.31$ Gyr age for the youngest known crustal ages would imply 
$t_{\rm LMO}=260$ Myr. This could suggest that our nominal calibration of leftovers is a factor of $\sim 
2$ too low, and that there were $\sim 8 \times 10^5$ $D>10$ km planetesimals at $t=42$ Myr. It could 
indicate that the effects of collisional grinding were not as severe as we assumed here (e.g., stronger 
$Q^*_{\rm D}$ required).

It has to be noted that the LMO solidification is defined here in a specific context: the melt 
layer has to be thin enough for the topographic and crustal thickness signatures of basins to survive
and be identifiable today. Miljkovi\'c et al. (2021) showed that impact basins experience extreme 
relaxation for melt layers as thin as $\sim 10$ km (the melt layer needs to be close to the surface, 
Miljkovi\'c et al. considered the melt layer in depths 10--50 km below the surface). Here we therefore 
provisionally interpret $T_{\rm LMO}$ as the time when the melt layer became thinner than 10~km. 
The youngest radiogenic crustal ages can post-date our $T_{\rm LMO}$, because they mark the time when 
the LMO solidification was brought to completion.

Oceanus Procellarum is the largest of the lunar maria, covering roughly 10\% of the total lunar surface.
It is uncertain whether Procellarum is or is not an impact-generated basin (e.g., Andrews-Hanna 2016). 
Here we estimate the probability of an impact of Ceres-class planetesimal on the Moon. Adopting the scaled 
asteroid-like distribution from Fig. \ref{grind} (green line), there would be $\sim 100$ $d > 1000$ km 
planetesimals for $t=42$ Myr, each having a 0.4\% of probability to impact on the Moon (Sect. 6). We can 
therefore roughly estimate that one $d > 1000$ km impact on the Moon for $t>42$ Myr is a $\sim40$\% 
probability event (i.e., it {\it can} happen). This is consistent with the results reported in 
Zhu et al. (2019b), who considered impactor diameters $\simeq 500$--1000 km and found similarly large 
probabilities.  

\section{Imbrium-era impacts}

Having established that (nearly) all lunar basins formed from impacts of leftover planetesimals
(only $\sim 0.7$ and $\sim 0.8$ $d>20$ km lunar impacts of asteroids and comets are expected to happen
for $t>190$ Myr, respectively), we now consider the Imbrium-era impacts. What are our constraints? Evidence
suggests that Imbrium formed at $T \simeq 3.92$ Ga ($t \simeq 650$~Myr; Zhang et al. 2019) 
by an impact of a $d \gtrsim 100$ km object (Miljkovi\'c et al. 2013, 2016; Schultz \& Crawford 2016). From the 
crater counts we know that the Orientale and Schr\"odinger basins formed after Imbrium (e.g., 
Orgel et al. 2018). These smaller basins were produced by $d \simeq 50$--64-km and $d \simeq 20$-km 
impactors, respectively (Miljkovi\'c et al. 2016, Johnson et al. 2016b). 

Having only two smaller basins with post-Imbrium formation ages is surprising. Adopting the asteroid belt 
size distribution, we find that there should be $\sim 7.4$ $d>20$-km impacts for every $d>100$-km impact. 
The expected number of $d>20$-km impacts would be even larger if we assumed a steeper size distribution 
(Minton et al. 2015). Our interpretation of this problem is that the Imbrium basin should have 
formed unusually late, by chance, relative to the expectation from the lunar impact chronology. In other 
words, statistically, the Imbrium basin formation needs to be an unusually late event. 

The relatively late formation of some lunar basins is expected in our model. We find, for
example, with the standard normalization considered here -- $4 \times 10^5$ $d>10$ km leftover 
planetesimals at $t=42$ Myr, and the asteroid-like size distribution for $d>10$ km -- that on average
$\simeq  2.3$ basins are expected to form for $t>600$ Myr ($T<3.97$~Ga; Fig. \ref{chrono}).
Given the strong decline of planetesimal impacts for $t > 600$ Myr, any basin-scale impacts are 
expected to happen just after 600 Myr. As for Imbrium, we estimate that there should be $\simeq 0.31$ 
$d>100$ km lunar impacts for $t>600$ Myr, and that the Imbrium formation at $t>600$ Myr should 
therefore be a $\simeq 23$\% probability event (from the standard Poisson statistics), or 15--35\%
considering the calibration uncertainty described in Sect. 5. 

We now ask how likely it is for Imbrium to form late ($t>600$ Myr, $d>100$ km) {\it and} have exactly 
two basins (Orientale and Schr\"odinger, $d>20$ km impactors) younger than Imbrium. This is a very 
restrictive conditions and the probability is therefore not expected to be high. But this is not the main 
point here. The main point is to understand whether a shorter or longer tail of impactors than the 
one obtained here, would fit data better. The condition is evaluated with our standard calibration 
of leftovers (see Nesvorn\'y et al. 2022 for other calibrations). 
We fix $\alpha=0.45$, vary $\tau$ in Eq. (\ref{flux3}), and generate a statistically 
large number of random impact sequences in each case. 

\begin{figure}
\epsscale{0.5}
%\plotone{imbrium.eps}
\plotone{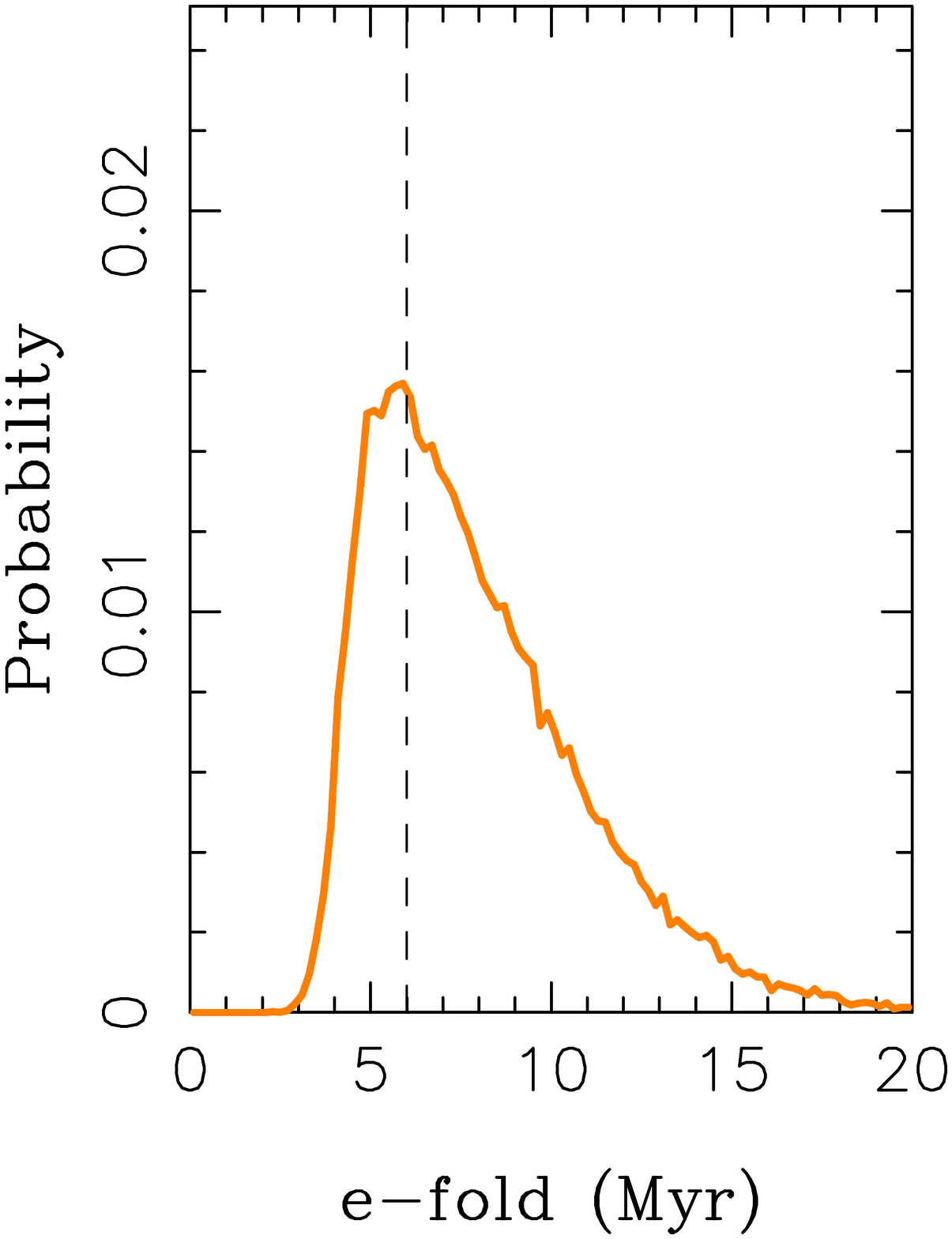}
\caption{The probability of matching the Imbrium-era constraints described in the main text as a function
of the impact flux decline $e$-fold $\tau$ ($\alpha=0.45$ and the standard calibration of leftover 
planetesimals). The vertical dashed line marks $\tau =6$ Myr obtained from our dynamical modeling. }
\label{imbrium}
\end{figure}

We find that the probability has a broad maximum around $\tau=6$ Myr (Fig. \ref{imbrium}), {\it which 
was the best-fit e-fold that we obtained for leftover planetesimals.} 
The probability drops for $\tau<5$ Myr because it becomes very difficult to obtain any late impacts in 
this case. It decreases for $\tau > 6$ Myr, because too many basins form after Imbrium if the decline 
of the impact flux is too slow. It thus cannot be argued that a very slow decline of impactors would 
help to explain the lunar basin record. Instead, the impact chronology obtained here ($\alpha=0.45$ and 
$\tau=6$ Myr) is (nearly) optimal to satisfy the Imbrium-era constraints. A similar argument can be 
made about the overall normalization of leftover planetesimals. The results are not affected by small 
changes of the impact chronology. For example, the probability peaks at $\tau \simeq 9$--12 Myr 
when we use $\alpha=0.5$ from Eq. (\ref{flux3}), whereas $\tau=12.2$~Myr was our best fit with 
$\alpha=0.5$ (Fig. \ref{left1}).  

The Imbrium-era impactors are often found to be stored on orbits with $a \simeq 1.3$-1.7 au, $e<0.2$,
and in the mean motion resonances with Mars (e.g., 6:5, 7:6, 11:9; inside or outside of the Mars orbit).
They either start with $a \simeq 1.3$--1.7 au at $t=0$ or are scattered to 1.3--1.7 au from $\lesssim 1$ 
au. The resonances provide a phase-protection mechanism against collisions with Mars. The orbits have low 
eccentricities such that they do not cross the orbit of Earth. The bodies stay in the resonances for hundreds 
of Myr, are eventually released to orbits with higher eccentricities, at which point they can impact.
This is an example of the ``storage places'' hypothesized by Wetherill (1975).  
We are only able to see this because our model self-consistently followed the growth of the terrestrial 
planets and impact profile of leftover planetesimals (Sects. 2 and 4.3).\footnote{If we would switch to 
the real orbits of planets at $t=42$ Myr, or at any other time during the simulation, the resonant 
planetesimals would be prematurely released and this would (presumably) produce a shorter impact tail.} 
  
%In the run with 1000 clones, the late impactors are: 119, 115, 98, 80, 38, 35, 27, 15, 14. 

\section{Lunar $N_{20}$ chronology}

\begin{figure}
%\epsscale{0.487}
%\plotone{ln20.eps}
%\epsscale{0.49}
%\plotone{ln1.eps}
\epsscale{1.2}
%\hspace*{0.5cm}\plotone{lunarchrono.pdf}
\hspace*{0.5cm}\plotone{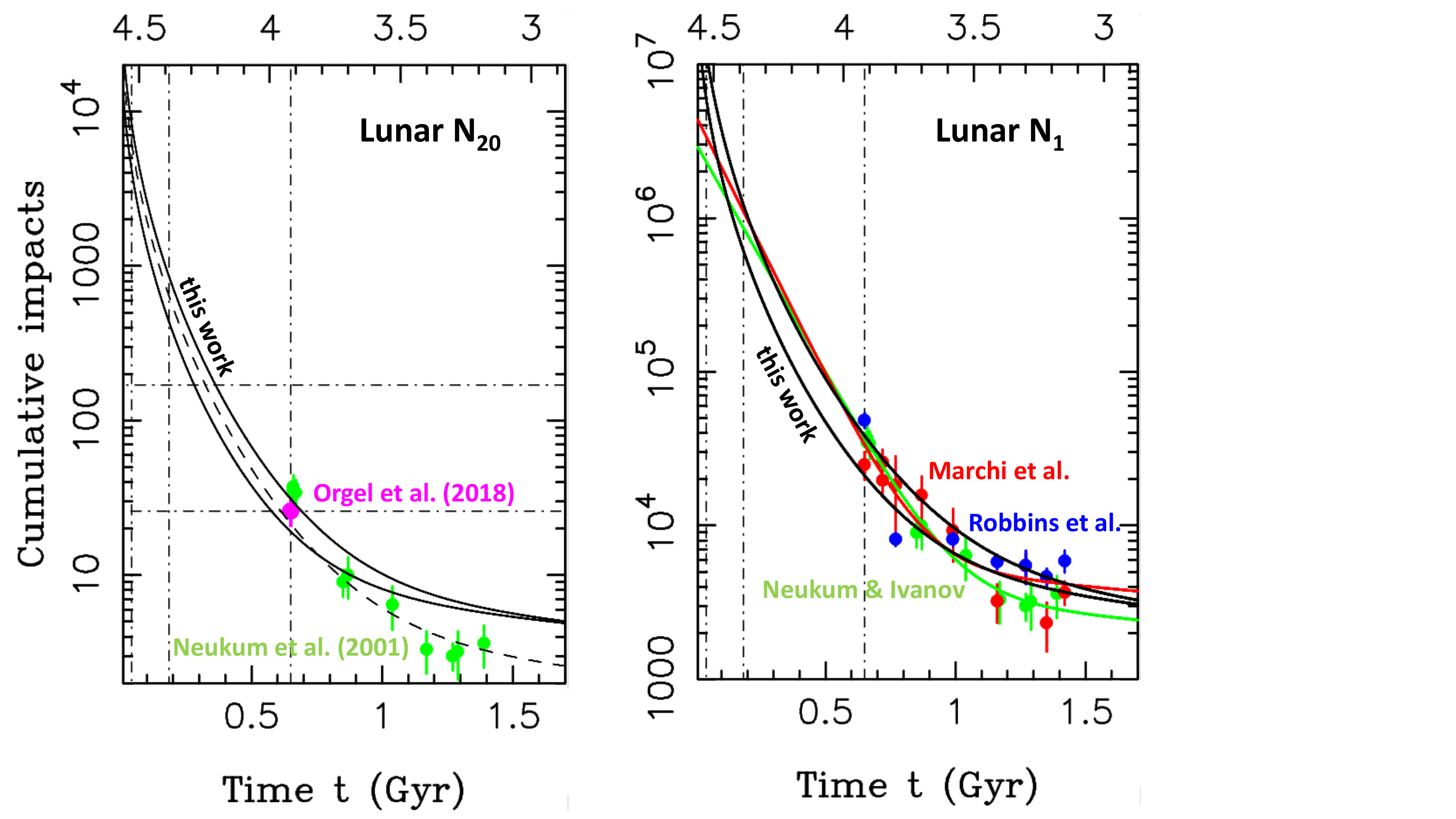}
\caption{Left: The lunar $N_{20}$ chronology. The two black lines show our model chronologies with minimal 
($2.6 \times 10^5$ $d>10$ km bodies at $t=42$ Myr) and maximal ($5.2 \times 10^5$ $d>10$ km bodies at $t=42$ Myr)
calibrations of leftover planetesimals. The purple dot is the density of $D>20$ km craters 
on Fra Mauro/Imbrium highlands (Fassett et al. 2012, Orgel et al. 2018). The green dots are the $N_{20}$ values inferred 
for different terrains from Neukum et al. (2001). The dashed line shows a modified model chronology 
where the constant term of the asteroid flux was reduced by a factor of 2. Right: The lunar $N_1$ chronologies 
from this work (black lines), Neukum \& Ivanov (1994) (green line and symbols), and Marchi et al. (2009) 
(red line and symbols). The crater counts from Robbins et al. (2014) are shown by blue dots. The 
estimated uncertainties of different measurements are shown by vertical bars. The numbers on the upper 
axes indicate time $T$ (in Gyr) measured looking back from today.}
\label{lns}
\end{figure}

To compare our model results with lunar chronologies we need to compute the cratering flux for small 
impactors. We start discussing the $N_{20}$ chronology (Fig. \ref{lns}A), where we have more 
confidence in the model results (given that no wild extrapolations to very small sizes are needed).
For the lunar impact speeds of leftover planetesimals and asteroids obtained here ($\sim 20$ km s$^{-1}$; 
Sects. 4.1 and 7), we assume that a $d=1$ km projectile makes a $D=20$~km lunar crater (see Morbidelli 
et al. (2018) for a discussion of the scaling laws). The overall mean impact velocity of planetesimals on 
the Moon is lower ($v_{\rm i}  \simeq 17$ km s$^{-1}$), but the mean value is mainly contributed by 
the very early impactors that have, in general, lower impact speeds (because they are on orbits 
with $a \sim 1$ au). The mean impact speeds of planetesimals for $t>200$ Myr are higher 
($v_{\rm i} \simeq 20$ km s$^{-1}$), and more similar to asteroids ($v_{\rm i} \simeq 21$ km s$^{-1}$; Sect. 
4.1). This gives some justification to using the same projectile size for both types of impactors. 
      
To compute $N_{20}(T)$ from our impact model, we assume that the size distribution of leftover planetesimals 
for $1<d<10$ km was similar to that of today's asteroid belt (Sect. 5). There are roughly $1.1\times10^6$ 
main-belt asteroids with $d>1$ km (Bottke et al. 2020) and $\simeq 8200$ with $d>10$ km (Mainzer et 
al. 2019). We therefore scale the planetesimal flux profile from $d>10$ km to $d>1$ km by a factor 
$1.1\times10^6/8200 = 134$. This gives $F(1\, {\rm km})=2.5 \times 10^6$ in Eq. (\ref{flux3}) for the 
whole lunar surface ($\tau=6$ Myr and $\alpha=0.45$ from Sect. 7). The same scaling factor is used 
for the early flux of asteroids (the first term in Eq. (\ref{flux})). The constant term in Eq. (\ref{flux}) 
for $d>1$ km is obtained from $d>1$ km NEAs at the present epoch. Morbidelli et 
al. (2020) estimated that the average time interval between $d>1$ km NEA impacts on Earth is $\simeq 0.75$ 
Myr (also see Harris \& D'Abramo 2015). We thus have $F_2(1\, {\rm km}) \simeq 1.3$ Myr$^{-1}$ in 
Eq. (\ref{flux}).    
 
Neglecting comets, which should only contribute at the $< 10$\% level, a simple $N_{20}(T)$ 
chronology function -- the number of accumulated $D>20$ km craters in area $10^6$ km$^2$ of the lunar 
surface since $T$ -- is given by
\begin{equation}
N_{20}(T) = a \exp [ -(t/6\, {\rm Myr})^{0.45}] + b \exp[ -( t / 65\, {\rm Myr} )^{0.6}] + c T  
\label{final}
\end{equation}  
with $t=4570-T$ ($t$ and $T$ given in Myr), $a = 6.7 \times 10^4$, $b=40$ and $c=1.7 \times 
10^{-3}$~Myr$^{-1}$. The first term in Eq. (\ref{final}) stands for leftover planetesimals,
the second and third for asteroids. The second term is much less important than the other two
and can be discarded for a rough estimate of $N_{20}$. Figure \ref{lns}A compares our model 
chronology with crater densities measured for different lunar terrains.
 
Fassett et al. (2012) reported $N_{20} = 30 \pm 5$ for the $\simeq 3.92$ Gyr old Fra Mauro/Imbrium highlands. 
This value was slightly revised in Orgel et al. (2018), giving $N_{20} = 26 \pm 5$ (purple symbol in 
Fig. \ref{lns}A), which nearly perfectly aligns with our model chronology obtained for the standard
calibration of terrestrial planetesimals. We cannot plot $N_{20}(T)$ for other basins from Fassett et al. (2012) 
and Orgel et al. (2018) in Fig. \ref{lns}A, because their radiometric age is unavailable or uncertain.
For reference, we also show data inferred from Neukum et al. (2001). For that, we take the $N_1$ 
values reported in Neukum et al. (2001) and scale them to $N_{20}$ with the Neukum's ``old'' 
production function ($N_1/N_{20} \simeq 1000$). We believe this is a reasonable approach because the 
crater counts on old surfaces were typically done for relatively large craters, and were extrapolated 
down to $N_{1}$ with the same production function. Reporting them as $N_{20}$ should therefore be
 equally valid.

The $N_{20}$ value inferred from Neukum et al. (2001) for the Fra Mauro/Imbrium highlands is slightly larger than 
the more recent crater counts discussed above; it plots near the upper limit of our leftover calibration 
range. The results for young lunar maria are more discrepant. Here the $N_{20}$ values inferred from  
Neukum et al. (2001) plot below our chronology function, indicating lower crater densities, and the 
difference becomes larger for younger lunar maria (Oceanum Procellarum, Mare Imbrium, Mare Crisium, 
Mare Fecundatis) than for the older ones (Taurus Littrow, Mare Tranquillitatis).

This may mean one of several things. At the face value, inferred $N_{20}$ densities could indicate 
that our chronology function should more steeply drop for $T<3.5$ Ga (the dashed line in Fig. \ref{lns}A
shows an example with $c=0.85 \times 10^{-3}$ Myr$^{-1}$). For that to work, however, the Earth impact 
rate of $d>1$ km asteroids at $T \simeq 3.1$--3.5 Ga would have to be lower, by a factor of $\sim 2$, 
than the current impact rate (e.g., Morbidelli et al. 2020; the current impact rate of large NEAs 
is known to a better than $\sim 10$\% precision), suggesting the number of impacts would 
have to increase at some point in the last 3 Gyr (Culler et al. 2000, Mazrouei et al. 2019). 

%Adopting a different factor when  scaling from $d>10$ km to $d>1$ km impactors in our model would 
%not change things, because our chronology function for $T<3.5$ Gyr is not sensitively tied to that 
%factor; it is instead calibrated on the Earth-impact rate of modern NEAs (Harris \& D'Abramo 2015, 
%Granvik et al. 2018, Morbidelli et al. 2020). 

Another possibility is that the $N_{20}$ values inferred from Neukum et al. (2001) for $T \simeq 3.1$--3.5 
Ga are systematically a factor of $\sim$2 lower. Note that the density of $d>20$ km craters on young 
lunar maria is very low and subject to small number statistics. We therefore extrapolated from smaller 
craters, which were actually counted with some confidence, to estimate $N_{20}$. But perhaps the actual 
production function is flatter for $D \gtrsim 5$ km than the one we adopted from Neukum et al. (2001), 
and would thus give a larger $N_{20}$ value.\footnote{Scaling $N_1$ from Neukum et al. (2001) to $N_{20}$ 
with the ``new'' production from Neukum \& Ivanov (1994) ($N_1/N_{20} \simeq 420$) would resolve the 
problem with the young lunar maria in Fig. \ref{lns}A, but this would move Imbrium's $N_{20}$ to a 
factor of $\simeq 3$ above $N_{20}$ from Fassett et al. (2012) and Orgel et al. (2018). Thus, in some 
sense, $N_1/N_{20}$ needs to be higher for Imbrium and lower for young lunar maria.} 
Alternatively, the crater counts reported in Neukum et al. (2001) for young lunar maria may be lower 
than the actual values. The results of Robbins et al. (2014) give some credit to this possibility, because 
their $N_1$ counts for young lunar maria (the \textit{actual} counts of $D>1$ km craters) are a factor 
of $\sim 1.5$--2 higher than $N_1$ from Neukum et al. (2001). If this difference propagates to $N_{20}$, 
it could be fully responsible for the problem shown in Fig. \ref{lns}A.   

Our chronology can be used to estimate the formation ages of lunar basins from $N_{20}$ counts. For 
example, the age of the Nectaris basin with $N_{\rm 20}=170$ per $10^6$ km$^2$ (Orgel et al. 2018), is 
estimated to be $T=4.21$--4.29 Gyr, where the uncertainty is dominated by the uncertain calibration
of the lunar chronology function in Eq. (\ref{final}) (here given for the two chronologies shown
by solid lines in in Fig. \ref{lns}A). For comparison, Orgel et al. (2018) suggested a younger, 
$T=4.17$-Gyr age for the Nectaris basin from the ``old'' Neukum chronology function.   

%Le Feuvre \& Wieczorek (2011) suggest that small impacts are in porous regime (make smaller craters)
%than large impacts (those penetrate the megaregolith, Hartman \& Morbidelli 2021). Could this resove 
%some of the discrepancy I have with $N_1/N_{20}$ scaling for different terrains? 

\section{Lunar $N_1$ chronology}
    
We have less confidence in extrapolating our model results to $N_1(T)$, because very small 
bodies are subject to a host of dynamical (e.g., radiation effects; Vokrouhlick\'y et al. 
2015) and physical (e.g., rotational spin-up and mass loss) effects that were not modeled 
in this work. The small body populations such as the main asteroid belt are also not well 
characterized for $d<1$ km. Some useful information can nevertheless be obtained by 
assuming that the impact flux of small asteroids was roughly constant  in the last $\sim 3$ Gyr 
and equal to the impact flux of modern NEAs. 

From the scaling laws we estimate that a 
$d \simeq 40$-m asteroid impactor would produce a $D \simeq 1$ km lunar crater.\footnote{The 
impactor size adopted here is intermediate between the one used in Morbidelli et al. (2018), 
$d=50$~m, and Marchi et al. (2022), $d=33$ m -- see the discussion of scaling laws in these papers.}  
There are $\sim 5 \times 10^5$ $d>40$-m NEAs (Bottke et al. 2020; debiased data from updates of 
Harris \& D'Abramo 2015). Assuming the usual impact probability $p_{\rm i}=1.5 \times 10^{-3}$ Myr$^{-1}$ 
of NEAs with the Earth (e.g., Morbidelli et al. 2020), and scaling things to the Moon, we 
obtain $c=1.0$ Myr$^{-1}$ (per $10^6$ km$^2$ of lunar surface).\footnote{This 
calibration works well to reproduce the crater counts on very young 
lunar terrains ($T<1$ Ga; Cone, Tycho, North and South Ray craters; Neukum et al. 2001, Marchi et 
al. 2009, Hiesinger et al. 2016), when the standard production functions are used to extrapolate the counts from very small 
craters to $N_1$. We do not discuss the young terrains in detail here, because we believe that 
a NEA-derived $N_1$ value for $T<1$ Ga is better constrained than the one obtained from counts 
of $D<100$-m craters.} This estimate is consistent with a recent analysis of planetary impacts 
from small NEAs (Nesvorn\'y et al. 2023).    
 
There are no useful constraints on the population of $d>40$ m leftover planetesimals. To obtain 
a rough estimate of $N_1(T)$ for $T>3.5$ Ga, we adopt $N_1/N_{20}=1400$ from Morbidelli et al. (2018) 
(also see Marchi et al. 2012). Note that this is just a reference ratio that turns out to give an 
acceptable fit to the $N_1$ data (Fig. \ref{lns}B). This does not mean, however, that the adopted 
value is strictly correct. For example, a somewhat higher calibration of leftover planetesimals 
(e.g., $5.2 \times 10^5$ $d>10$-km planetesimals at $t=42$ Myr) would still fit the data in Fig. 
\ref{lns}A) quite well. If that calibration is adopted, $N_1/N_{20} \sim 1000$ -- consistent 
with the ``old'' production function from Neukum et al. (2001) -- would work well in Fig.~\ref{lns}B.

Neglecting comets, a simple $N_{1}(T)$ chronology function -- the number of accumulated $D>1$ km 
craters in the area $10^6$ km$^2$ of lunar surface since $T$ -- is given by
\begin{equation}
N_1(T) = a \exp [ -(t/6\, {\rm Myr})^{0.45}] + b \exp[ -( t / 65\, {\rm Myr})^{0.6}] + c T  
\label{final2}
\end{equation}  
with $t=4570-T$ ($t$ and $T$ in Myr), $a = 9.4 \times 10^7$, $b=5.6 \times 10^4$ and $c=1.0$ Myr$^{-1}$ 
(again, the second term is much less important than the other two and can be neglected). The $c$ 
coefficient is only $\simeq 16$\% higher than the one in Eq. (\ref{ni94}) (Neukum et al. 2001; 
the one Eq. (\ref{ni94}) needs to be multiplied 
by $10^3$ to have the number of craters per Myr per 10$^6$ km$^2$, which are the units used here). 
The main difference with respect to Eq. (\ref{ni94}) in Neukum et al. (2001) is that the first term 
in Eq. (\ref{ni94}) is an exact exponential function, whereas here we have a stretched exponential 
in Eq. (\ref{final2}). 

Figure \ref{lns}B compares our $N_1(T)$ chronology, with all the caveats mentioned above, to various 
crater counts and chronologies from Neukum et al. (2001) and Marchi et al. (2009). The main focus 
in Fig. \ref{lns}B is the data from Robbins et al. (2014) (blue dots in Fig. \ref{lns}B), which are 
the {\it actual} counts of $D \sim 1$ km craters for Apollo and Luna landing sites; all other ``data'' 
points are the extrapolations from counts of larger craters.\footnote{We do not show the chronology 
function from Robbins et al. (2014), but note that their data-driven fit has a very different shape.}

The $N_1$ crater counts from Robbins et al. (2014) are slightly higher, in general, that the extrapolations
from Neukum et al. (2001) and Marchi et al. (2009) (except for the old Mare Tranquillitatis age). They 
plot near the higher end of our chronology range (i.e., near the higher end of planetesimal 
calibration), but the agreement is reasonably good. While this gives some justification to the choice
of parameters described above, we caution that other choices would potentially work as well. In particular, 
there is a degeneracy between the leftover planetesimal calibration and the $N_1/N_{20}$ factor. If, 
for example, the NEA population were taken as a guide, and assuming that $d=40$ m ($d=1$ km) impactors 
make $D=1$ km ($D=20$ km) craters, we would find $N_1/N_{20} \sim 540$, a factor of 2.6 below the 
$N_1/N_{20}$ value adopted above. To obtain $N_1/N_{20} \sim 1400$ from the NEA population, one would 
have to assume that a $d=25$--30-m impactor makes $D=1$ km lunar crater.  

Comparing $N_{20}$ from Fassett et al. (2012) with $N_1$ from Neukum et al. (2001), both for Fra 
Mauro/Imbrium highlands, gives $N_1/N_{20}\simeq 1400$. With $N_1$ for Fra Mauro/Imbrium 
highlands from Robbins et al. (2014),
we get $N_1/N_{20}\simeq 1800$. Playing the same game for the young lunar maria, where the $N_1$ counts
are secure (Robbins et al. 2014), but the $N_{20}$ counts require an extrapolation (Neukum et al. 2001),    
gives $N_1/N_{20} \sim 700$. Overall, these values may indicate that there was a trend of $N_1/N_{20}$ 
with $T$, with the younger terrains showing lower $N_1/N_{20}$ values than the old terrains. If this 
is interpreted in terms of the crater scaling laws, it would hint on a time dependence of the 
impactor-to-crater ratio (e.g., related to the depth of porous megaregolith; Le Feuvre \& Wieczorek 2011). 
If this is interpreted in terms of the size distribution of impactors, we would speculatively infer that 
the size distribution of leftover planetesimals below $d \simeq 1$ km was (slightly) steeper than 
that of modern NEAs.

Additional $N_1$ calibration data point comes from the analysis of Chang'e-5 samples by Che et al. (2021),
who found $N_1(T)=1200$--1800 (per $10^6$ km$^{2}$) for a $T=1963 \pm 57$ Ma old lunar terrain. Using the standard 
$N_1(T)$ chronology from Eq. (\ref{final2}), we find $N_1(T) \simeq 1980$ for $T=1963$ Ma -- only a slightly 
higher value than $N_1$ reported by Che et al. (2021). This is consistent with the observation made in 
Che et al. (2021) that the existing chronology curves, which are calibrated on very young lunar terrains 
($T<1$ Ga; Cone, Tycho, North and South Ray craters; Neukum et al. 2001, Marchi et al. 2009, Hiesinger
et al. 2012), fall above the $N_1$ value obtained from Chang'e-5 samples. This may indicate that the impact 
flux at $T \simeq 2$ Ga was slightly lower (by $\simeq20$\%) than the one estimated here.

\section{Archean spherule beds}

When a large impactor strikes the Earth, it produces a vapor-rich ejecta plume containing numerous 
small melt droplets, most of which rise above the atmosphere. As the plume cools down, glassy 
spherules form and fall back, producing a global layer that can be several millimeters (for a 
Chicxulub-sized impact event) to many centimeters thick (Johnson \& Melosh 2012). The late Archean (2.5--3.5 Ga)
spherule beds are thicker than those associated with the 66 million years old, 180 km wide Chicxulub crater -- 
estimated $d \sim 10$ km impactor (Collins et al. 2020) -- and should have therefore been produced 
by $d>10$ km impactors (Bottke et al. 2012, Johnson et al. 2016a). Some $\sim 16$ spherule beds 
have been found in the late Archean (e.g., Marchi et al. 2021), although preservation biases and 
incomplete sampling may be an issue. At least some of these layers may have been produced by very 
large, $d \sim 50$ km impactors.   

\begin{figure}
%\epsscale{0.487}
%\plotone{earth1.eps}
%\epsscale{0.49}
%\plotone{earth2.eps}
\epsscale{1.2}
%\hspace*{0.5cm}\plotone{earth.pdf}
\hspace*{0.5cm}\plotone{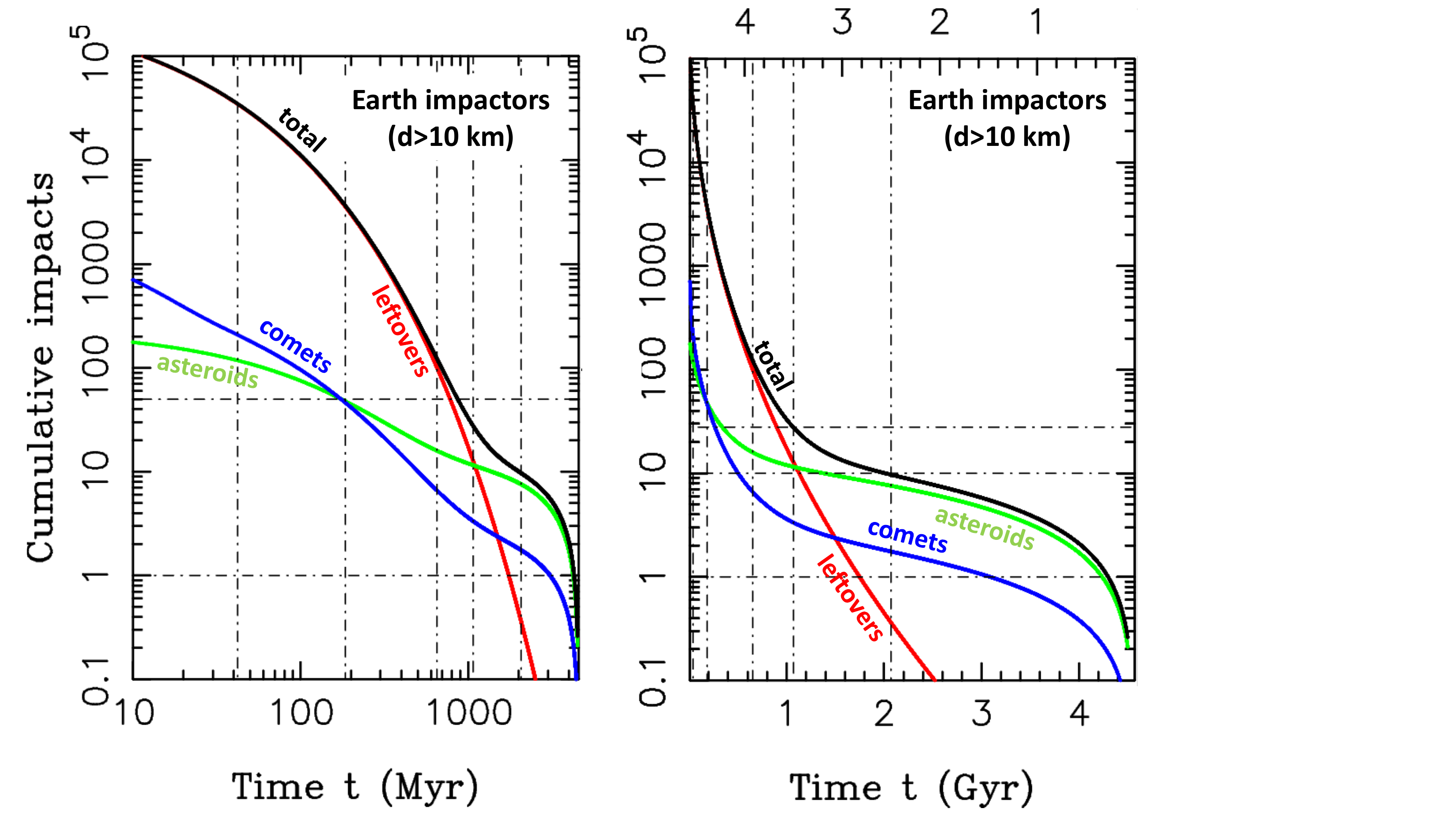}
\caption{Impacts of $d>10$ km bodies on the Earth. The vertical dash-dotted lines show the Moon formation 
in our model ($t=42$~Myr), estimated start of the known lunar basin record ($t \simeq 190$ or 
$T \simeq 4.38$ Ga), Imbrium formation ($t \simeq 650$ Myr or $T\simeq3.92$ Ga), and the beginning/end 
of the late Archean period ($T = 2.5$--3.5 Ga). The numbers on the upper axis of the left panel indicate 
time $T$ (in Gyr) measured looking back from today.}
\label{earth}
\end{figure}

The impact flux of $d>10$ km bodies on the Earth is shown in Fig. \ref{earth}. We find $\simeq 20$ 
$d>10$-km impacts on the Earth for $T=2.5$--3.5 Ga. This is similar to the number of known
spherule beds in the late Archean period.\footnote{The known Archean spherule beds occur in two
  distinct time intervals, 2.4--2.7 Ga and 3.2--3.5 Ga (e.g., Lowe \& Byerly 1986, Byerly et al. 2002,
  Glass \& Simonson 2013, Lowe et al. 2014, Schultz et al. 2017, Ozdemir et al. 2019; see
  Marchi et al. 2021 for a recent review). If they are indicative
of the average flux in the late Archean, there should be $\sim 14$ additional spherule beds 
at 2.7--3.2 Ga. If so, the number of $d>10$ km impacts in our model would represent $\sim 60$\%
of the total number of spherule beds. For reference, some $\sim 43$ $d>7$ km impacts on the Earth 
are expected in our model at 2.5--3.5 Ga (scaled from $d>10$ km to $d>7$ km with the size distribution 
of main belt asteroids).} The leftover planetesimals and main-belt asteroids contribute
equally to impacts in this time frame ($\sim 10$ impacts each). Whereas the asteroid impacts were 
more uniformly spread over late Archean, nearly all planetesimal impacts should have happened 
for $T>3$ Ga. The model predicts that $\sim 10$ and $\sim 2.2$ $d>10$-km asteroids should have 
impacted the Earth in the last 2.5 and 0.6 Gyr, respectively. Assuming that the number ratio of 
$d>10$-km to $d>50$-km impactors is 8.9, as inferred from the size distribution of main belt 
asteroids, the model implies $\sim 2$ $d>50$ km impactors in late Archean. 

\section{Martian crater chronology}

\begin{figure}
%\epsscale{0.487}
%\plotone{mars1.eps}
%\epsscale{0.49}
%\plotone{mars2.eps}
\epsscale{1.2}
%\hspace*{0.5cm}\plotone{mars.pdf}
\hspace*{0.5cm}\plotone{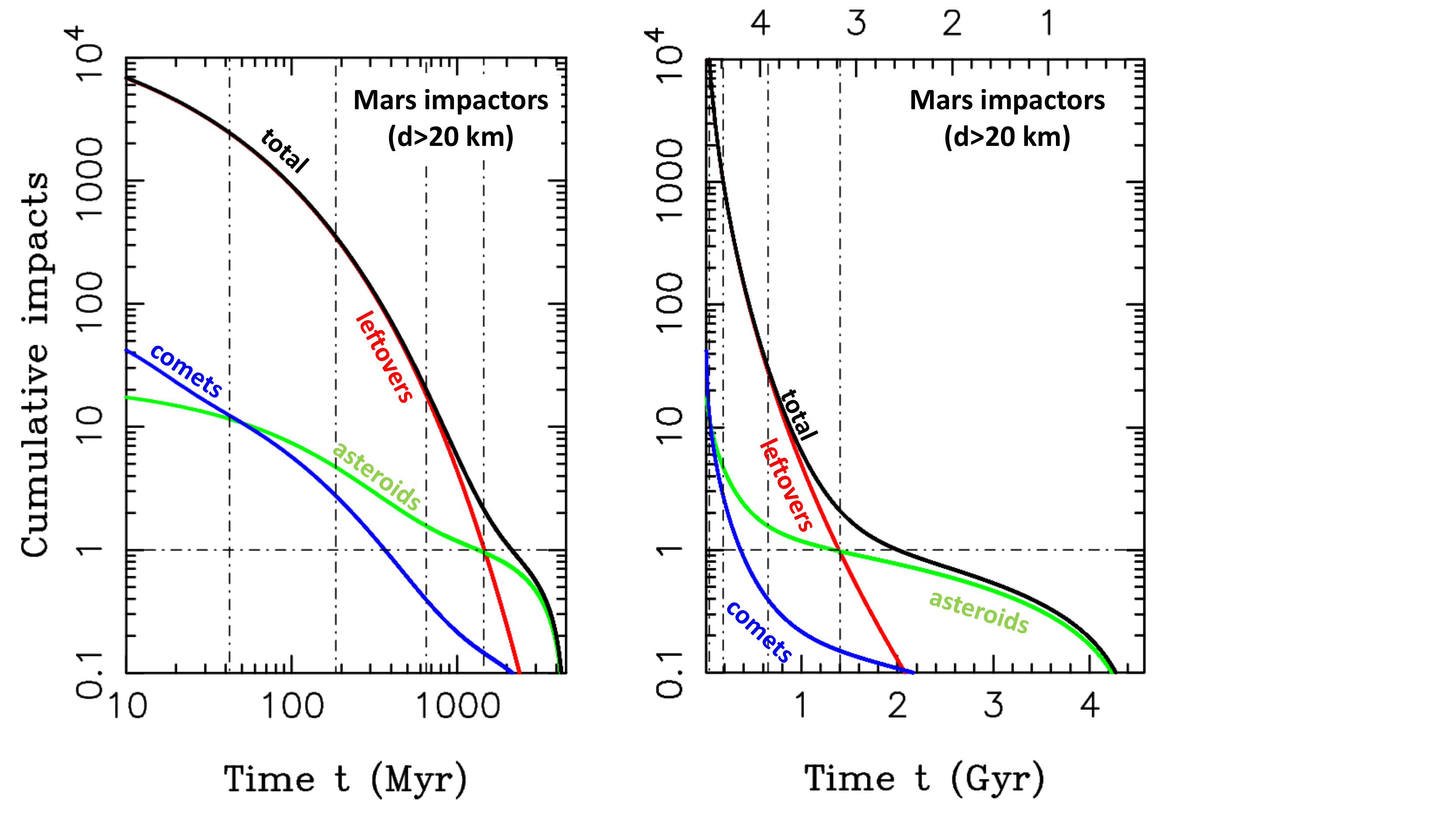}
\caption{Impacts of $d>20$ km bodies on Mars. The plot shows the accumulated number of impacts {\it since} 
time $t$ (i.e., on the surface with age $T$), where $t=0$ ($T=4.57$ Ga) is the birth of the solar system, 
and $t \simeq 4.57$ Gyr ($T=0$) is the present time. The vertical dash-dotted lines show the Moon 
formation in our model ($t=42$~Myr), estimated start of the known lunar basin record ($t \simeq 190$ or 
$T \simeq 4.38$ Ga), Imbrium formation ($t \simeq 650$ Myr or $T\simeq3.92$ Ga), and transition from the 
planetesimal-dominated to asteroid-dominated impact stages for Mars ($t \simeq 1.4$ Gyr or $T \simeq 3.2$ 
Ga). The profiles were normalized to having $4 \times 10^5$ $d>10$-km planetesimals at $t=42$ Myr, and 
$r_{\rm out}=1.5$ au. The numbers on the upper axis of the left panel indicate 
time $T$ (in Gyr) measured looking back from today.} 
\label{mars}
\end{figure}

The impact flux of $d>20$ km impactors on Mars is shown in Fig. \ref{mars}. The Martian impact profile 
is different from that of the Moon, Earth and Venus. The impact profile is more extended in time and has 
a longer tail of late impacts. This is simply because leftover planetesimals at $\sim 1.5$ au live longer 
and can impact later. The transition from the planetesimal-dominated to asteroid-dominated impact stages 
thus probably occurred later for Mars than for the Moon ($t \simeq 1.4$ Gyr vs. $t \simeq 1.1$ Gyr).
The impact flux of leftover planetesimals on Mars is more sensitive to $r_{\rm out}$ than that of 
any other terrestrial world. The results shown in Fig. \ref{mars} were obtained for our nominal case
with $r_{\rm out}=1.5$ au (and $4 \times 10^5$ $d>10$-km planetesimals at $t=42$ Myr). The overall 
number of planetesimal impacts on Mars is $\simeq1.4$ times lower for $r_{\rm out}=1.25$ au, and $\simeq 2$ lower
for $r_{\rm out}=1$ au. This would reduce the planetesimal contribution and shift the transition back
in time.

%Following the method described in Sect. 10, the $N_{20}(T)$ chronology (per $10^6$ km$^2$; Fig. 
%\ref{mars2}--left panel) for Mars can be given as
%\begin{equation}
%N_{20}(T) = a \exp [ -(t/6\, {\rm Myr})^{0.4}] + c T  
%\label{marsn20}
%\end{equation}  
%with $t=4570-T$ ($t$ and $T$ given in Myr), $a = 2.8 \times 10^4$ and $c=1.6 \times 10^{-3}$~Myr$^{-1}$. 
%The first term in Eq. (\ref{marsn20}) stands for the leftover planetesimals, the second stands for NEAs. 
%This chronology takes into account the dependence of the scaling laws on surface gravity and impact 
%speed. Combining these effects we find that a $d \simeq 1.4$-km impactor is needed to produce a 
%$D = 20$-km Martian crater. The difference between $d=1$ km (for lunar surface) and $d \simeq 1.4$-km, 
%when convolved with the size distribution of impactors, contributes by a factor of $\sim 2$ to reduce
%the $N_{20}$ for Mars (relative to lunar $N_{20}$). 

Following the method described in Sect. 11, the $N_{1}(T)$ chronology (per $10^6$ km$^2$; Fig.~\ref{mars2}) 
for Mars can be given as
\begin{equation}
N_{1}(T) = a \exp [ -(t/6\, {\rm Myr})^{0.42}] + c T  
\label{marsn1}
\end{equation}  
with $a = 3.7 \times 10^7$ and $c=0.41$~Myr$^{-1}$. The second term in Eq. (\ref{final2}) -- corresponding to 
the early impact flux of asteroids -- in not included here for simplicity.  
Here we truncated the planetesimal disk at $r_{\rm out}=1.5$~au (at $t=0$).    
The contribution of leftover planetesimals is $\simeq 2$ times lower 
for $r_{\rm out}=1$ au; hence $a = 1.9 \times 10^7$ for $r_{\rm out}=1$ au. The asteroid branch -- factor 
$c$ in Eq. (\ref{marsn1}) -- is rescaled from lunar $c=1.0$ Myr$^{-1}$ (Eq. \ref{final2}) using input 
from the dynamical models of modern NEAs (e.g., Granvik et al. 2018). The models indicate $R_{\rm b} \simeq 1.2$ 
for small NEAs (Nesvorn\'y et al. 2023), where $R_{\rm b}$ is the number ratio of Mars-over-Moon impacts 
normalized to a unit surface area; Hartmann \& Neukum 2001). Note that this value is much lower than 
the one adopted in Hartmann (2005) and Marchi (2021), $R_{\rm b} \simeq 2.6$.

We assume the impact velocities 14--15 km s$^{-1}$ for Mars and 19--20 km s$^{-1}$ for the Moon, fold in 
the effect of different surface gravities of the Moon and Mars, and estimate that Mars requires a 
$\sim 1.4$ times larger impactor than the Moon to create a $D=1$ km crater (Johnson et al. 2016a). 
Our reference asteroid size distribution (Harris \& D'Abramo 2015, Bottke et al. 2020) has a steep 
slope for $20<d<50$ m with the  cumulative power index $\simeq 3.2$. The Martian $N_{1}(T)$ in Eq. 
(\ref{marsn1}) is therefore penalized by a factor of $1.4^{3.2} \simeq 2.9$ relative to the lunar chronology.
We use the same penalization for the planetesimal and asteroid branches. 
 
\begin{figure}
%\epsscale{0.5}
%\plotone{mars_n1_comp.eps}
\epsscale{1.2}
%\hspace*{4.0cm}\plotone{marschrono.pdf}
\hspace*{4.0cm}\plotone{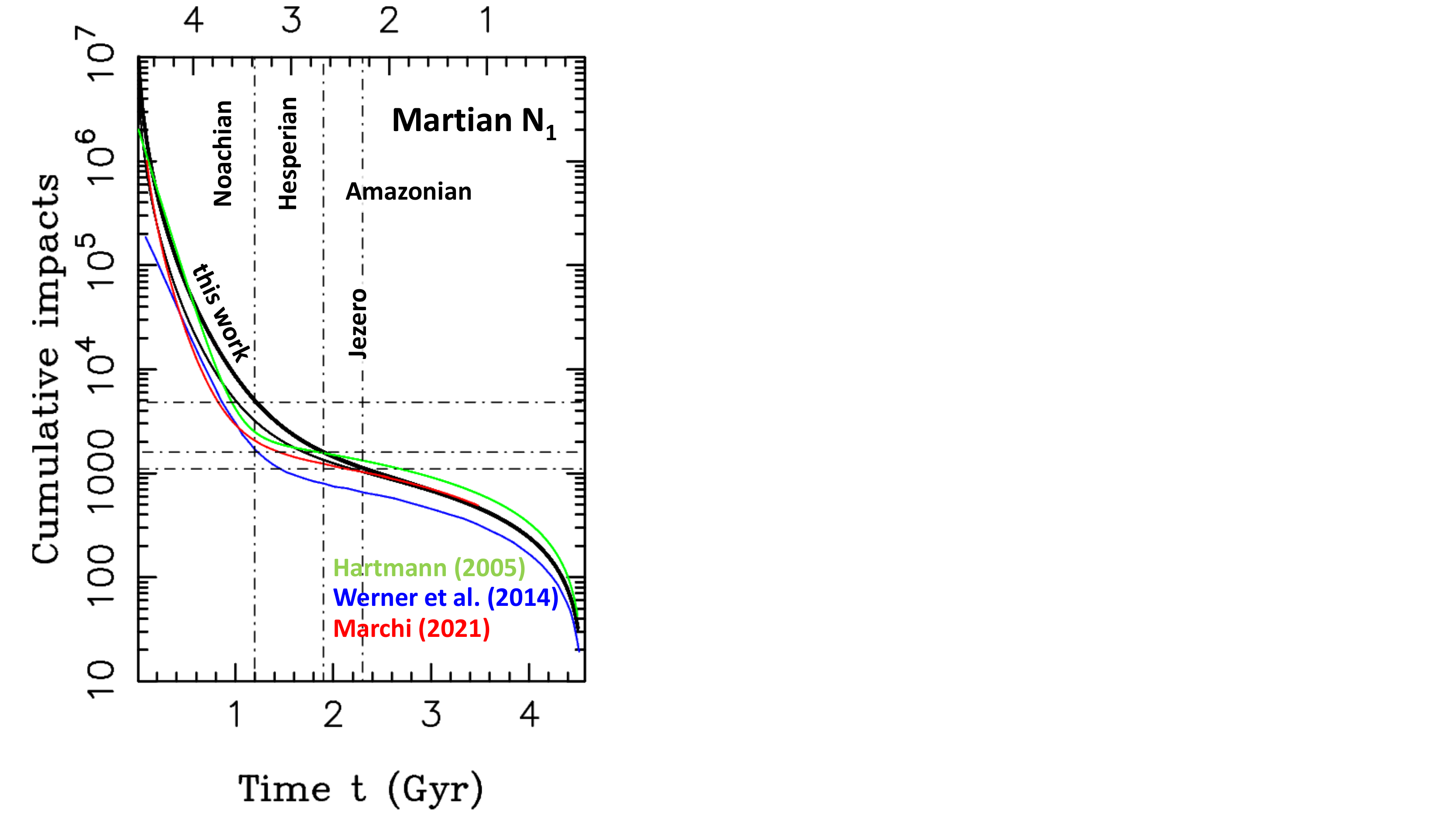}
\caption{The $N_{1}(T)$ chronologies for Mars. The bold and thin black lines show our Martian chronologies for 
$r_{\rm out}=1.5$ au and $r_{\rm out}=1$ au, respectively. For comparison, we also show the Martian chronologies 
from Hartmann (2005, green line), Werner et al. (2014, blue), and Marchi (2021, red; early instability/NEA scaling). 
The horizontal dot-dashed lines are the crater counts for Noachian/Hesperian and Hesperian/Amazonian boundaries, 
and for the Jezero crater terrains. The vertical dashed lines are our age estimates for these units in the 
model with $r_{\rm out}=1.5$ au. The ages inferred with $r_{\rm out}=1$ au are $\sim 200$ Myr older.
The numbers on the upper axis indicate time $T$ (in Gyr) measured looking back from today.} 
\label{mars1}
\end{figure}

Figure \ref{mars1} compares the Martian chronology obtained here with the chronologies from Hartmann (2005),
Werner et al. (2014) and Marchi (2021). We plot two chronologies from Eq. (\ref{marsn1}),
one for $r_{\rm out} = 1$ au (thin black line) and $r_{\rm out} = 1.5$ au (thick black line), to illustrate 
the dependence on the initial planetesimal profile. There is a significant difference between the two 
(the planetesimal branch of $N_1(T)$ is a factor of $\simeq 2$ lower for $r_{\rm out} = 1$~au). This is bad 
news for the prospect of accurate dating: the age estimates derived from the two chronologies differ by 
up to $\sim 200$ Myr for $T>2$ Gyr. But this can be good news for the prospect of constraining the radial 
extension  of the planetesimal disk from the Martian crater record -- assuming that the radiometric age 
of an old Martian terrain with known $N_1$ will be measured in the near future (e.g., NASA Mars 2020).           

The asteroid branches ($T<2.5$ Ga) of the Hartmann (2005) and Werner et al. (2014) chronologies are 
a factor of $\sim 1.5$ higher and lower, respectively, that the one derived here. We find a good agreement 
with the chronology of Marchi (2021) for $T<2$ Ga.\footnote{This is somewhat a coincidence because Marchi 
(2021) adopted $R_{\rm b}=2.6$ for asteroids, a steeper size distribution for $d=20$--50 m and an additional 
0.81 reduction factor from Popova et al. (2003).} Looking back in time, the Hartmann, Werner, and Marchi 
chronologies continue relatively flat to $T > 3$ Ga, where they show a sharp bend upward. Our chronologies 
instead connect to the planetesimal branch and start raising already at $T \sim 2.5$ Ga. 

Our best age estimate for the Noachian/Hesperian and Hesperian/Amazonian boundaries is 3.4--3.6 Ga and 
and 2.6--2.9 Ga (the range given here for $r_{\rm out}=1$--1.75 au and the standard calibration of 
leftover planetesimals). Here we adopt the crater densities $N_1=4.8 \times 10^3$ and $N_1=1.6 \times 10^3$, 
both per $10^6$ km$^2$, as defining the two boundaries (Tanaka 1986, Hartmann 2005). The Jezero crater -- 
relevant to the NASA Mars 2020 mission -- is estimated to be 2.2--2.5 Gyr old for $N_1=1.1 \times 10^3$ per 
$10^6$ km$^2$ from Warner et al. (2020) or 2.4--2.7 Gyr old for $N_1=1.5 \times 10^3$ per $10^6$ km$^2$ 
from Shahrzad et al. (2019).

For illustrative purposes, following the traditional approach (e.g., Hartmann \& Neukum 2001, Hartmann 2005, 
Werner et al. 2014, Marchi 2021), we also approximately rescale our lunar chronology from Eq. (\ref{final2}) 
to Mars. For that, we use $R_{\rm b}=1$ (roughly applicable for $r_{\rm out}=1.25$--1.75 au) for the 
planetesimal branch and $R_{\rm b} \simeq 1.2$ for the asteroid branch (Nesvorn\'y et al. 2023), and 
penalize Martian $N_{1}(T)$ by the factor of 2.9 to account for the impact velocity and surface gravity 
difference. For comparison, Marchi (2021) used $R_{\rm b}=0.5$ for planetesimals from Morbidelli et al. (2018), which is 
smaller than our value, presumably because their terrestrial formation model was effectively run with 
$r_{\rm out} \sim 1$ au. Indeed we find $R_{\rm b}=0.63$ for $r_{\rm out} \sim 1$ au (statistics based on 
all impacts). This is consistent with a smaller planetesimal contribution for disks truncated at $r_{\rm out} 
\sim 1$ au (for Mars, not in general -- the planetesimal branch for the Earth is only slightly steeper for 
$r_{\rm out} = 1$ au than for $r_{\rm out} = 1.5$ au, and the overall number of terrestrial impacts remains 
nearly the same -- again adopting $4 \times 10^5$ $d>10$-km planetesimals at $t=42$ Myr, independently 
of $r_{\rm out}$). Figure \ref{mars2} shows that the Mars chronology function is more extended in time than 
the lunar chronology, implying that $R_{\rm b}$ is time dependent. {\it Applying the lunar chronology function 
to Mars can lead to inaccurate age estimates that can differ, by up to $\sim 500$ Myr, from the age estimates 
obtained from the accurate Martian chronology.}  

Scaling from the most densely cratered surfaces of the Moon and Mars, Bottke \& Norman (2017) estimated 
$\sim 200$ $D>150$ km craters over the whole lunar surface, and $\sim 500$ $D>150$ km craters over the 
whole Martian surface. From the scaling laws we find that a $d \simeq 10$ km impactor is needed 
to produce a $D=150$ km crater on the Moon, and a $d \simeq 13$ km impactor is needed to make a 
$D = 150$ km crater on Mars (Holsapple \& Housen 2007, Johnson 2016, Morbidelli et al. 2018).   
In Sect. 8, we estimated $\sim 180$ $d>10$ km lunar impacts for $T < 4.38$ Ga, which is consistent
with Bottke \& Norman (2017), assuming that the lunar surface recorded $D>150$ km craters since 
$T \simeq 4.38$ Ga (presumably the LMO solidification time). To obtain $\sim 500$ $D>150$ km 
craters for Mars with our model-derived impact flux, we infer that the Martian surface would have to 
record $D>150$ km craters since $T \simeq 4.27$ Ga for $r_{\rm out}=1.5$ au, and $T \simeq 4.35$ Ga 
for $r_{\rm out}=1$ au. Morbidelli et al. (2018) noted the same problem and suggested a global resurfacing 
event at $T \sim 4.4$ Ga, perhaps associated with the formation of the Borealis basin. There is currently 
no evidence for the late formation of Borealis. Robbins (2022) proposed that the formation of Borealis basin 
could have kept the surface warm enough for long enough to prevent large craters/basins from forming 
for an extended period of time. The volcanic and fluvial resurfacing of early Mars (Hartmann \& Neukum 2001) 
would have to be extremely powerful to globally erase $D>150$-km craters for $T>4.3$--4.4~Ga.

%Cox et al. (2022) found evidence for high-pressure shock effects in a 4.45-Gyr old zircon
%from the matrix of NWA (Northwest Africa) 7034. These findings provide direct evidence of 
%large impacts that persisted on Mars for $T\lesssim 4.48$ Ga.    

%The impact flux profiles do not depend much on weighting. Planetesimals starting at 0.5--1 au
%produce a slightly stronger decline of impact flux but the difference is not large with $\tau=8$--10 Myr
%(instead of 12 Myr). Mars continues having a slightly weaker decline with $\tau \simeq 13$--14 but
%the statistics is not great for Mars. 

\section{Discussion}

We find that the terrestrial--zone planetesimals were the dominant source of lunar impactors for 
$T>3.5$ Ga, and that asteroids were the dominant source of impactors for $T<3.5$ Ga. This is in 
line with the findings of Morbidelli et al. (2018).\footnote{Brasser et al. (2020), instead, 
suggested that the early impact record was dominated by bodies from the E-belt.} The leftovers 
are expected to evolve collisionally (Bottke et al. 2007)\footnote{Unlike Bottke et al. 
(2007), we find that the effects of collisional grinding were not strong enough to prevent 
the leftover planetesimals from producing the dominant share of lunar basins.} and reach an 
asteroid--belt--like size distribution in only $\sim 20$ Myr after the first solar system solids. 
The size distribution of lunar impactors in the last 3.5 Gyr should be similar to that of modern 
NEAs. The modern NEAs evolve from the asteroid belt by size-dependent radiation processes which 
favor mobility of very small bodies (Vokrouhlick\'y et al. 2015). They therefore have slightly 
steeper size distribution for $d<10$ km than the main belt. This explains why the crater size 
distribution on ancient lunar craters is related to the main belt asteroids, and why the modern 
impactors have a NEA--like size distribution (Str\"om et al. 2005). Head et al. (2010) suggested 
that the transition between the two populations of impactors happened $\sim 3.5$ Ga, which is 
what we find here from dynamical modeling. 

Minton et al. (2015) pointed out that the asteroid-like size distribution of early lunar impactors 
would produce too many mega-basins ($D>1200$ km) and suggested that the size distribution of 
impactors below $d \simeq 100$ km was somewhat steeper than that of today's asteroid belt. 
Johnson et al. (2016a) reiterated that point and proposed that the lunar basin data and Archean 
spherule beds could best be fit with a main-belt--like size distribution of impactors for 
$d\lesssim50$ km, and a steeper slope for $d \gtrsim 50$ km. Here we find that the ancient impactors were 
leftover planetesimals from the terrestrial planet zone ($r \lesssim 1.5$ au), not asteroids. This makes 
it easier to understand any inferred differences. Note, however, that the size-distribution break at 
$d \sim 50$ km would imply fewer Imbrium impacts, and would diminish the probability of the late 
Imbrium formation in our model.  

We showed that the lunar crater record is consistent with having $(2.6$--$5.2) \times 10^5$ 
$d>10$~km planetesimals in the terrestrial planet zone ($\sim 0.5$--1.5 au) at $\sim 50$ Myr after 
the first solar system solids. The collisional evolution in the first $\sim 20$ Myr would have been 
stronger for higher initial planetesimal mass and weaker for lower initial mass. We therefore 
cannot predict, from the collisional modeling alone, the initial planetesimal mass. All that we 
can say is that there was at least $\sim 0.1$ $M_{\rm Earth}$ in planetesimals to start with. This 
is significant, however, because it shows that there {\it was} a large population of planetesimals 
to start with. The formation models where the terrestrial planets grow from cm-size pebbles (e.g., 
Johansen et al. 2021) do not postulate any large planetesimal population in the terrestrial 
planet zone (they do not exclude it either).     

Using the lunar basin record as a constraint on the LMO solidification (Sect. 8), we found that the 
LMO should have lasted to $T = 4.36$--4.42 Ga, roughly 160--210 Myr after the first solar 
system solids, and $\sim 110$--160 Myr after the Moon-forming impact in our model (Sect. 2). Morbidelli 
et al. (2018) reached similar conclusions based on considerations related to the HSEs in the 
lunar mantle. They considered the possibility that HSEs were sequestered from the mantle of the Earth 
during magma ocean crystallization, due to iron sulfide exsolution (Rubie et al. 2016), and showed that 
this likely affected the Moon as a well (if the lunar mantle overturn is taken into account; 
Elkins-Tanton et al. 2011). The HSE would accumulate in the lunar mantle only after the LMO crystallization, 
estimated to happen $\sim 100$--150 Myr after the Moon formation. This would correspond to 
$\sim 150$--200~Myr after the solar system solids if the Moon formed at $t \sim 50$~Myr. 

Figure \ref{noble}A shows the HSE constraints for the Earth, Mars, and Moon. Assuming chondritic 
composition of impactors it has been inferred that these worlds accreted $3 \times 10^{25}$~g, $1.6 \times 
10^{24}$ g, and $1.5 \times 10^{22}$ g during the Late Veneer (i.e., after their differentiation is fully 
over such that the accreted HSEs do not sink to the core). Assuming 30\% retention of impactor mass for 
the Moon (Zhu et al. 2019a), we find that the LMO would have to solidify at $t \sim 200$ Myr ($T 
\sim 4.37$ Ga) to explain lunar HSEs. This is in an excellent agreement with the results of Morbidelli 
et al. (2018). The Earth differentiation should have ended within $\sim10$ Myr after the 
Moon-forming impact (e.g., Elkins-Tanton et al. 2011). We find that the accreted mass for $t>50$ Myr
is a factor of $\sim 3$ too low to explain Earth's HSEs (it would be a factor of $\sim 10$ too low
if the Moon-forming impact happened at $t \sim 120$ Myr; Maurice et al. 2020, Kruijer et al. 2021).
This shows the need for the accretion of very large planetesimals ($\sim 1000$--3000 km; Bottke et al. 
2010, Marchi et al. 2014). The contribution of very large planetesimals to Earth's HSEs was not accounted 
for in Fig. \ref{noble}A, because we adopted a steep size distribution for $d>100$ km (Fig. \ref{grind}). The 
stochastic accretion of very 
large planetesimals could help to explain the large difference in the HSE content between the 
Moon and Earth ($\sim 2000$; Bottke et al. 2010).\footnote{A great share of this difference is 
explained by the smaller accretional cross-section of the Moon (factor of $\sim20$, focusing 
included), long-lived lunar LMO (factor of $\sim10$), and lower retention factor of the Moon
(factor of $\sim 3$; Zhu et al. 2019a).} 

If Mars accreted and differentiated early ($t \lesssim 10$ Myr, Dauphas \& Pourmand 2011, Marchi 
et al. 2020), it would have accreted $\sim 5 \times 10^{25}$ g in our model during the Late Veneer
(in small planetesimals). This is a factor of $\sim 3$ higher than the late addition of chondritic material
inferred from Martian meteorites (shergottite-nakhlite-chassigny, SNC; Marchi et al. 2020). It is 
possible that the average mantle abundance of HSEs is underestimated, perhaps because the average 
mantle abundance may be difficult to establish from (heterogeneous) SNC meteorites, or perhaps 
because our SNC collection is not fully representative of the Martian mantle. 
The retention of impactor mass could contribute as well. In Fig. \ref{noble}A,
we assumed a 100\% retention factor for Mars but at least some work indicates that the retention 
factor may be lower (e.g., $\sim 60$\% in Artemieva \& Ivanov 2004). Finally, as we already 
discussed, the impact flux of planetesimals on Mars is lower when the disk of terrestrial 
planetesimals is truncated at the lower orbital radius. For example, if $r_{\rm out} \simeq 1$ au,
instead of $r_{\rm out} = 1.5$ au that we used in Fig. \ref{noble}A, the accreted mass would be 
reduced by a factor of $\simeq 2$. 
    
The cometary impact profile is more extended in time than previously thought (Morbidelli et al. 
2018). This is a consequence of the much improved statistic of simulations reported in Sect. 
4.2. With the standard comet calibration, accounting for spontaneous comet disruptions, and 
assuming that the Moon-forming impact happened $\sim 50$ Myr ($T \sim 4.52$ Ga) after the birth 
of the solar system, we find that the Earth would have accreted $\sim 1.5 \times 10^{22}$~g of 
cometary material for $T<4.52$ Ga (Fig. \ref{noble}B). This is consistent with comets being the 
source of noble gasses in the Earth atmosphere (Marty et al. 2016). The mass accreted in 
cometary material would be larger if: (i) some comets fade instead of disrupting (Fig. \ref{noble}B), 
(ii) the instability happened later, or (iii) the Moon formed earlier. In fact, the noble gas 
argument can be used to roughly constrain the delay between the instability and Moon formation, $\Delta t$.
From Fig. \ref{noble}B we infer $20 < \Delta t < 60$ Myr, with the exact value depending on the 
physical lifetime of comets (Sect. 4.2). This suggests, if the instability happened very early 
($t < 10$ Myr; Clement et al. 2018, Liu et al. 2022), that the Moon must have formed early as 
well ($t < 70$ Myr or $T>4.5$ Gyr; Thiemens et al. 2019).

We reported the impact profiles of leftover planetesimals from the best terrestrial planet formation 
simulation from Nesvorn\'y et al. (2021a). This falls short, given that only one case was tested, 
to understand the possible variability of the impact profiles when different assumptions are made. 
This should be the focus of future investigations. If the variability is large, it would perhaps 
be possible to rule out/confirm some specific setups based on the lunar crater constraints. For 
example, the initially very massive asteroid belt (Clement et al. 2019; $\sim 1$ $M_{\rm Earth}$), 
would presumably lead to a much larger (a factor of $>100$?) contribution of asteroid impactors
to the early impact record. We tested the effects of the radial profile of the planetesimal disk 
in the terrestrial planet zone and found that the cases with the outer disk edge at $r \sim 1$--1.75 au 
work quite well to match different constraints (with the mass $\gtrsim 0.1$ $M_{\rm Earth}$ 
in planetesimals at the time of the gas disk dispersal). Whereas this is broadly consistent with at 
least some planetesimal formation models (e.g., Izidoro et al. 2021, Morbidelli et al. 2022), the 
planetesimal population needs to be better characterized (e.g., from the radiometrically calibrated 
Martian chronology) before a more specific inference can be made.  

%The piece-wise derivation of $N(>\!\!D)$ from young ($T<1$ Gyr) and old surfaces ($T=3$--4 Gyr) in the NI94 model
%is not ideal because we have empirical evidence that the crater distribution on the oldest and youngest surfaces
%is different. For example, Str\"om et al. (2005) pointed out that the crater record on the oldest terrains
%(Lunar Highlands) is virtually identical in size distribution to the present main belt asteroids (MBAs), whereas
%the younger craters show steeper size distribution similar to the present NEAs.
%Such a difference is expected because modern NEAs are supplied from the main belt by size-dependent radiation
%effects (the Yarkovsky effect; Bottke et al. 2006, Vokrouhlick\'y et al. 2015). In the model presented here,
%the early impactors were leftover planetesimals that collisionally evolved in $t<20$ Myr to have the
%asteroid-belt--like size distribution (Sect. X).

\section{Comparison with previous work}

The originality of the present paper consists in explaining the lunar crater record with a physical model 
that is firmly tied to the terrestrial planet formation and observational constraints. We show that lunar 
basins were produced by impacts of leftover planetesimals that formed in the terrestrial planet zone -- a 
notable shift from previous theories that invoked E-belt, main-belt asteroids and comets. This has important 
implications for planet formation: our work constrains the initial mass of planetesimals and identifies 
planetesimal accretion as the preferred mode of the terrestrial planet growth.

The results address several critical scientific issues, some of which date back to the Apollo program,
including the late formation of Imbrium and Orientale, and provide context for the interpretation of 
classical/empirical impact chronologies. The model impact flux of planetesimals and asteroids on the 
Earth matches the number of spherule beds found in the late Archean. Cometary impactors provide the 
right amount of noble gases for the Earth atmosphere. In a broader context, our work highlights the 
importance of impacts for the early Earth.

Here we relate our work to two previous publications (Bottke et al. 2007, Brasser et al. 2020) that 
considered the same general subject -- impacts in the early Solar System -- but ended up favoring opposing 
conclusions. We discuss how various assumptions and approximations adopted in these works may have affected 
the results.  Bottke et al. (2007), hereafter B07, studied lunar impacts of leftover planetesimals. They 
accounted for the collisional evolution of planetesimals and showed how the planetesimal population is 
reduced by (disruptive) collisions, and that the reduction factor is greater when the initial population is larger.
This is the notion of the 'convergent collisional evolution' that we use in this work to calibrate 
the population of planetesimal impactors from new \texttt{Boulder} simulations. 

B07 concluded that the leftover planetesimals \textit{cannot} produce basin-scale impacts 
on the Moon during the Imbrium era, because the collisional and dynamical decline of planetesimal 
impactors was presumably too quick; by 3.9 Ga, the planetesimal population was reduced by a huge factor 
in their simulations. Comparing their impact profile with the ones obtained here, we 
identify the main reason behind this: the planetesimal population in B07 dynamicaly dropped 
by two orders of magnitude in 100 Myr and transitioned to a more steady decline after 100 Myr (their 
figure 5). We do not find any such strong initial trend in our work. 
The problem in question is most likely related to the approximate nature of initial conditions in
B07, where the present-day near-Earth asteroids were used as a proxy for 
terrestrial planetesimals. We note that modern NEAs have relatively short dynamical lifetimes,
which may be the main reason behind the fast decline seen in B07 in the first 100 Myr. 
For comparison, we determine the impact flux from accurate simulations of the terrestrial 
planet formation (Sect. 2), and find that the late lunar impactors were stored on Mars-crossing orbits 
at 1.2--1.7 au (Sect. 9).

Brasser et al. (2020), hereafter B20, developed a dynamical model for lunar impactors with four different 
components: E-belt, asteroid belt, comets and leftover planetesimals.\footnote{Brasser et al. (2021)
followed the implications of the model developed in B20 for HSEs. 
We do not comment on Brasser et al. (2021) here, because B20 is 
more closely related to our work.} The general methodology used in B20. 
is similar to that used in the present work (numerical integrations used to construct the impact model).
The main methodology-related difference between our work and B20 is that the B20 
model is uncalibrated and can (and must, see below, because it does not fit the lunar record) be 
adjusted to satisfy constraints. Our model is more rigid. All impactor populations in our model 
were calibrated from independent means (Sects. 4 and 5). Our model can either fit or not fit 
the lunar record -- there is no middle ground or room for data-driven adjustment. 

Some of the main results in B20 were presented in terms of $N_{20}$ in their Sect.  
5.2 (their Figs. 5 and 8). Here, B20 adjusted their model to match the Neukum's lunar 
chronology. This was done by assigning weights to leftover planetesimals and E-belt. The 
best-fit contribution of planetesimals was found to be zero; leftovers were therefore concluded to 
be only a minor source of lunar impacts. E-belt was assigned a weight equal to $\sim 10$ (to fit 
the Neukum chronology). This implies that the E-belt population would have to be increased by a 
factor of $\sim 10$ to fit the existing data. B20 proceeded by explaining why this choice may be reasonable.

This can be related to the present work where we found the dominant role of leftover planetesimals -- 
with the standard calibration, no need for any (arbitrary) enhancement. We considered E-belt as an 
extension of the asteroid belt (this is what the letter E stands for; Bottke et al. 2012) and assumed 
that it is not justified to arbitrarily increase its initial population if things do not work (Sect.
4.1). There is no obvious reason for why the initial number density of asteroids immediately below 
2 au should have discontinuously increased from that immediately above 2 au. B20 argued 
that the probability of Rheasilvia basin formation on Vesta would be too low if the E-belt population 
were not enhanced, but this is simply not the case. In a work dedicated to Vesta's cratering, Roig 
\& Nesvorn\'y (2020) determined a $\sim 50$\% probability to form Rheasilvia over 4.5 Gyr (with 
standard assumptions). The Hungaria population, which represents an important constraint on the E-belt 
(Bottke et al. 2012), was ignored in B20.

B20's asteroid flux is $\sim$10--20 times higher (as explicitly noted in their Sect. 5.2) 
than the one obtained here, where we calibrate asteroid impactors on the number of large main belt asteroids 
observed at the present time (and their orbital distribution; Nesvorn\'y et al. 2017a). This represents 
an anchor. We use (forward) numerical simulations of asteroids, tie the results to the anchor at 
the present time, and compute the historical impact flux from that.  This is a well-defined calibration 
method. B20 `calibrated' the E-belt impactors on the lunar crater data, but that presumes that the 
E-belt was the dominant source of lunar impactors in the first place.

The contribution of asteroids to lunar impacts was simulated in B20 by assuming that all 
planets formed and remained on their current orbits. This ignored dynamical effects of the outer 
planet migration/instability. When the outer planet migration/instability is accounted for, the 
main asteroid belt is depleted by a much larger factor and the impact profile is different (more 
asteroid impactors early on). B20 therefore underestimated the contribution of the main 
belt to early lunar impacts (by a factor $\sim 100$ relative to the enhanced E-belt; compare 
Fig. 8 in B20 to Fig. 10 in Nesvorn\'y et al. 2017a). 

The outer planet migration/instability is needed to reduce the population of inner belt asteroids 
and match the orbital structure of the asteroid belt (Nesvorn\'y et al. 2017a). It is not explained in 
B20 how the lunar impact flux of main belt asteroids was calibrated, and we can only 
speculate that B20 used the total population that was somewhat similar to that of the 
present asteroid belt. In any case, the simulation in B20 were not run long enough (only 
1 Gyr, not 4.5 Gyr) to demonstrate whether their results were consistent with the present-day asteroid 
belt. 

The lunar $N_{20}$ flux of main-belt asteroids was shown to drop below $10^{-7}$ by 3.8 Ga in B20 
(their Figs. 5 and 8). This indicates that their model would predict $N_{20} < 10^{-7}$ today. For comparison, 
the present-day lunar $N_{20}$ -- as inferred from NEAs and radiometrically dated young lunar surfaces 
--  is estimated $\simeq 10^{-6}$. This shows, even if the E-belt is enhanced by a factor of $\sim10$ 
in B20 to fit the Neukum chronology for $T>3.5$ Ga, the model is discrepant with the Neukum 
chronology for $T<3.5$ Ga (by at least an order of magnitude; the E-belt is not a major source of lunar 
impacts at the present epoch).  

The flux of cometary impactors in B20 drops by roughly three orders of magnitude from 4.5 Ga 
to 4.3 Ga. This can be contrasted with our results where cometary impacts decline by only a factor of 
$\simeq 30$ over the same interval. This is most likely caused by the small number statistics in B20.
We struggled with this problem ourselves and had to repeatedly increase the resolution of cometary 
reservoirs to get things right. Eventually, we simulated 100 million comets in Sect. 4.2 (cloning 
included). This happens because the impact probability of comets on the Moon is very low ($<10^{-7}$ 
per one body starting in the outer disk). When an insufficient resolution is used, very few comets 
evolve to 1 au late in the simulations and the impact probability computed from the \"Opik method is 
practically zero. B20 results for comets are in direct contradiction to the observed population of 
present-day Jupiter-family comets, by many orders in magnitude.

The planetesimal population in B20 was taken from simulations published in Brasser et al. 
(2016). They grouped together planetesimal populations obtained in different formation simulations, 
replaced the model-generated terrestrial planets with the real ones, and added outer planets. This is not 
ideal because the solar system is `not an average of many trial cases' -- it is just one 
realization of the initial conditions. That is why, in our work, we focus on the simulations 
from Nesvorn\'y et al. (2021) that best fit various constraints, including planetary masses and orbits, 
asteroid belt, the Moon-forming impact, etc. The outer planets were included at the start of our simulations 
(early migration/instability accounted for) and the inner planets did not need to be 
replaced -- practically all their attributes produced in the selected model were accurate.

The collisional evolution of leftover planetesimals was ignored in B20. This is another original
component of our work, where the collisional evolution is used to calibrate the population of leftover 
planetesimals. The planetesimal population is unconstrained if the collisional evolution is not accounted for. 
In B20, we presume that the planetesimal population was fixed from the HSE constraints (Morbidelli et al. 
2018). If so, this would not be consistent with the E-belt dominance in the impact record, as advocated later 
in their article. 

The overall lunar impact flux of planetesimals is underestimated by a factor of $\sim 10$ in B20 and the 
impact profile shown for planetesimals in their Fig. 5 is much too steep: it drops by $\sim 4$ orders of 
magnitude from 4.5 Ga to 3.8 Ga. This is roughly a factor of 10 faster decline than we obtain in the 
present work. Again, this is probably related to the insufficient statistics in B20 (we have $\sim 4$ 
times better statistics for planetesimals; the need for large statistics was established by convergence 
studies). If the decline were really that fast, the leftover planetesimals could not produce the 
Imbrium-era basins, including Imbrium itself. B20 followed planetesimals for only 0.5 Gyr,  which did not 
allow them to address the Imbrium formation. In their section 4.2.5., B20 suggested that the Imbrium event -- 
for which an E-belt impactor was assumed to be responsible  -- cannot be ruled at $>95$\% confidence. This 
would imply, if their statement is taken at a face value, that Imbrium was a $\sim 5$\% probability 
event in their model -- and that's already when the E-belt is enhanced by a factor of $\sim10$ in B20 
(their nominal model would indicate the $\sim 0.5$\% probability). For comparison, we estimate that the 
Imbrium formation from leftover planetesimals was a $\sim15$--35\% probability event, and explain that many more 
younger/smaller lunar basins than Imbrium would be expected (only two are known: Orientale and 
Schr\"odinger), if the probability were much larger than that (Sect. 9). 

B20 did not consider the terrestrial impact flux during the late Archean, and the general impact 
record in the inner solar system in the last 3.5 Gyr. It is therefore difficult to establish whether 
their results would satisfy the additional constraints considered in the present work. 

\section{Summary}

\begin{figure}
%\epsscale{0.5}
%\plotone{mars_n1.eps}
\epsscale{1.2}
%\hspace*{4.0cm}\plotone{marschrono2.pdf}
\hspace*{4.0cm}\plotone{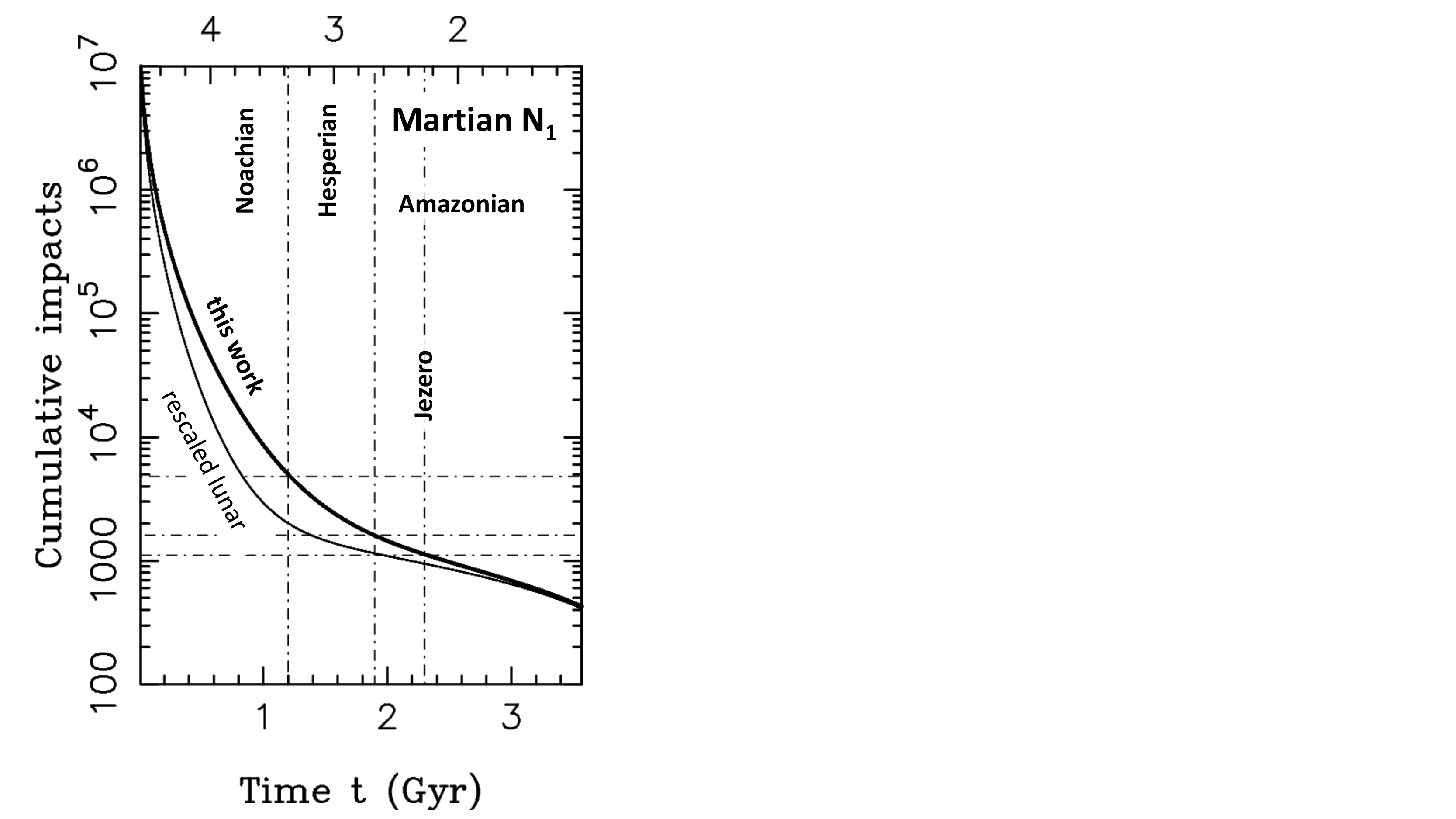}
\caption{A comparison of lunar and Martian chronologies. The thin solid line is the lunar $N_1(T)$ chronology 
rescaled to Mars as described in Sect. 13. The bold solid line is the Martian chronology inferred
in this work for the standard calibration of leftover planetesimals and the original planetesimal disk 
truncated at 1.5 au (Eq. \ref{marsn1}). The horizontal dot-dashed lines are the crater counts for 
Noachian/Hesperian and Hesperian/Amazonian boundaries, and for the Jezero crater terrains. The vertical dashed 
lines are our age estimates for these units. The ages inferred from the rescaled lunar chronology would 
be $\sim 400$--500 Myr older. The numbers on the upper axis indicate time $T$ (in Gyr) measured looking back from today.} 
\label{mars2}
\end{figure}

\begin{figure}
%\epsscale{0.49}
%\plotone{hses.eps}
%\plotone{noble.eps}
\epsscale{1.2}
%\hspace*{0.5cm}\plotone{noble.pdf}
\hspace*{0.5cm}\plotone{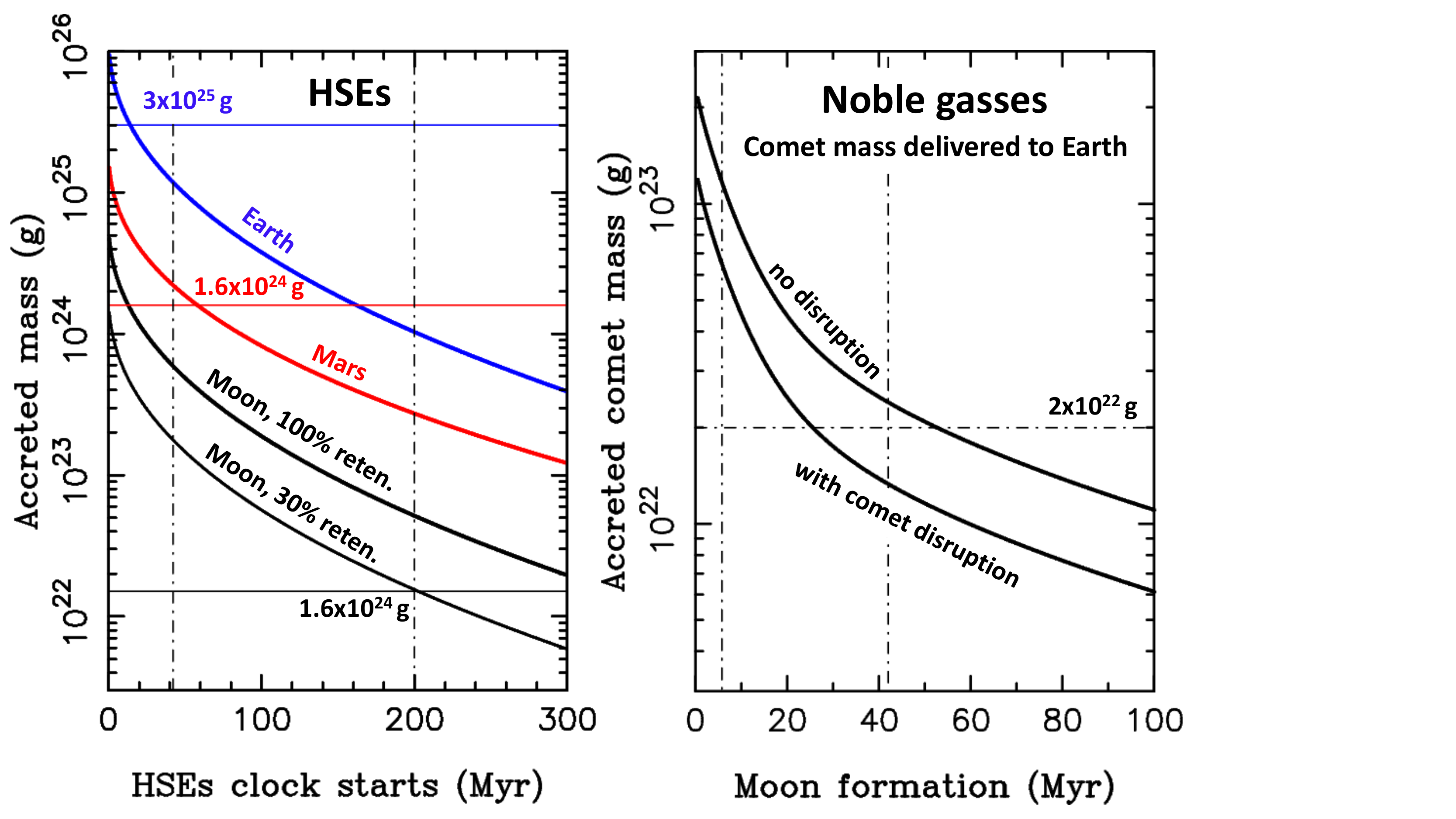}
\caption{Panel A. The total mass -- leftovers, 
asteroids, and comets combined -- accreted by the Earth (blue line), Mars (red), and Moon (black). We show
two lines for the Moon: the upper one for the 100\% retention and the lower one for the 30\% retention (Zhu 
et al. 2019a). The horizontal lines show constraints from the measured HSE abundances assuming that the HSEs
were delivered in chondritic proportion: the total accreted masses $3 \times 10^{25}$~g, $1.6 \times 10^{24}$ g,
and $1.5 \times 10^{22}$ g for the Earth, Mars, and Moon, respectively. The vertical dot-dashed lines are
the Moon-forming impact ($t=42$ Myr) and estimated time of LMO solidification ($t \sim 200$ Myr).
Panel B. The total mass of comets delivered to the Earth. The upper curve 
shows the profile when spontaneous comet disruptions are ignored. The bottom curve is the cometary flux reduced 
by a factor of 1.8 to approximately account for disruptions of $\sim100$-km-class comets (which presumably 
bring the bulk of accreted mass). The vertical dot-dashed lines are the instability ($t=5.8$ Myr) and Moon-forming
impact ($t=42$~Myr) in our model. The horizontal dot-dashed line corresponds to $2 \times 10^{22}$ g, as required 
to deliver noble gasses in the Earth atmosphere (Marty et. al. 2016).} 
\label{noble}
\end{figure}

We developed an accurate dynamical model that accounts for all major sources of impactors in the inner solar system 
(leftover planetesimals, asteroids and comets). Each impactor population was calibrated from independent means 
(not using crater records; Sect 3).\footnote{Except that we used the crater record to infer at least 
$\gtrsim 0.1$ $M_{\rm Earth}$ in planetesimals in the terrestrial zone at the time of the gas disk dispersal.} 
Here we summarize the main results:
\begin{itemize}
\item The leftover planetesimals produced most lunar impacts in the first 1.1 Gyr ($t<1.1$ Gyr or $T>3.5$ Ga).
Asteroids were the main source of impacts in the last 3.5 Gyr. The transition from leftover planetesimals to asteroids 
has been imprinted in the crater size distributions (Str\"om et al. 2005, Head et al. 2010, Orgel et al. 2018). 
The comet contribution to the crater record is found to be insignificant (for the early instability case 
adopted here). This would suggest that isotopic and other signatures of comets may be difficult to find on the 
lunar surface.
\item Some $500$ $d>20$-km planetesimals from the terrestrial planet zone (0.5--1.5 au) are expected to 
impact the Moon since its formation. The early crater record must have been erased plausibly because the lunar 
surface was unable to support basin-scale impact structures. The $\sim 50$ known lunar basins formed after $t \simeq 
160$--215 Myr ($T \lesssim 4.36$--4.41 Ga). This is consistent with the long-lived LMO (lunar magma ocean; 
Elkins-Tanton et al. 2011, Morbidelli et al. 2018, Zhu et al. 2019a). The Nectaris basin is estimated to be 
$T = 4.21$--4.29 Gyr old (from crater counts). The South Pole--Aitken basin should date back to $T = 4.36$--4.41 Ga.
\item About two lunar basins are expected to form for $t \gtrsim 650$ Myr ($T \lesssim 3.92$ Ga). The Imbrium 
basin formation ($T \simeq 3.92$~Ga, $d>100$ km impactor) is estimated to happen with a $\simeq 23$\% probability 
in our model. Imbrium should have formed unusually late, relative to the expectations from the lunar impact 
chronology, to have only two smaller/younger basins than Imbrium (Orientale and Schr\"odinger); there would be 
many more younger/smaller basins otherwise (Sect. 9).  
\item The Imbrium-era impactors were leftover planetesimals that were stored on orbits with $a \simeq 1.3$-1.7 au, 
$e<0.2$, and in the mean motion resonances with Mars (e.g., 6:5, 7:6, 11:9). The resonances provided a 
phase-protection mechanism against collisions with Mars. The orbits had low eccentricities such that they 
did not cross the orbit of Earth. The bodies stayed in the resonances for hundreds of Myr and were eventually 
released to orbits with higher eccentricities. This is an example of the ``storage places'' hypothesized 
by Wetherill (1975).  
\item The lunar and Martian chronologies can be given as a sum of two terms: the stretched exponential
function (the leftover planetesimal branch) and a constant (the asteroid or NEA branch). This is similar to the 
classical (empirical) crater chronologies of Neukum et al. (2001) and Hartmann \& Neukum (2001), except that 
the cratering rate profile in the first $\sim 1$ Gyr had a longer tail than the exact exponential. 
This can lead to modest, $\sim 50$ Myr differences in the estimates of lunar basin ages.
\item The Martian chronology is found to have a slower decline at late times than the chronologies of the 
Moon, Earth and Venus. This is a consequence of long dynamical lifetimes of bodies at $\sim 1.5$ au -- these 
bodies are more likely to produce late impacts on Mars. Applying the lunar chronology function to Mars can 
lead to age estimates that can differ, by up to $\sim 500$ Myr, from the age estimates obtained from 
the accurate Martian chronology (Fig. \ref{mars2}A). 
\item The Noachian/Hesperian and Hesperian/Amazonian boundaries are found to be $T=3.4$--3.6 Ga and 
and $T=2.6$--2.9 Ga, respectively (the range given here for $r_{\rm out}=1$--1.75 au and the standard 
calibration of leftover planetesimals). The Jezero crater is estimated to be 2.2--2.5 Gyr old for 
$N_1=1.1 \times 10^3$ per $10^6$ km$^2$ or 2.4--2.7 Gyr old for $N_1=1.5 \times 10^3$ per $10^6$ km$^2$.
The asteroid branch of the Martian chronology ($T \lesssim 2$ Ga) is found to be similar to the chronology
developed in Marchi (2021).  
\item Our model predicts $\simeq 20$ $d>10$-km impacts on the Earth for $T=2.5$--3.5 Ga. This is similar 
to the number of known spherule beds in the late Archean (Bottke et al. 2012, Johnson et al. 2016, Marchi 
et al. 2021). Both the leftover planetesimals and main-belt asteroids contribute to impacts in this time 
interval. Whereas the asteroid impacts were more uniformly spread over the late Archean, nearly all 
planetesimal impacts should have happened before 3 Ga.
\item The cometary impact profile is more extended in time than thought previously. For example, assuming 
the instability at $t<10$ Myr, 90\% of impacts happened in the first $\simeq 55$ Myr, and 99\% of impacts 
happened in the first $\simeq 370$ Myr. With the Moon-forming impact at $t \sim 50$ Myr ($T \sim 4.52$ Ga), 
we find that the Earth should have accreted $\sim 1.5$--$2.5 \times 10^{22}$ g of cometary material for 
$T<4.52$ Ga. This is consistent with comets being the source of noble gasses in the Earth atmosphere 
(Marty et al. 2016).
\item To explain lunar HSEs, we find -- in agreement with Morbidelli et al. (2018) and our lunar basin 
results (Sect. 8) -- that the LMO should have solidified at $t \sim 200$ Myr after the first solar system 
solids ($T \sim 4.37$ Ga). The mass accreted by the Earth for $t>50$ Myr is a factor of $\sim 3$ too low 
to explain Earth's HSEs. This shows the need for the stochastic late accretion of very large planetesimals 
(Bottke et al. 2010). 
\item Mars would have accreted a factor of $\sim 3$ more HSEs for $t>10$ Myr in our nominal model (the 
standard calibration of planetesimals and $r_{\rm out}=1.5$ au) than what is inferred from the SNC meteorites 
(Marchi et al. 2020). It is possible that: (i) the average abundance of HSEs in the Martian mantle 
is underestimated (e.g., the SNC meteorites are not representative), (ii) the impactor-retention factor
of Mars is smaller than 100\%, and/or (iii) the accreted mass was lower because the ring of terrestrial
planetesimals was originally confined to 0.5--1 au.

\end{itemize}

\acknowledgements

The work of D.N. was supported by the NASA Solar System Workings program. F.R. acknowledges support from 
the Brazilian National Council of Research - CNPq. The work of D.V. was 
supported by the Czech Science Foundation (grant number 21-110585). R.D. acknowledges support from the
NASA Emerging Worlds program, grant 80NSSC21K0387. A.M. received funding from the European Research 
Council (ERC) under the European Union's Horizon 2020 research and innovation program (grant No. 
101019380 HolyEarth). We thank both reviewers for detailed and helpful comments.

\clearpage

\clearpage

\clearpage

%\clearpage
%gr.job35_fig1 in Clement/ on vega
%\begin{figure}
%\epsscale{0.6}
%\plotone{switch.eps}
%\caption{The orbital distribution of bodies at $t=42$ Myr.}
%\label{switch}
%\end{figure}

\clearpage
%gr.distr_job35_start.f and end in LHB/ on laptop
%\begin{figure}
%\epsscale{0.49}
%\plotone{start.eps}
%\plotone{end.eps}
%\caption{The orbital distribution of bodies at $t=100$ Myr (left) and $t=700$ Myr (right).}
%\label{distr}
%\end{figure}

\clearpage

\clearpage

\clearpage

\clearpage

\clearpage

\clearpage

\clearpage

%\clearpage
%\begin{figure}
%\epsscale{0.49}
%\plotone{chrono.eps}
%\plotone{chrono_nolog_moon.eps}
%% gr.impact_leftovers6-7.f in LHB/ on laptop
%\caption{Impact flux of $d>20$ km bodies on the Moon.}
%\label{chrono2}
%\end{figure}

\clearpage

%\clearpage
%\begin{figure}
%\epsscale{0.5}
%\plotone{basin.eps}
%\caption{The chronology functions for impacts of $d>1$ km (solid line) and $d>20$ km (dashed)
%  bodies on the Moon. The plot illustrates how the chronology function depends on
%  impactor/crater size. How normalized?}
%\label{basin}
%\end{figure}

%\clearpage
%\begin{figure}
%\epsscale{0.49}
%\plotone{basin2.eps}
%\plotone{basin3.eps}
%\caption{Comparison of Neukum's chronology with our $N_{20}$ and the chronology with different calibrations
%of leftovers (the bold line shows the nominal distribution).}
%\label{basin2}
%\end{figure}

\clearpage

\clearpage

%\clearpage
%\begin{figure}
%\epsscale{0.49}
%\plotone{chrono_venus.eps}
%\plotone{chrono_nolog_venus.eps}
%\caption{Impact flux of $d>20$ km bodies on Venus.}
%\label{chrono4}
%\end{figure}

\clearpage

\clearpage

\clearpage

\clearpage


\begin{thebibliography}{}

\bibitem[Andrews-Hanna(2016)]{2016LPICo1911.6061A} Andrews-Hanna, J.~C.\ 2016, New Views of the Moon 2, 1911, 6061

\bibitem[Artemieva \& Ivanov(2004)]{2004Icar..171...84A} Artemieva, N. \& Ivanov, B.\ 2004, Icarus, 171, 84. doi:10.1016/j.icarus.2004.05.003

\bibitem[Barboni et al.(2017)]{2017LPI....48.1900B} Barboni, M., Boehnke, P., Keller, C.~B., et al.\ 2017, 48th Annual Lunar and Planetary Science Conference

\bibitem[Benz \& Asphaug(1999)]{1999Icar..142....5B} Benz, W. \& Asphaug, E.\ 1999, Icarus, 142, 5. doi:10.1006/icar.1999.6204

\bibitem[Bernstein et al.(2004)]{2004AJ....128.1364B} Bernstein, G.~M., Trilling, D.~E., Allen, R.~L., et al.\ 2004, \aj, 128, 1364. doi:10.1086/422919

\bibitem[Bierhaus et al.(2018)]{2018M&PS...53..638B} Bierhaus, E.~B. and 6 colleagues 2018.\ Secondary craters and ejecta across the solar system: Populations and effects on impact-crater-based chronologies.\ Meteoritics and Planetary Science 53, 638–671. doi:10.1111/maps.13057

\bibitem[Bottke \& Norman(2017)]{2017AREPS..45..619B} Bottke, W.~F. \& Norman, M.~D.\ 2017, Annual Review of Earth and Planetary Sciences, 45, 619. doi:10.1146/annurev-earth-063016-020131

\bibitem[Bottke et al.(1994)]{1994Icar..107..255B} Bottke, W.~F., Nolan, M.~C., Greenberg, R., et al.\ 1994, Icarus, 107, 255. doi:10.1006/icar.1994.1021

\bibitem[Bottke et al.(2005)]{2005Icar..175..111B} Bottke, W.~F., Durda, D.~D., Nesvorn{\'y}, D., et al.\ 2005, Icarus, 175, 111. doi:10.1016/j.icarus.2004.10.026

\bibitem[Bottke et al.(2007)]{2007Icar..190..203B} Bottke, W.~F., Levison, H.~F., Nesvorn{\'y}, D., et al.\ 2007, Icarus, 190, 203. doi:10.1016/j.icarus.2007.02.010

\bibitem[Bottke et al.(2010)]{2010Sci...330.1527B} Bottke, W.~F., Walker, R.~J., Day, J.~M.~D., et al.\ 2010, Science, 330, 1527. doi:10.1126/science.1196874

\bibitem[Bottke et al.(2012)]{2012Natur.485...78B} Bottke, W.~F., Vokrouhlick{\'y}, D., Minton, D., et al.\ 2012, \nat, 485, 78. doi:10.1038/nature10967

\bibitem[Bottke et al.(2020)]{2020AJ....160...14B} Bottke, W.~F., Vokrouhlick{\'y}, D., Ballouz, R.-L., et al.\ 2020, \aj, 160, 14. doi:10.3847/1538-3881/ab88d3

\bibitem[Bouvier et al.(2007)]{2007GeCoA..71.1583B} Bouvier, A., Blichert-Toft, J., Moynier, F., et al.\ 2007, \gca, 71, 1583. doi:10.1016/j.gca.2006.12.005

\bibitem[Brasser et al.(2016)]{2016ApJ...821...75B} Brasser, R., Matsumura, S., Ida, S., et al.\ 2016, \apj, 821, 75. doi:10.3847/0004-637X/821/2/75

\bibitem[Brasser et al.(2020)]{2020Icar..33813514B} Brasser, R., Werner, S.~C., \& Mojzsis, S.~J.\ 2020, Icarus, 338, 113514. doi:10.1016/j.icarus.2019.113514

\bibitem[Brasser et al.(2021)]{2021Icar..36114389B} Brasser, R., Mojzsis, S.~J., Werner, S.~C., et al.\ 2021, Icarus, 361, 114389. doi:10.1016/j.icarus.2021.114389

\bibitem[Bro{\v{z}} et al.(2021)]{2021NatAs...5..898B} Bro{\v{z}}, M., Chrenko, O., Nesvorn{\'y}, D., et al.\ 2021, Nature Astronomy, 5, 898. doi:10.1038/s41550-021-01383-3

\bibitem[Burkhardt et al.(2008)]{2008GeCoA..72.6177B} Burkhardt, C., Kleine, T., Bourdon, B., et al.\ 2008, \gca, 72, 6177. doi:10.1016/j.gca.2008.10.023

\bibitem[Byerly et al.(2002)]{2002Sci...297.1325B} Byerly, G.~R., Lowe, D.~R., Wooden, J.~L., Xie, X.\ 2002.\ An Archean Impact Layer from the Pilbara and Kaapvaal Cratons.\ Science 297, 1325–1327. doi:10.1126/science.1073934
  
\bibitem[Canup(2012)]{2012Sci...338.1052C} Canup, R.~M.\ 2012, Science, 338, 1052. doi:10.1126/science.1226073

\bibitem[Canup et al.(2021)]{2021arXiv210302045C} Canup, R.~M., Righter, K., Dauphas, N., et al.\ 2021, arXiv:2103.02045

\bibitem[Chambers \& Wetherill(1998)]{1998Icar..136..304C} Chambers, J.~E. \& Wetherill, G.~W.\ 1998, Icarus, 136, 304. doi:10.1006/icar.1998.6007

\bibitem[Che et al.(2021)]{2021Sci...374..887C} Che, X. and 24 colleagues 2021.\ Age and composition of young basalts on the Moon, measured from samples returned by Chang{\textquoteright}e-5.\ Science 374, 887–890. doi:10.1126/science.abl7957

\bibitem[Chen \& Jewitt(1994)]{1994Icar..108..265C} Chen, J. \& Jewitt, D.\ 1994, Icarus, 108, 265. doi:10.1006/icar.1994.1061

\bibitem[Clement et al.(2018)]{2018Icar..311..340C} Clement, M.~S., Kaib, N.~A., Raymond, S.~N., et al.\ 2018, Icarus, 311, 340. doi:10.1016/j.icarus.2018.04.008

\bibitem[Clement et al.(2019)]{2019AJ....157...38C} Clement, M.~S., Raymond, S.~N., \& Kaib, N.~A.\ 2019, \aj, 157, 38. doi:10.3847/1538-3881/aaf21e

\bibitem[Cohen et al.(2000)]{2000Sci...290.1754C} Cohen, B.~A., Swindle, T.~D., \& Kring, D.~A.\ 2000, Science, 290, 1754. doi:10.1126/science.290.5497.1754

\bibitem[Collins et al.(2020)]{2020NatCo..11.1480C} Collins, G.~S. and 7 colleagues 2020.\ A steeply-inclined trajectory for the Chicxulub impact.\ Nature Communications 11. doi:10.1038/s41467-020-15269-x
  
\bibitem[Culler et al.(2000)]{2000Sci...287.1785C} Culler, T.~S., Becker, T.~A., Muller, R.~A., et al.\ 2000, Science, 287, 1785. doi:10.1126/science.287.5459.1785

\bibitem[Dauphas \& Pourmand(2011)]{2011Natur.473..489D} Dauphas, N. \& Pourmand, A.\ 2011, \nat, 473, 489. doi:10.1038/nature10077

\bibitem[Deienno et al.(2018)]{2018ApJ...864...50D} Deienno, R., Izidoro, A., Morbidelli, A., et al.\ 2018, \apj, 864, 50. doi:10.3847/1538-4357/aad55d

\bibitem[Deienno et al.(2019)]{2019ApJ...876..103D} Deienno, R., Walsh, K.~J., Kretke, K.~A., et al.\ 2019, \apj, 876, 103. doi:10.3847/1538-4357/ab16e1

\bibitem[DeMeo \& Carry(2013)]{2013Icar..226..723D} DeMeo, F.~E. \& Carry, B.\ 2013, Icarus, 226, 723. doi:10.1016/j.icarus.2013.06.027

\bibitem[Di Sisto et al.(2009)]{2009Icar..203..140D} Di Sisto, R.~P., Fern{\'a}ndez, J.~A., \& Brunini, A.\ 2009, Icarus, 203, 140. doi:10.1016/j.icarus.2009.05.002

\bibitem[Dones et al.(2009)]{2009sfch.book..613D} Dones, L., Chapman, C.~R., McKinnon, W.~B., et al.\ 2009, Saturn from Cassini-Huygens, 613. doi:10.1007/978-1-4020-9217-6\_19
  
\bibitem[Elkins-Tanton et al.(2011)]{2011E&PSL.304..326E} Elkins-Tanton, L.~T., Burgess, S., \& Yin, Q.-Z.\ 2011, Earth and Planetary Science Letters, 304, 326. doi:10.1016/j.epsl.2011.02.004

\bibitem[Fassett et al.(2012)]{2012JGRE..117.0H06F} Fassett, C.~I., Head, J.~W., Kadish, S.~J., et al.\ 2012, Journal of Geophysical Research (Planets), 117, E00H06. doi:10.1029/2011JE003951

\bibitem[Fraser et al.(2014)]{2014ApJ...782..100F} Fraser, W.~C., Brown, M.~E., Morbidelli, A., et al.\ 2014, \apj, 782, 100. doi:10.1088/0004-637X/782/2/100

\bibitem[Fraser et al.(2013)]{2014ApJ...782..100G} Glass, B. P. \& Simonson, B. M. 2013, Distal Impact Ejecta Layers: A Record of Large Impacts
  in Sedimentary Deposits, Springer.
  
\bibitem[Gomes et al.(2004)]{2004Icar..170..492G} Gomes, R.~S., Morbidelli, A., \& Levison, H.~F.\ 2004, Icarus, 170, 492. doi:10.1016/j.icarus.2004.03.011

\bibitem[Gomes et al.(2005)]{2005Natur.435..466G} Gomes, R., Levison, H.~F., Tsiganis, K., et al.\ 2005, \nat, 435, 466. doi:10.1038/nature03676

\bibitem[Granvik et al.(2018)]{2018Icar..312..181G} Granvik, M., Morbidelli, A., Jedicke, R., et al.\ 2018, Icarus, 312, 181. doi:10.1016/j.icarus.2018.04.018

\bibitem[Harris \& D'Abramo(2015)]{2015Icar..257..302H} Harris, A.~W. \& D'Abramo, G.\ 2015, Icarus, 257, 302. doi:10.1016/j.icarus.2015.05.004

\bibitem[Hartmann(1966)]{1966Icar....5..406H} Hartmann, W.~K.\ 1966, Icarus, 5, 406. doi:10.1016/0019-1035(66)90054-6

\bibitem[Hartmann(2005)]{2005Icar..174..294H} Hartmann, W.~K.\ 2005.\ Martian cratering 8: Isochron refinement and the chronology of Mars.\ Icarus 174, 294–320. doi:10.1016/j.icarus.2004.11.023


\bibitem[Hartmann(2019)]{2019Geosc...9..285H} Hartmann, W.~K.\ 2019, Geosciences, 9, 285. doi:10.3390/geosciences9070285

\bibitem[Hartmann \& Neukum(2001)]{2001SSRv...96..165H} Hartmann, W.~K. \& Neukum, G.\ 2001, \ssr, 96, 165. doi:10.1023/A:1011945222010

\bibitem[Hartmann et al.(2007)]{2007Icar..186...11H} Hartmann, W.~K., Quantin, C., \& Mangold, N.\ 2007, Icarus, 186, 11. doi:10.1016/j.icarus.2006.09.009

\bibitem[Hartung(1974)]{1974Metic...9..349H} Hartung, J.~B.\ 1974, Meteoritics, 9, 349

\bibitem[Head et al.(2010)]{2010Sci...329.1504H} Head, J.~W., Fassett, C.~I., Kadish, S.~J., et al.\ 2010, Science, 329, 1504. doi:10.1126/science.1195050

\bibitem[Hiesinger et al.(2012)]{2012JGRE..117.0H10H} Hiesinger, H. and 6 colleagues 2012.\ How old are 
young lunar craters?\ Journal of Geophysical Research (Planets) 117. doi:10.1029/2011JE003935

\bibitem[Hiesinger et al.(2016)]{2016LPICo1911.6036H} Hiesinger, H., van der Bogert, C.~H., Pasckert, J.~H., 
Plescia, J.~B., Robinson, M.~S.\ 2016.\ Impact Chronology of the Moon -- Results from the Lunar 
Reconnaissance Orbiter Camera (LROC).\ New Views of the Moon, 2, 1911.

\bibitem[Hiesinger et al.(2020)]{2020LPI....51.2045H} Hiesinger, H., van der Bogert, C.~H., Iqbal, W., 
Gebbing, T.\ 2020.\ The Lunar Chronology: A Status Report.\ 51st Annual Lunar and Planetary Science Conference.

\bibitem[Holsapple \& Housen(2007)]{2007Icar..187..345H} Holsapple, K.~A. \& Housen, K.~R.\ 2007, Icarus, 187, 345. doi:10.1016/j.icarus.2006.08.029

\bibitem[Huang et al.(2018)]{2018GeoRL..45.6805H} Huang, Y.-H., Minton, D.~A., Zellner, N.~E.~B., Hirabayashi, M., Richardson, J.~E., Fassett, C.~I.\ 2018.\ No Change in the Recent Lunar Impact Flux Required Based on Modeling of Impact Glass Spherule Age Distributions.\ Geophysical Research Letters 45, 6805–6813. doi:10.1029/2018GL077254

\bibitem[Izidoro et al.(2021)]{2021NatAs.tmp..262I} Izidoro, A., Dasgupta, R., Raymond, S.~N., et al.\ 2021, Nature Astronomy. doi:10.1038/s41550-021-01557-z

\bibitem[Jacobson \& Morbidelli(2014)]{2014RSPTA.37230174J} Jacobson, S.~A. \& Morbidelli, A.\ 2014, Philosophical Transactions of the Royal Society of London Series A, 372, 0174. doi:10.1098/rsta.2013.0174

\bibitem[Jacobson et al.(2017)]{2017E&PSL.474..375J} Jacobson, S.~A., Rubie, D.~C., Hernlund, J., et al.\ 2017, Earth and Planetary Science Letters, 474, 375. doi:10.1016/j.epsl.2017.06.023

\bibitem[Johansen et al.(2021)]{2021SciA....7..444J} Johansen, A., Ronnet, T., Bizzarro, M., et al.\ 2021, Science Advances, 7, eabc0444. doi:10.1126/sciadv.abc0444

\bibitem[Johnson and Melosh(2012)]{2012Natur.485...75J} Johnson, B.~C., Melosh, H.~J.\ 2012.\ Impact spherules as a record of an ancient heavy bombardment of Earth.\ Nature 485, 75–77. doi:10.1038/nature10982
  
\bibitem[Johnson et al.(2016)]{2016Icar..271..350J} Johnson, B.~C., Collins, G.~S., Minton, D.~A., et al.\ 2016a, Icarus, 271, 350. doi:10.1016/j.icarus.2016.02.023

\bibitem[Johnson et al.(2016)]{2016Sci...354..441J} Johnson, B.~C., Blair, D.~M., Collins, G.~S., et al.\ 2016b, Science, 354, 441. doi:10.1126/science.aag0518

%\bibitem[Joy et al.(2012)]{2012Sci...336.1426J} Joy, K.~H., Zolensky, M.~E., Nagashima, K., et al.\ 2012, Science, 336, 1426. do%i:10.1126/science.1219633

\bibitem[Kenyon \& Bromley(2001)]{2001AJ....121..538K} Kenyon, S.~J. \& Bromley, B.~C.\ 2001, \aj, 121, 538. doi:10.1086/318019

\bibitem[Kirchoff et al.(2021)]{2021Icar..35514110K} Kirchoff, M.~R., Marchi, S., Bottke, W.~F., et al.\ 2021, Icarus, 355, 114110. doi:10.1016/j.icarus.2020.114110

\bibitem[Kleine \& Walker(2017)]{2017AREPS..45..389K} Kleine, T. \& Walker, R.~J.\ 2017, Annual Review of Earth and Planetary Sciences, 45, 389. doi:10.1146/annurev-earth-063016-020037

\bibitem[Kring \& Cohen(2002)]{2002JGRE..107.5009K} Kring, D.~A. \& Cohen, B.~A.\ 2002, Journal of Geophysical Research (Planets), 107, 5009. doi:10.1029/2001JE001529

\bibitem[Leinhardt \& Stewart(2012)]{2012ApJ...745...79L} Leinhardt, Z.~M. \& Stewart, S.~T.\ 2012, \apj, 745, 79. doi:10.1088/0004-637X/745/1/79

\bibitem[Le Feuvre \& Wieczorek(2011)]{2011Icar..214....1L} Le Feuvre, M. \& Wieczorek, M.~A.\ 2011, Icarus, 214, 1. doi:10.1016/j.icarus.2011.03.010

\bibitem[Levison \& Duncan(1994)]{1994Icar..108...18L} Levison, H.~F. \& Duncan, M.~J.\ 1994, Icarus, 108, 18. doi:10.1006/icar.1994.1039

\bibitem[Levison \& Duncan(1997)]{1997Icar..127...13L} Levison, H.~F. \& Duncan, M.~J.\ 1997, Icarus, 127, 13. doi:10.1006/icar.1996.5637

\bibitem[Levison et al.(2001)]{2001Icar..151..286L} Levison, H.~F., Dones, L., Chapman, C.~R., et al.\ 2001, Icarus, 151, 286. doi:10.1006/icar.2001.6608

\bibitem[Levison et al.(2002)]{2002Sci...296.2212L} Levison, H.~F., Morbidelli, A., Dones, L., et al.\ 2002, Science, 296, 2212. doi:10.1126/science.1070226

\bibitem[Li et al.(2021)]{2021Natur.600...54L} Li, Q.-L., Zhou, Q., Liu, Y., et al.\ 2021, \nat, 600, 54. doi:10.1038/s41586-021-04100-2

\bibitem[Liu et al.(2022)]{2022Natur.604..643L} Liu, B., Raymond, S.~N., \& Jacobson, S.~A.\ 2022, \nat, 604, 643. doi:10.1038/s41586-022-04535-1

\bibitem[Lowe and Byerly(1986)]{1986Geo....14...83L} Lowe, D.~R., Byerly, G.~R.\ 1986.\ Early Archean silicate spherules of probable impact origin, South Africa and Western Australia.\ Geology 14, 83.

\bibitem[Lowe et al.(2014)]{2014Geo....42..747L} Lowe, D.~R., Byerly, G.~R., Kyte, F.~T.\ 2014.\ Recently discovered 3.42-3.23 Ga impact layers, Barberton Belt, South Africa: 3.8 Ga detrital zircons, Archean impact history, and tectonic implications.\ Geology 42, 747–750. doi:10.1130/G35743.1

\bibitem[Mainzer et al.(2019)]{2019PDSS..251.....M} Mainzer, A.~K., Bauer, J.~M., Cutri, R.~M., et al.\ 2019, NASA Planetary Data System. doi:10.26033/18S3-2Z54

\bibitem[Marchi(2021)]{2021AJ....161..187M} Marchi, S.\ 2021, \aj, 161, 187. doi:10.3847/1538-3881/abe417

\bibitem[Marchi et al.(2009)]{2009AJ....137.4936M} Marchi, S., Mottola, S., Cremonese, G., et al.\ 2009, \aj, 137, 4936. doi:10.1088/0004-6256/137/6/4936

\bibitem[Marchi et al.(2012)]{2012E&PSL.325...27M} Marchi, S., Bottke, W.~F., Kring, D.~A., et al.\ 2012, Earth and Planetary Science Letters, 325, 27. doi:10.1016/j.epsl.2012.01.021

\bibitem[Marchi et al.(2014)]{2014Natur.511..578M} Marchi, S., Bottke, W.~F., Elkins-Tanton, L.~T., et al.\ 2014, \nat, 511, 578. doi:10.1038/nature13539

\bibitem[Marchi et al.(2020)]{2020SciA....6.2338M} Marchi, S., Walker, R.~J., \& Canup, R.~M.\ 2020, Science Advances, 6, eaay2338. doi:10.1126/sciadv.aay2338

\bibitem[Marchi et al.(2021)]{2021NatGe..14..827M} Marchi, S., Drabon, N., Schulz, T., et al.\ 2021, Nature Geoscience, 14, 827. doi:10.1038/s41561-021-00835-9

\bibitem[Marty et al.(2016)]{2016E&PSL.441...91M} Marty, B., Avice, G., Sano, Y., et al.\ 2016, Earth and Planetary Science Letters, 441, 91. doi:10.1016/j.epsl.2016.02.031

\bibitem[Maurice et al.(2020)]{2020SciA....6.8949M} Maurice, M., Tosi, N., Schwinger, S., et al.\ 2020, Science Advances, 6, eaba8949. doi:10.1126/sciadv.aba8949

\bibitem[Mazrouei et al.(2019)]{2019Sci...363..253M} Mazrouei, S., Ghent, R.~R., Bottke, W.~F., et al.\ 2019, Science, 363, 253. doi:10.1126/science.aar4058

\bibitem[Merle et al.(2013)]{2013LPI....44.1833M} Merle, R.~E., Nemchin, A.~A., Grange, M.~L., et al.\ 2013, 44th Annual Lunar and Planetary Science Conference

\bibitem[Meyer et al.(2010)]{2010Icar..208....1M} Meyer, J., Elkins-Tanton, L., \& Wisdom, J.\ 2010, Icarus, 208, 1. doi:10.1016/j.icarus.2010.01.029

\bibitem[Miljkovi{\'c} et al.(2013)]{2013Sci...342..724M} Miljkovi{\'c}, K., Wieczorek, M.~A., Collins, G.~S., et al.\ 2013, Science, 342, 724. doi:10.1126/science.1243224
  
\bibitem[Miljkovi{\'c} et al.(2016)]{2016JGRE..121.1695M} Miljkovi{\'c}, K., Collins, G.~S., Wieczorek, M.~A., et al.\ 2016, Journal of Geophysical Research (Planets), 121, 1695. doi:10.1002/2016JE005038

\bibitem[Miljkovi{\'c} et al.(2021)]{2021NatCo..12.5433M} Miljkovi{\'c}, K., Wieczorek, M.~A., Laneuville, M., et al.\ 2021, Nature Communications, 12, 5433. doi:10.1038/s41467-021-25818-7

\bibitem[Minton et al.(2015)]{2015Icar..247..172M} Minton, D.~A., Richardson, J.~E., \& Fassett, C.~I.\ 2015, Icarus, 247, 172. doi:10.1016/j.icarus.2014.10.018

\bibitem[Morbidelli et al.(2001)]{2001M&PS...36..371M} Morbidelli, A., Petit, J.-M., Gladman, B., et al.\ 2001, MAPS, 36, 371. doi:10.1111/j.1945-5100.2001.tb01880.x

\bibitem[Morbidelli et al.(2009)]{2009Icar..204..558M} Morbidelli, A., Bottke, W.~F., Nesvorn{\'y}, D., et al.\ 2009a, Icarus, 204, 558. doi:10.1016/j.icarus.2009.07.011

\bibitem[Morbidelli et al.(2009)]{2009Icar..202..310M} Morbidelli, A., Levison, H.~F., Bottke, W.~F., et al.\ 2009b, Icarus, 202, 310. doi:10.1016/j.icarus.2009.02.033

\bibitem[Morbidelli et al.(2012)]{2012E&PSL.355..144M} Morbidelli, A., Marchi, S., Bottke, W.~F., et al.\ 2012, Earth and Planetary Science Letters, 355, 144. doi:10.1016/j.epsl.2012.07.037

\bibitem[Morbidelli et al.(2018)]{2018Icar..305..262M} Morbidelli, A., Nesvorny, D., Laurenz, V., et al.\ 2018, Icarus, 305, 262. doi:10.1016/j.icarus.2017.12.046

\bibitem[Morbidelli et al.(2020)]{2020Icar..34013631M} Morbidelli, A., Delbo, M., Granvik, M., et al.\ 2020, Icarus, 340, 113631. doi:10.1016/j.icarus.2020.113631

\bibitem[Morbidelli et al.(2022)]{2022NatAs...6...72M} Morbidelli, A., Bailli{\'e}, K., Batygin, K., et al.\ 2022, Nature Astronomy, 6, 72. doi:10.1038/s41550-021-01517-7

\bibitem[Nakajima \& Stevenson(2014)]{2014Icar..233..259N} Nakajima, M. \& Stevenson, D.~J.\ 2014, Icarus, 233, 259. doi:10.1016/j.icarus.2014.01.008

\bibitem[Nesvorn{\'y}(2018)]{2018ARA&A..56..137N} Nesvorn{\'y}, D.\ 2018, \araa, 56, 137. doi:10.1146/annurev-astro-081817-052028

\bibitem[Nesvorn{\'y} \& Vokrouhlick{\'y}(2019)]{2019Icar..331...49N} Nesvorn{\'y}, D. \& Vokrouhlick{\'y}, D.\ 2019, Icarus, 331, 49. doi:10.1016/j.icarus.2019.04.030

\bibitem[Nesvorn{\'y} et al.(2017)]{2017AJ....153..103N} Nesvorn{\'y}, D., Roig, F., \& Bottke, W.~F.\ 2017a, \aj, 153, 103. doi:10.3847/1538-3881/153/3/103

\bibitem[Nesvorn{\'y} et al.(2017)]{2017ApJ...845...27N} Nesvorn{\'y}, D., Vokrouhlick{\'y}, D., Dones, L., et al.\ 2017b, \apj, 845, 27. doi:10.3847/1538-4357/aa7cf6

\bibitem[Nesvorn{\'y} et al.(2019)]{2019AJ....158..132N} Nesvorn{\'y}, D., Vokrouhlick{\'y}, D., Stern, A.~S., et al.\ 2019, \aj, 158, 132. doi:10.3847/1538-3881/ab3651

\bibitem[Nesvorn{\'y} et al.(2020)]{2020AJ....160...46N} Nesvorn{\'y}, D., Vokrouhlick{\'y}, D., Alexandersen, M., et al.\ 2020, \aj, 160, 46. doi:10.3847/1538-3881/ab98fb

\bibitem[Nesvorn{\'y} et al.(2021)]{2021Icar..36814621N} Nesvorn{\'y}, D., Bottke, W.~F., \& Marchi, S.\ 2021a, Icarus, 368, 114621. doi:10.1016/j.icarus.2021.114621

\bibitem[Nesvorn{\'y} et al.(2021)]{2021AJ....161...50N} Nesvorn{\'y}, D., Roig, F.~V., \& Deienno, R.\ 2021b, \aj, 161, 50. doi:10.3847/1538-3881/abc8ef

\bibitem[Nesvorn{\'y} et al.(2022)]{2022ApJ...941L...9N} Nesvorn{\'y}, D., Roig, F.~V., Vokrouhlick{\'y}, D., et al.\ 2022, \apjl, 941, L9. doi:10.3847/2041-8213/aca40e

\bibitem[Nesvorn{\'y} et al.(2023)]{2022ApJ...941L...10N} Nesvorn{\'y}, D. et al., submitted to AJ.

\bibitem[Neukum \& Ivanov(1994)]{1994hdtc.conf..359N} Neukum, G. \& Ivanov, B.~A.\ 1994, Hazards Due to Comets and Asteroids, 359

\bibitem[Neukum et al.(2001)]{2001SSRv...96...55N} Neukum, G., Ivanov, B.~A., \& Hartmann, W.~K.\ 2001, \ssr, 96, 55. doi:10.1023/A:1011989004263

\bibitem[Neumann et al.(2015)]{2015SciA....1E0852N} Neumann, G.~A., Zuber, M.~T., Wieczorek, M.~A., et al.\ 2015, Science Advances, 1, e1500852. doi:10.1126/sciadv.1500852

\bibitem[Orgel et al.(2018)]{2018JGRE..123..748O} Orgel, C., Michael, G., Fassett, C.~I., et al.\ 2018, Journal of Geophysical Research (Planets), 123, 748. doi:10.1002/2017JE005451

\bibitem[Ozdemir et al.(2019)]{2019M&PS...54.2203O} Ozdemir, S., Schulz, T., van Acken, D., Luguet, A., Reimold, W.~U., Koeberl, C.\ 2019.\ Meteoritic highly siderophile element and Re-Os isotope signatures of Archean spherule layers from the CT3 drill core, Barberton Greenstone Belt, South Africa.\ Meteoritics and Planetary Science 54, 2203–2216. doi:10.1111/maps.13234

\bibitem[Popova et al.(2003)]{2003M&PS...38..905P} Popova, O., Nemtchinov, I., \& Hartmann, W.~K.\ 2003, MAPS, 38, 905. doi:10.1111/j.1945-5100.2003.tb00287.x

\bibitem[Raymond \& Nesvorny(2020)]{2020arXiv201207932R} Raymond, S.~N. \& Nesvorny, D.\ 2020, arXiv:2012.07932

\bibitem[Robbins(2014)]{2014E&PSL.403..188R} Robbins, S.~J.\ 2014, Earth and Planetary Science Letters, 403, 188. doi:10.1016/j.epsl.2014.06.038

\bibitem[Robbins(2022)]{2022PSJ.....3..274R} Robbins, S.~J.\ 2022, PSJ, 3, 274. doi:10.3847/PSJ/aca282

\bibitem[Roig \& Nesvorn{\'y}(2015)]{2015AJ....150..186R} Roig, F. \& Nesvorn{\'y}, D.\ 2015, \aj, 150, 186. doi:10.1088/0004-6256/150/6/186

\bibitem[Roig \& Nesvorn{\'y}(2020)]{2020AJ....160..110R} Roig, F. \& Nesvorn{\'y}, D.\ 2020, \aj, 160, 110. doi:10.3847/1538-3881/aba750

\bibitem[Rubie et al.(2016)]{2016Sci...353.1141R} Rubie, D.~C., Laurenz, V., Jacobson, S.~A., et al.\ 2016, Science, 353, 1141. doi:10.1126/science.aaf6919

\bibitem[Ryder(1990)]{1990EOSTr..71..313R} Ryder, G.\ 1990, EOS Transactions, 71, 313. doi:10.1029/90EO00086

\bibitem[Ryder(2002)]{2002JGRE..107.5022R} Ryder, G.\ 2002, Journal of Geophysical Research (Planets), 107, 5022. doi:10.1029/2001JE001583
  
\bibitem[Schultz \& Crawford(2016)]{2016Natur.535..391S} Schultz, P.~H. \& Crawford, D.~A.\ 2016, \nat, 535, 391. doi:10.1038/nature18278

\bibitem[Schultz \& Crawford(2017)]{2016Natur.535..392S} Schulz, T. et al. 2017, New constraints on the Paleoarchean meteorite bombardment
  of the Earth—Geochemistry and Re–Os isotope signatures of the BARB5 ICDP drill core from the Barberton Greenstone Belt, South Africa.
  Geochim. Cosmochim. Acta 211, 322–340.
  
\bibitem[Shahrzad et al.(2019)]{2019GeoRL..46.2408S} Shahrzad, S., Kinch, K.~M., Goudge, T.~A., et al.\ 2019, \grl, 46, 2408. doi:10.1029/2018GL081402

\bibitem[Spudis(1993)]{1993gmri.book.....S} Spudis, P.~D.\ 1993, The Geology of Multi-ring Impact Basins, ISBN 0521261031, Cambridge University Press, 1993. Camb. Planet. Sci. Ser., Vol. 8

\bibitem[St{\"o}ffler \& Ryder(2001)]{2001SSRv...96....9S} St{\"o}ffler, D. \& Ryder, G.\ 2001, \ssr, 96, 9. doi:10.1023/A:1011937020193

%\bibitem[Stoffler(2006)]{2006RvMG...60..519S} St{\"o}ffler, D.\ 2006.\ Cratering History and Lunar Chronology.\ Reviews in Miner%alogy and Geochemistry 60, 519–596. doi:10.2138/rmg.2006.60.05

\bibitem[Strom et al.(2005)]{2005Sci...309.1847S} Strom, R.~G., Malhotra, R., Ito, T., et al.\ 2005, Science, 309, 1847. doi:10.1126/science.1113544

\bibitem[Tanaka(1986)]{1986JGR....91E.139T} Tanaka, K.~L.\ 1986, \jgr, 91, E139. doi:10.1029/JB091iB13p0E139

\bibitem[Tera et al.(1974)]{1974E&PSL..22....1T} Tera, F., Papanastassiou, D.~A., \& Wasserburg, G.~J.\ 1974, Earth and Planetary Science Letters, 22, 1. doi:10.1016/0012-821X(74)90059-4

\bibitem[Thiemens et al.(2019)]{2019NatGe..12..696T} Thiemens, M.~M., Sprung, P., Fonseca, R.~O.~C., et al.\ 2019, Nature Geoscience, 12, 696. doi:10.1038/s41561-019-0398-3

\bibitem[Vokrouhlick{\'y} et al.(2015)]{2015aste.book..509V} Vokrouhlick{\'y}, D., Bottke, W.~F., Chesley, S.~R., et al.\ 2015, Asteroids IV, 509. doi:10.2458/azu\_uapress\_9780816532131-ch02

\bibitem[Vokrouhlick{\'y} et al.(2019)]{2019AJ....157..181V} Vokrouhlick{\'y}, D., Nesvorn{\'y}, D., \& Dones, L.\ 2019, \aj, 157, 181. doi:10.3847/1538-3881/ab13aa

\bibitem[Warner et al.(2020)]{2020GeoRL..4789607W} Warner, N.~H., Schuyler, A.~J., Rogers, A.~D., et al.\ 2020, \grl, 47, e89607. doi:10.1029/2020GL089607

\bibitem[Weidenschilling et al.(1997)]{1997Icar..128..429W} Weidenschilling, S.~J., Spaute, D., Davis, D.~R., et al.\ 1997, Icarus, 128, 429. doi:10.1006/icar.1997.5747

\bibitem[Werner et al.(2014)]{2014Sci...343.1343W} Werner, S.~C., Ody, A., \& Poulet, F.\ 2014, Science, 343, 1343. doi:10.1126/science.1247282

\bibitem[Wetherill(1975)]{1975LPSC....6.1539W} Wetherill, G.~W.\ 1975, Lunar and Planetary Science Conference Proceedings, 2, 1539

\bibitem[Wetherill(1990)]{1990AREPS..18..205W} Wetherill, G.~W.\ 1990, Annual Review of Earth and Planetary Sciences, 18, 205. doi:10.1146/annurev.ea.18.050190.001225

\bibitem[Wilhelms et al.(1987)]{1987ghm..book.....W} Wilhelms, D.~E., McCauley, J.~F., \& Trask, N.~J.\ 1987, Washington : U.S. G.P.O. ; Denver, CO (Federal Center, Box 25425, Denver 80225) : For sale by the Books and Open-file Reports Section, U.S. Geological Survey, 1987.

\bibitem[Williams \& Cieza(2011)]{2011ARA&A..49...67W} Williams, J.~P. \& Cieza, L.~A.\ 2011, \araa, 49, 67. doi:10.1146/annurev-astro-081710-102548

\bibitem[Wisdom \& Holman(1991)]{1991AJ....102.1528W} Wisdom, J. \& Holman, M.\ 1991, \aj, 102, 1528. doi:10.1086/115978

\bibitem[Wishard et al.(2022)]{2022DPS....5410203W} Wishard, C., Minton, D., Singh, J.\ 2022.\ Size-dependent Decay of Accretion Populations.\ AAS/Division for Planetary Sciences Meeting Abstracts.
  
\bibitem[Youdin \& Goodman(2005)]{2005ApJ...620..459Y} Youdin, A.~N. \& Goodman, J.\ 2005, \apj, 620, 459. doi:10.1086/426895

\bibitem[Zahnle et al.(2003)]{2003Icar..163..263Z} Zahnle, K., Schenk, P., Levison, H., et al.\ 2003, Icarus, 163, 263. doi:10.1016/S0019-1035(03)00048-4

\bibitem[Zellner and Delano(2015)]{2015GeCoA.161..203Z} Zellner, N.~E.~B., Delano, J.~W.\ 2015.\ $^{40}$Ar/$^{39}$Ar ages of lunar impact glasses: Relationships among Ar diffusivity, chemical composition, shape, and size.\ Geochimica et Cosmochimica Acta 161, 203–218. doi:10.1016/j.gca.2015.04.013

\bibitem[Zhang et al.(2019)]{2019JGRE..124.3205Z} Zhang, B., Lin, Y., Moser, D.~E., et al.\ 2019, Journal of Geophysical Research (Planets), 124, 3205. doi:10.1029/2019JE005992

\bibitem[Zhu et al.(2019)]{2019Natur.571..226Z} Zhu, M.-H., Artemieva, N., Morbidelli, A., et al.\ 2019a, \nat, 571, 226. doi:10.1038/s41586-019-1359-0

\bibitem[Zhu et al.(2019)]{2019JGRE..124.2117Z} Zhu, M.-H., W{\"u}nnemann, K., Potter, R.~W.~K., et al.\ 2019b, Journal of Geophysical Research (Planets), 124, 2117. doi:10.1029/2018JE005826

\bibitem[Zhu et al.(2021)]{2021NatAs...5.1286Z} Zhu, M.-H., Morbidelli, A., Neumann, W., et al.\ 2021, Nature Astronomy, 5, 1286. doi:10.1038/s41550-021-01475-0

\end{thebibliography}
\end{document}